\documentclass[twocolumn]{aastex631}

\newcommand\omicron{o}

\usepackage{CJK}
\usepackage{float}
\usepackage[absolute]{textpos}
\usepackage{amsmath}
\usepackage{mathrsfs}
\usepackage{txfonts}
\usepackage{bm}

\shorttitle{TMRT Orion KL Q-band Survey} 
\begin{document}
\begin{CJK*}{UTF8}{gbsn}
\title{A Q-band line survey towards Orion KL using the Tianma radio telescope}
\author[0000-0001-8315-4248]{Xunchuan Liu (刘 训川)}
\thanks{liuxunchuan@shao.ac.cn}
\affiliation{Shanghai Astronomical Observatory, Chinese Academy of Sciences, Shanghai 200030, PR China}
\author[0000-0002-5286-2564]{Tie Liu}
\thanks{liutie@shao.ac.cn}
\affiliation{Shanghai Astronomical Observatory, Chinese Academy of Sciences, Shanghai 200030, PR China}
\author[0000-0003-3540-8746]{Zhiqiang Shen}
\thanks{zshen@shao.ac.cn}
\affiliation{Shanghai Astronomical Observatory, Chinese Academy of Sciences, Shanghai 200030, PR China}
\author{Sheng-Li Qin}
\affiliation{Department of Astronomy, Yunnan University, Kunming, 650091, PR China}
\author[0000-0003-4506-3171]{Qiuyi Luo}
\affiliation{Shanghai Astronomical Observatory, Chinese Academy of Sciences, Shanghai 200030, PR China}
\author[0000-0002-8691-4588]{Yu Cheng}
\affiliation{National Astronomical Observatory of Japan, 2-21-1 Osawa, Mitaka, Tokyo, 181-8588, Japan}
\affiliation{Shanghai Astronomical Observatory, Chinese Academy of Sciences, Shanghai 200030, PR China}
\author[0000-0002-2826-1902]{Qilao Gu}
\affiliation{Shanghai Astronomical Observatory, Chinese Academy of Sciences, Shanghai 200030, PR China}
\author[0000-0002-1466-3484]{Tianwei Zhang}
\affiliation{I. Physikalisches Institut, Universit{\"a}t zu K{\"o}ln, Z{\"u}lpicher Stra{\ss}e 77, 50937 K{\"o}ln, Germany}
\author{Fengyao Zhu}
\affiliation{Center for Intelligent Computing Platforms, Zhejiang Laboratory, Hangzhou, 311100, PR China}
\author{Sheng-Yuan Liu}
\affiliation{Institute of Astronomy and Astrophysics, Academia Sinica, Roosevelt Road, Taipei 10617, Taiwan (R.O.C)}
\author{Xing Lu}
\affiliation{Shanghai Astronomical Observatory, Chinese Academy of Sciences, Shanghai 200030, PR China}
\author{Rongbing Zhao}
\affiliation{Shanghai Astronomical Observatory, Chinese Academy of Sciences, Shanghai 200030, PR China}
\author{Weiye Zhong}
\affiliation{Shanghai Astronomical Observatory, Chinese Academy of Sciences, Shanghai 200030, PR China}
\author{Yajun Wu}
\affiliation{Shanghai Astronomical Observatory, Chinese Academy of Sciences, Shanghai 200030, PR China}
\author[0000-0003-3520-6191]{Juan Li}
\affiliation{Shanghai Astronomical Observatory, Chinese Academy of Sciences, Shanghai 200030, PR China}
\author{Zhang Zhao}
\affiliation{Shanghai Astronomical Observatory, Chinese Academy of Sciences, Shanghai 200030, PR China}
\author{Jinqing Wang}
\affiliation{Shanghai Astronomical Observatory, Chinese Academy of Sciences, Shanghai 200030, PR China}
\author{Qinghui Liu}
\affiliation{Shanghai Astronomical Observatory, Chinese Academy of Sciences, Shanghai 200030, PR China}
\author{Bo Xia}
\affiliation{Shanghai Astronomical Observatory, Chinese Academy of Sciences, Shanghai 200030, PR China}
\author{Bin Li}
\affiliation{Shanghai Astronomical Observatory, Chinese Academy of Sciences, Shanghai 200030, PR China}
\author{Li Fu}
\affiliation{Shanghai Astronomical Observatory, Chinese Academy of Sciences, Shanghai 200030, PR China}
\author{Zhen Yan}
\affiliation{Shanghai Astronomical Observatory, Chinese Academy of Sciences, Shanghai 200030, PR China}
\author{Chao Zhang}
\affiliation{Shanghai Astronomical Observatory, Chinese Academy of Sciences, Shanghai 200030, PR China}
\author{Lingling Wang}
\affiliation{Shanghai Astronomical Observatory, Chinese Academy of Sciences, Shanghai 200030, PR China}
\author{Qian Ye}
\affiliation{Shanghai Astronomical Observatory, Chinese Academy of Sciences, Shanghai 200030, PR China}
\author[0000-0002-8149-8546]{Ken'ichi Tatematsu}
\affiliation{Nobeyama Radio Observatory, National Astronomical Observatory of Japan, National Institutes of Natural Sciences, 462-2 Nobeyama, Minamimaki, Minamisaku, Nagano 384-1305, Japan}
\author[0000-0003-3343-9645]{Hongli Liu}
\affiliation{Department of Astronomy, Yunnan University, Kunming, 650091, PR China}
\author{Hsien Shang}
\affiliation{Institute of Astronomy and Astrophysics, Academia Sinica, Roosevelt Road, Taipei 10617, Taiwan (R.O.C)}
\author[0000-0001-5950-1932]{Fengwei Xu}
\affiliation{Kavli Institute for Astronomy and Astrophysics, Peking University, 5 Yiheyuan Road, Haidian District, Beijing 100871, PR China}
\author{Chin-Fei Lee}
\affiliation{Institute of Astronomy and Astrophysics, Academia Sinica, Roosevelt Road, Taipei 10617, Taiwan (R.O.C)}
\author[0000-0002-5682-2906]{Chao Zhang}
\affiliation{Institute of Astronomy and Astrophysics, School of Mathematics and Physics, Anqing Normal University, Anqing, China}
\author[0000-0002-2338-4583]{Somnath Dutta}
\affiliation{Institute of Astronomy and Astrophysics, Academia Sinica, Roosevelt Road, Taipei 10617, Taiwan (R.O.C)}

\begin{abstract}
We have conducted a line survey towards Orion KL using the 
Q-band receiver of Tianma 65 m radio telescope (TMRT), 
covering 34.8--50 GHz with a velocity resolution between 0.79 km s$^{-1}$ and 0.55 km s$^{-1}$ respectively.
The observations reach a sensitivity on the level of 1-8 mK, proving 
that the TMRT is sensitive for conducting deep line surveys.
In total, 597 Gaussian features are extracted. 
Among them, 177 radio recombination lines (RRLs) are identified, including 126, 40 and 11 RRLs of hydrogen,
helium and carbon, with a 
maximum $\Delta n$ of 16, 7, and 3, respectively. 
The carbon RRLs are confirmed to originate from photodissociation regions
with a $V_{\rm LSR}\sim$9  km s$^{-1}$.
In addition, 371 molecular transitions of 53 molecular species are identified.
Twenty-one molecular species of this survey were not firmly detected in the Q band by  
\citet{2017A&A...605A..76R}, including
species such as
H$_2$CS, HCOOH, C$_2$H$_5$OH, H$_2^{13}$CO, 
H$_2$CCO, CH$_3$CHO, CH$_2$OCH$_2$, HCN $\varv_2=1$, and CH$_3$OCHO $\varv_t=1$.
In particular, the vibrationally excited states of  ethyl cyanide (C$_2$H$_5$CN $\varv$13/$\varv$21) 
are for the first time  firmly detected in the Q band.
NH$_3$ (15,15) and (16,16) are identified, and they are so far the highest  transitions  of the NH$_3$ inversion lines 
detected towards Orion KL. 
All the identified lines can be reproduced by a radiative transfer model. 
\end{abstract}

\section{Introduction} \label{sec:intro}
The spectral line survey is one of the best ways to study the physical and astrochemical
properties of astronomical objects.
An unbiased wide-band line survey  is generally time-consuming,
and thus usually focuses toward the most representative objects of its kind, such as the
Orion KL, IRC +10216, W51, TMC-1 and Sgr B2
\citep[e.g.,][]{1984A&A...130..227J,1993ApJS...86..211B,2004PASJ...56...69K,2013A&A...559A..47B,2017A&A...606A..74Z,2020ApJ...900L..10M,
2021A&A...652L...9C,2021A&A...645A..37T,2022A&A...658A..39P}.
Among those well studied objects, the Orion KL is probably the most classical
one. It is the closest high-mass star formation region 
from us \citep[$\sim$414 pc;][]{2007A&A...474..515M} with
particularly rich chemical and dynamic properties \citep[e.g.,][]{2014A&A...567A..95E}.
The Orion KL has been targeted by tens of line surveys  over the past few decades,
especially in the millimeter and submillimeter bands with frequency $\nu>70$ GHz \citep[e.g.,][]{1984A&A...130..227J,
1989ApJS...70..539T,1997ApJS..108..301S,Schilke_2001,2005ApJS..156..127C,2012pmcu.reptE...1P,2003A&A...407..589W,
2010A&A...517A..96T,2013A&A...556A.143E}.
In the millimeter and submillimeter bands, the spectrum of 
Orion KL is crowded with numerous molecular lines.

Emission of the rotational transitions of heavy species are  expected in lower frequency
bands thanks to their small rotational constants. Those low-frequency lines tend to be optically thinner
and not severely blended.
Another  benefit of lower frequency bands is that the intensities of radio recombination lines
(RRLs) are stronger and thus easier to be detected \citep{2002ASSL..282.....G}.  
However there are only a few line surveys towards Orion KL at lower frequency bands in 
contrast to the ample surveys at frequencies above 70 GHz, as mentioned earlier.

\citet{2015A&A...581A..48G} conducted a radio K-band ($\sim$1.3 cm) line survey covering 
the frequency range between 17.9 and 26.2 GHz using the Effelsberg 100 m telescope.
The K-band spectrum of Orion KL was found to be dominated by RRLs, which contribute
164 emission lines among the 261 detected ones.

Although the line surveys in Q band ($\sim$40 GHz, 7.5 mm) located between the RRL dominant centimeter bands
and the optically thick molecular lines dominant (sub)millimeter bands
could be very helpful in studying the RRLs and emission lines of complex organic molecules
(COMs) simultaneously, very few surveys have been published.
\citet{2009ApJ...691.1254G} conducted a line survey (from 42.3 to 43.6 GHz) using the GBT 100 m
optimized for the SiO maser emission of Orion KL.  
\citet{2017A&A...605A..76R} conducted a line survey (from 41.5 to 50 GHz) using the DSS-54 antenna 
with a diameter of 34 m. 
They modeled the Q-band spectrum of Orion KL combining their survey and other surveys.
The model predicted emission lines of organic molecules such as  H$_2^{13}$CO, CH$_2$OCH$_2$
and C$_2$H$_5$CN $\varv$13/$\varv$21, but they
were not detected or only marginally detected
limited by the aperture size, sensitivity and spectral resolution ($\sim$180 kHz) of
the survey of \citet{2017A&A...605A..76R}.
The carbon RRLs are usually blended with helium RRLs, and only several carbon RRLs were spectrally 
resolved by \citet{2017A&A...605A..76R}. 
Thus, a deeper  Q-band line survey towards the Orion KL with wider 
frequency coverage, better sensitivity, higher spatial and spectral resolution
is extremely valuable for complementing surveys at other bands 
for a comprehensive modeling of the physical and chemical properties of Orion KL. 

Employing the Tianma 65 m radio telescope (TMRT) of  Shanghai Astronomical Observatory,
we conducted a Q-band line survey covering 34.8--50 GHz towards Orion KL.
This is the first systematic line survey of the TMRT.
This survey reaches a sensitivity on the level of 1-8 mK with a frequency resolution of $\sim$90 kHz. 
In this work, we present preliminary results
of the TMRT Q-band survey towards Orion KL.
The paper is structured as follows: In Sect. \ref{secobs} we briefly
introduce the equipment 
and observations setup adopted by this survey, as well as the data reduction process.
In Sect. \ref{seclineid} we describe the procedure for the identification of lines.
In Sect. \ref{modelfit_sec} a simple radiative transfer model is fitted to
reproduce the observed Q-band emission lines of Orion KL, including both
RRLs and molecular lines.
In Sect. \ref{secdis}, we discuss RRLs and some individual species detected in this survey. 
Sect. \ref{secsummary} provides a summary.
  
\begin{figure}[!thp]
\includegraphics[width=0.9\linewidth]{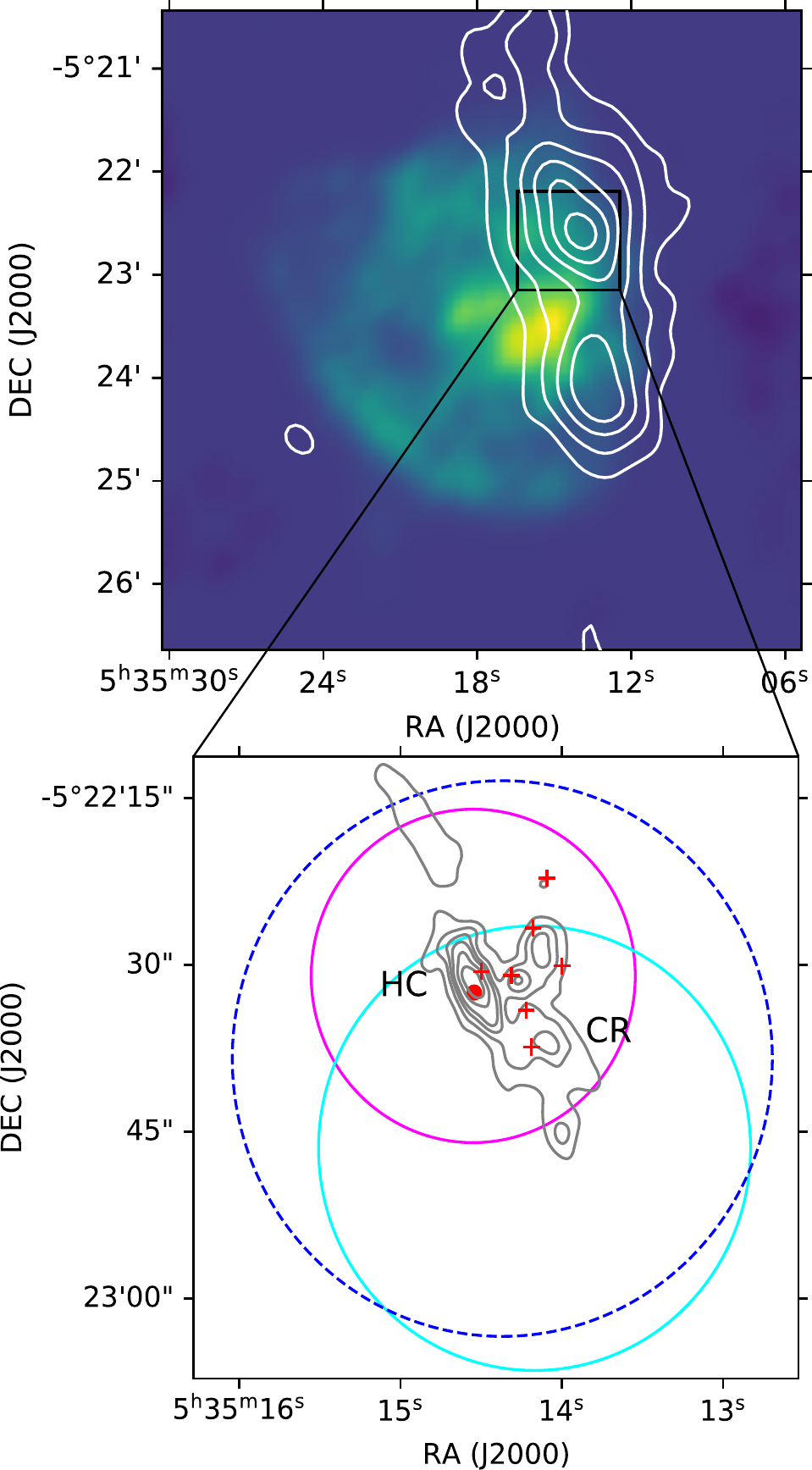}
\caption{Upper: Contours of SCUBA 850 $\mu$m dust emission \citep{2008ApJS..175..277D} 
overlaid on the 6 cm VLA continuum image.
Lower: Continuum map of Orion KL at 230 GHz from the ALMA-SV line survey.
The hot core region and the compact ridge are indicated by HC and CR, respectively.
The red crosses represent the infrared clumps \citep{2004AJ....128..363S}. 
The red dot marks HC(S) \citep{2013ApJ...770..142N}.
The purple, cyan, and blue circles represent the beams from this survey,  
\citet{2015A&A...581A..48G}, and  \citet{2017A&A...605A..76R}, respectively.
The blue circle is dashed since \citet{2017A&A...605A..76R} did not mentioned their targeting center,
and hence, only the size of the blue circle is meaningful.\label{zoom_inout}
}
\end{figure}

\section{Observation and data reduction} \label{secobs}

\begin{figure*}[!htp]
\includegraphics[width=0.99\linewidth]{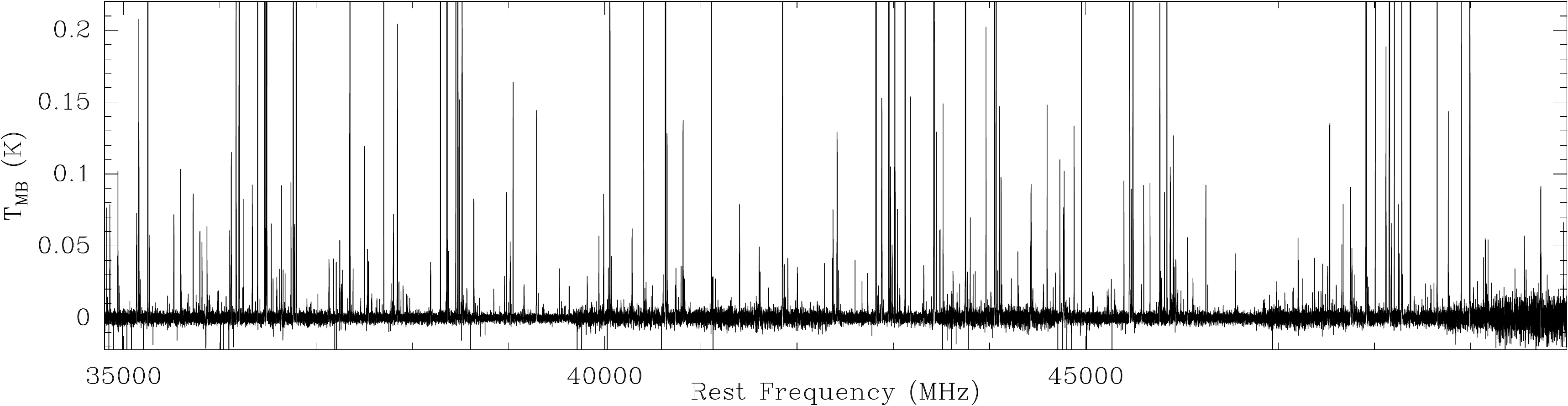}
\caption{The overview of the Orion KL spectrum in Q band observed by the TMRT.\label{fig_overall_spe}}
\end{figure*}
\begin{figure}[!htb]
\includegraphics[width=0.95\linewidth]{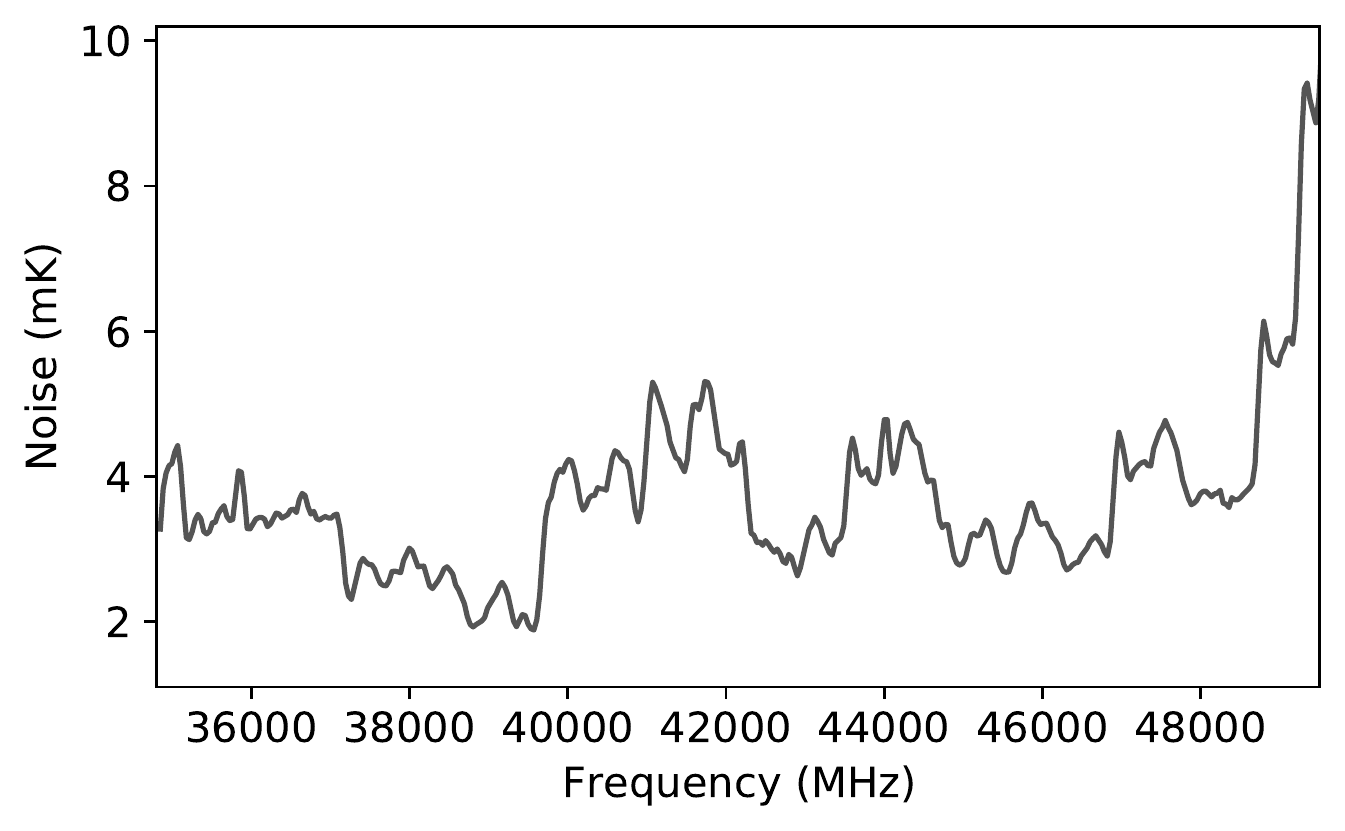}
\caption{The rms noise of the spectrum. The line features and bad channels (Sect. \ref{sect_linefeature}) have been masked out
before calculating the rms noise. \label{noise_freq}}
\end{figure}
\subsection{TMRT}
The observations were carried out using the TianMa Radio Telescope (TMRT) of the Shanghai
Astronomical Observatory\footnote{\url{http://english.shao.cas.cn/sbysys/}}. 
The TMRT is a 65-m diameter fully steerable radio telescope located in a western 
suburb of Shanghai, China. Receivers from  L to Q band are available, covering a frequency
range of 1--50 GHz.
The Q-band receiver provides the highest frequency coverage of Tianma 65 m.
Before this survey, the Q-band receiver of Tianma 65 m 
was not fully used for single-dish scientific observations,
compared to the receivers of C/Ku/K bands \citep{2016ApJ...824..136L,2017A&A...606A..74Z,2019MNRAS.488..495W,
2019A&A...627A.162W,2021SCPMA..6479511X,2022A&A...658A.140L}.

The Q-band receiver is a two-beam dual-polarization (LCP and RCP) cryogenic receiver, covering a
frequency range of 35--50 GHz \citep{2018RAA....18...44Z}.
The two beams can not work simultaneously at present for spectral line observation limited by the backend.
Only beam 2 was employed during our observations.
The receiver noise temperatures are roughly 30--40 K, and the
system temperature ranges from 60 K to 150 K depending on the
frequency and weather conditions \citep{2018RAA....18...44Z}. The full-width at half-maximum of the 
primary beam is $\sim$30\arcsec~at 40 GHz.
Pointing was conducted every two hours.
The pointing accuracy is better than 5 arcsec.

For spectral line observations,
an FPGA-based spectrometer based upon the design of the Versatile GBT Astronomical Spectrometer (VEGAS) 
was employed as the Digital backend system \citep[DIBAS;][]{2012AAS...21944610B}.
Twenty-nine observing modes (Mode 1--29) with different frequency bandwidths and resolutions are available. 
For our observations, Mode 2 was adopted, which provides a set of independent frequency banks with a bandwidth 
of 1500 MHz each. Three banks were designed for TMRT but only two of them are 
available at present. Each frequency bank provides data of both left-hand circular polarization (LCP)
and right-hand circular polarization (RCP).
For each polarization, a frequency bank has 16384 channels, 
corresponding to a frequency resolution of 91.553 kHz ($\sim$0.69 km s$^{-1}$ at 40 GHz).

For calibration, the signal of noise diodes was injected lasting for one second within each two-second period.
The temperatures of the noise diodes are $\sim$18 K and $\sim$12 K for the LCP
and RCP, respectively.
The aperture efficiency ($\eta_A$) at Q band has a dependence on elevation (el) following \citep{2017AcASn..58...37W} 
\begin{equation}
\eta_A = p_0+p_1{\rm el}+p_2{\rm el^2}+p_3{\rm el^3},
\end{equation}
where $p_0$, $p_1$, $p_2$ and $p_3$ are 6.33$\times$10$^{-2}$,
3.47$\times$10$^{-3}$, 4.11$\times$10$^{-4}$, and -5.72$\times$10$^{-6}$,
respectively and el is in degrees.
The efficiency decreases significantly for small and large elevation because of gravity deformation.
The telescope has an active surface control utilizing actuators to compensate for gravity deformation
in the main reflector during observations \citep{2018ITAP...66.2044D}.
It makes the Q-band aperture efficiency constant (0.5$\pm$0.1) for elevation within 15\degr--80\degr~\citep{2018RAA....18...44Z}.
Under the typical weather condition of TMRT in winter with an air pressure of 1000 mbar and a water vapor density of 8 g cm$^{-3}$,
the zenith atmospheric opacity ranges from 0.07 to 0.35 in the Q band \citep[35--50 GHz;][]{2017AcASn..58...37W}.
The main beam efficiency is $\sim$0.60 and depends on the 
elevation and frequency.
Calibration uncertainties are estimated to be
within 20\%.

\begin{table*}[!thb]
\centering
\caption{Detected molecular species of this survey $^{(1)}$.\label{allspeciestab}}
\begin{tabular}{lllllll}
\hline
\textcolor{blue}{CS}        & $^{34}$SO & \textcolor{blue}{HC$^{13}$CCN} & CH$_3$CN $\varv_t=1$ & \textcolor{red}{NH$_2$CHO} & H$_2^{13}$CO & \textcolor{blue}{SiO}\\
\textcolor{blue}{$^{13}$CS} & \textcolor{blue}{OCS} & \textcolor{blue}{HCC$^{13}$CN} & HCN $\varv_2=1$ & \textcolor{blue}{CH$_3$OH} & H$_2$CCO & \textcolor{blue}{SiO $\varv=1$}\\
\textcolor{blue}{$^{33}$CS} & O$^{13}$CS & HC$_3$N $\varv_6=1$ & \textcolor{blue}{C$_2$H$_3$CN} & \textcolor{blue}{$^{13}$CH$_3$OH} & CH$_3$CHO & \textcolor{blue}{SiO $\varv=2$}\\
\textcolor{blue}{$^{34}$CS} & \textcolor{blue}{OC$^{34}$S} & \textcolor{blue}{HC$_3$N $\varv_7=1$} & \textcolor{blue}{C$_2$H$_5$CN} & \textcolor{blue}{A-CH$_3$OH $\varv_t=1$} & \textcolor{blue}{CH$_3$OCHO} & \textcolor{blue}{$^{29}$SiO}\\
CCS                         & \textcolor{blue}{SO$_2$} & HC$_3$N $\varv_7=2$ & CH$_2$H$_5$CN $\varv_{13}/\varv_{21}$ & \textcolor{blue}{E-CH$_3$OH $\varv_t=1$} & \textcolor{blue}{CH$_3$OCHO $\varv_t=1$} & \textcolor{blue}{$^{30}$SiO}\\
HCS$^+$                     & \textcolor{blue}{$^{34}$SO$_2$} & \textcolor{blue}{HC$_5$N} & NH$_3$ & C$_2$H$_5$OH & CH$_2$OCH$_2$ & \\
H$_2$CS & \textcolor{blue}{HC$_3$N}           & \textcolor{red}{CH$_3$NH$_2$} & \textcolor{blue}{NH$_2$D} & HCOOH & \textcolor{blue}{CH$_3$OCH$_3$} & \\
SO                          & \textcolor{blue}{H$^{13}$CCCN}           & CH$_3$CN & \textcolor{blue}{HNCO} & \textcolor{blue}{H$_2$CO} & CH$_3$COCH$_3$ & \\
\hline
\end{tabular}\\
{\raggedright
$^{(1)}$ The species in blue have transitions detected and spectrally resolved by \citet{2017A&A...605A..76R}. 
Those in red means their transitions are marginally
detected or highly blended. 
}
\end{table*}

\subsection{Observation}
Our observations towards Orion KL were conducted  during March 9th to 29th, 2022. 
The targeted position is RA(J2000)=05:35:14.55, DEC(J2000)=$-$05:22:31.0 (Fig. \ref{zoom_inout}).
Position switching observation mode was adopted, with the off points 0.5\degr~away (in azimuth direction) from the target,
and integrating 2 minutes in each position (on/off).
Spectra of failed observations (which are wrongly calibrated 
with abnormal system temperatures larger than 1000 K) were discarded.
For each frequency bank, its  frequency coverage  in sky frequency scale was fixed.
The frequency of the local oscillator (LO) did not change during the observation for each frequency setup.  
The spectrum of each on-off repeat was corrected from the topocentric frame to the frame of local standard of rest (LSR)
during data processing.  
Banks were shifted in frequency to cover 34.5--50 GHz, but always leaving an overlap of $>$300 MHz between two adjacent configurations.
For each frequency setup, a telescope time of 3--10 hours was consumed, 
depending on the weather conditions.

\subsection{Data reduction} \label{sec_dr}
Combining all scans of observations from all frequency setups, 
a full frequency coverage between 34.8  and 50 GHz was achieved.
The spectra were then chopped into segments of 100 MHz in bandwidth, and
Gildas/CLASS\footnote{\url{https://www.iram.fr/IRAMFR/GILDAS/}} 
was adopted to fit and subtract the spectral baselines
for each segment.
We  combined all the segments weighted by their noise levels to obtain
a Q-band spectrum of Orion KL. 
We further converted the frequency of the spectrum 
from the frame of LSR to the rest frame of Orion KL assuming a systematic velocity of Orion KL ($V_{\rm LSR}$)  of 6 km/s, through
\begin{equation}
f_{\rm final} = f_{\rm LSR}\left(1+\frac{V_{\rm LSR}}{c}\right).
\end{equation}
Here, $c$ is the light speed,  $f_{\rm LSR}$ is the spectral frequency in  the frame of LSR, and 
$f_{\rm final}$ is the frequency of the final spectrum.
The frequencies related to Orion KL are always referred to $f_{\rm final}$ throughout this work.
The final spectrum is shown in Fig. \ref{fig_overall_spe}.
The rms noise of the final spectrum (in $T_{\rm MB}$ scale with a frequency resolution of 91 kHz) 
ranges from 1.8 mK to 8 mK, with a mean value of 4.2 mK and a
standard deviation of 1.2 mK (Fig. \ref{noise_freq}).

\begin{figure*}
\centering
\includegraphics[width=0.99\linewidth]{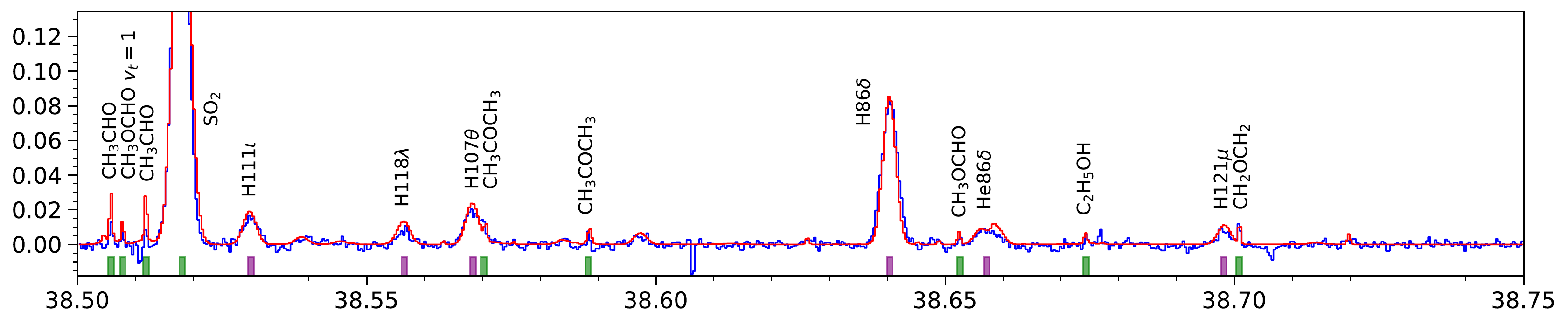}
\includegraphics[width=0.99\linewidth]{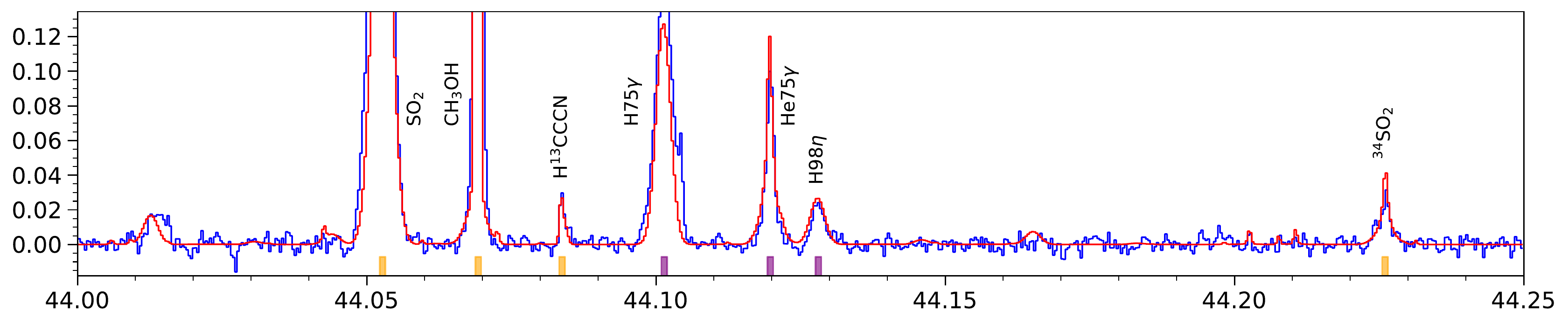}
\includegraphics[width=0.99\linewidth]{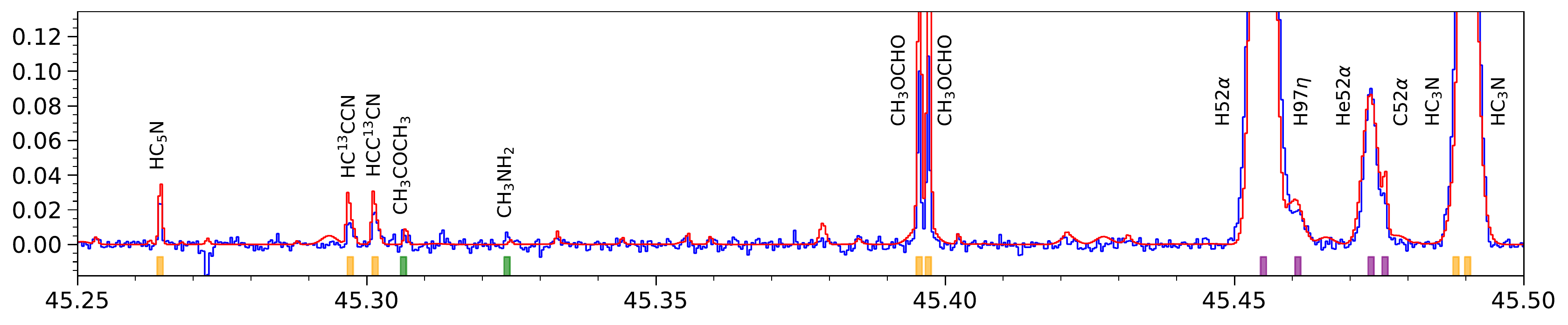}
\includegraphics[width=0.99\linewidth]{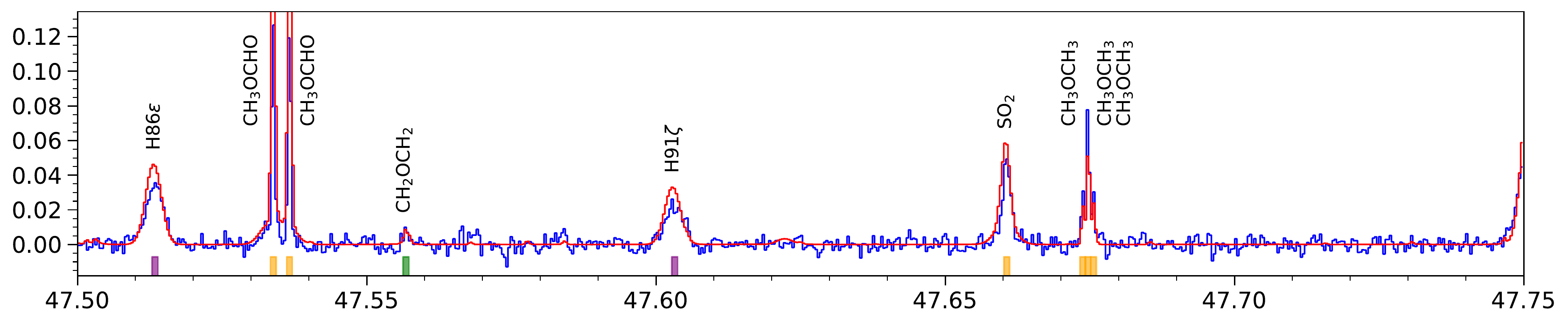}
\caption{The example zoom-in spectra of the Orion KL. The complete figure set (61 images) covering the
34.8--50 GHz is available in the online journal. 
The blue line is the Orion KL spectrum observed by the TMRT 65 m,
which has been smoothed to have a frequency resolution of 364 kHz ($\sim$2.8 km s$^{-1}$ at 40 GHz). 
The red line represents the results of model fitting (Sect. \ref{modelfit_sec}). 
The purple strips denote the detected RRLs. The yellow strips 
denote the molecular lines which have also been detected and resolved by \citet{2017A&A...605A..76R}. The green strips denotes the 
molecular lines detected by TMRT 65 m but have not been detected by \citet{2017A&A...605A..76R}.
The red strips denote the lines of C$_2$H$_5$CN $\varv$13/$\varv$21. The gray strips mark the U lines.
The lines of SiO and its isotopologues are not modeled.
The $x$ axis is rest frequency in  unit of GHz (with a Doppler correction applied adopting a $V_{\rm LSR}$ of 6 km s$^{-1}$). The $y$ axis is $T_{\rm MB}$ in units of K.
\label{specset_example}}
\end{figure*}

The example zoom-in spectra chopped into subbands of 250 MHz is displayed in Fig. \ref{specset_example}.
The complete figure set (61 images) is shown in Fig. \ref{allspectra}.
Fig. \ref{fig_comparerizzo} shows a fraction of the final spectrum from this survey and from 
\citet{2017A&A...605A..76R} for comparison.

\begin{table*}
    \centering
    \caption{Model parameters \label{model_pars_table}}
    \begin{tabular}{llllllclllllll}
    \hline
    \multicolumn{1}{|l}{Species$^{(1)}$} & size$^{(2)}$ & $T_{\rm ex}$ & $N_{\rm tot}^{(3)}$ & $\Delta V$ & $V_{\rm lsr}$ & &
    Species & size & $T_{\rm ex}$ & $N_{\rm tot}$ & $\Delta V$ & \multicolumn{1}{l|}{$V_{\rm lsr}$} \\
    \multicolumn{1}{|l}{}       & (\arcsec)& (K)          & (cm$^{-2}$)   & (km s$^{-1}$) &(km s$^{-1}$)& &
           & (\arcsec)& (K)          & (cm$^{-2}$)   & (km s$^{-1}$) & \multicolumn{1}{l|}{(km s$^{-1}$)}\\
    \hline
\multicolumn{1}{|l}{CS}                  & 30       & 100      &      9.6e+14 & 15.0     & \multicolumn{1}{l|}{7.0     }  &    &
\multicolumn{1}{|l}{$^{34}$SO$_2$}       & 30       & 50       &      1.5e+14 & 4.0      & \multicolumn{1}{l|}{7.0     }  \\
\multicolumn{1}{|l}{}                    & 10       & 100      &      1.2e+16 & 4.0      & \multicolumn{1}{l|}{8.5     }  &    &
\multicolumn{1}{|l}{}                    & 30       & 200      &      6.0e+14 & 10.0     & \multicolumn{1}{l|}{7.0     }  \\
\multicolumn{1}{|l}{C$^{34}$S}           & 30       & 100      &      6.0e+13 & 15.0     & \multicolumn{1}{l|}{7.0     }  &    &
\multicolumn{1}{|l}{}                    & 30       & 100      &      6.0e+14 & 25.0     & \multicolumn{1}{l|}{7.5     }  \\
\multicolumn{1}{|l}{}                    & 10       & 100      &      8.4e+14 & 4.0      & \multicolumn{1}{l|}{8.5     }  &    &
\multicolumn{1}{|l}{HCN $\varv_2=1$}     & 10       & 200      &      1.8e+17 & 4.0      & \multicolumn{1}{l|}{7.0     }  \\
\multicolumn{1}{|l}{$^{13}$CS}           & 30       & 100      &      2.4e+13 & 15.0     & \multicolumn{1}{l|}{7.0     }  &    &
\multicolumn{1}{|l}{HC$_3$N}             & 15       & 100      &      1.9e+14 & 3.0      & \multicolumn{1}{l|}{9.0     }  \\
\multicolumn{1}{|l}{}                    & 10       & 100      &      3.0e+14 & 4.0      & \multicolumn{1}{l|}{8.5     }  &    &
\multicolumn{1}{|l}{}                    & 10       & 100      &      3.0e+14 & 7.0      & \multicolumn{1}{l|}{5.5     }  \\
\multicolumn{1}{|l}{C$^{33}$S}           & 10       & 100      &      6.0e+13 & 4.0      & \multicolumn{1}{l|}{12.0    }  &    &
\multicolumn{1}{|l}{}                    & 10       & 100      &      6.0e+14 & 15.0     & \multicolumn{1}{l|}{5.5     }  \\
\multicolumn{1}{|l}{HCS$^+$}             & 15       & 100      &      1.7e+13 & 2.0      & \multicolumn{1}{l|}{10.0    }  &    &
\multicolumn{1}{|l}{}                    & 20       & 100      &      1.2e+14 & 25.0     & \multicolumn{1}{l|}{6.0     }  \\
\multicolumn{1}{|l}{H$_2$CS}             & 15       & 100      &      1.1e+15 & 2.0      & \multicolumn{1}{l|}{8.5     }  &    &
\multicolumn{1}{|l}{H$^{13}$CCCN}        & 15       & 100      &      1.4e+13 & 4.0      & \multicolumn{1}{l|}{10.0    }  \\
\multicolumn{1}{|l}{SO}                  & 10       & 100      &      1.2e+17 & 15.0     & \multicolumn{1}{l|}{6.0     }  &    &
\multicolumn{1}{|l}{}                    & 10       & 100      &      2.4e+13 & 7.0      & \multicolumn{1}{l|}{5.5     }  \\
\multicolumn{1}{|l}{}                    & 10       & 100      &      2.2e+17 & 25.0     & \multicolumn{1}{l|}{9.0     }  &    &
\multicolumn{1}{|l}{HC$^{13}$CCN}        & 15       & 100      &      1.4e+13 & 4.0      & \multicolumn{1}{l|}{10.0    }  \\
\multicolumn{1}{|l}{$^{34}$SO}           & 10       & 100      &      4.8e+15 & 15.0     & \multicolumn{1}{l|}{6.0     }  &    &
\multicolumn{1}{|l}{}                    & 10       & 100      &      2.4e+13 & 7.0      & \multicolumn{1}{l|}{5.5     }  \\
\multicolumn{1}{|l}{}                    & 10       & 100      &      9.6e+15 & 25.0     & \multicolumn{1}{l|}{9.0     }  &    &
\multicolumn{1}{|l}{HCC$^{13}$CN}        & 15       & 100      &      1.4e+13 & 4.0      & \multicolumn{1}{l|}{10.0    }  \\
\multicolumn{1}{|l}{OCS}                 & 30       & 100      &      1.2e+15 & 3.0      & \multicolumn{1}{l|}{8.0     }  &    &
\multicolumn{1}{|l}{}                    & 10       & 100      &      2.4e+13 & 7.0      & \multicolumn{1}{l|}{5.5     }  \\
\multicolumn{1}{|l}{}                    & 30       & 50       &      7.2e+14 & 10.0     & \multicolumn{1}{l|}{6.0     }  &    &
\multicolumn{1}{|l}{HC$_3$N $\varv_6=1$} & 10       & 150      &      2.4e+15 & 7.0      & \multicolumn{1}{l|}{5.5     }  \\
\multicolumn{1}{|l}{}                    & 30       & 100      &      7.2e+14 & 25.0     & \multicolumn{1}{l|}{5.0     }  &    &
\multicolumn{1}{|l}{HC$_3$N $\varv_7=1$} & 10       & 150      &      2.4e+15 & 7.0      & \multicolumn{1}{l|}{5.5     }  \\
\multicolumn{1}{|l}{OC$^{34}$S}          & 30       & 100      &      6.0e+13 & 3.0      & \multicolumn{1}{l|}{8.0     }  &    &
\multicolumn{1}{|l}{}                    & 10       & 100      &      3.6e+15 & 25.0     & \multicolumn{1}{l|}{6.0     }  \\
\multicolumn{1}{|l}{}                    & 30       & 50       &      3.6e+13 & 10.0     & \multicolumn{1}{l|}{6.0     }  &    &
\multicolumn{1}{|l}{HC$_3$N $\varv_7=2$} & 10       & 150      &      2.4e+15 & 7.0      & \multicolumn{1}{l|}{5.5     }  \\
\multicolumn{1}{|l}{O$^{13}$CS}          & 30       & 100      &      2.4e+13 & 3.0      & \multicolumn{1}{l|}{8.0     }  &    &
\multicolumn{1}{|l}{HC$_5$N}             & 10       & 100      &      3.6e+13 & 4.0      & \multicolumn{1}{l|}{8.5     }  \\
\multicolumn{1}{|l}{}                    & 30       & 50       &      2.4e+13 & 10.0     & \multicolumn{1}{l|}{6.0     }  &    &
\multicolumn{1}{|l}{CH$_3$CN}            & 10       & 200      &      4.8e+15 & 6.0      & \multicolumn{1}{l|}{7.0     }  \\
\multicolumn{1}{|l}{SO$_2$}              & 30       & 50       &      3.0e+15 & 4.0      & \multicolumn{1}{l|}{7.0     }  &    &
\multicolumn{1}{|l}{}                    & 10       & 100      &      6.0e+15 & 20.0     & \multicolumn{1}{l|}{8.0     }  \\
\multicolumn{1}{|l}{}                    & 30       & 200      &      6.0e+15 & 10.0     & \multicolumn{1}{l|}{7.0     }  &    &
\multicolumn{1}{|l}{CH$_3$CN $\varv_t=1$} & 10       & 200      &      4.8e+15 & 6.0      & \multicolumn{1}{l|}{6.0     }  \\
\multicolumn{1}{|l}{}                    & 30       & 100      &      1.9e+16 & 25.0     & \multicolumn{1}{l|}{7.5     }  &    &
\multicolumn{1}{|l}{C$_2$H$_3$CN}        & 5        & 320      &      3.6e+14 & 6.0      & \multicolumn{1}{l|}{5.0     }  \\
\hline
\end{tabular}\\
\vspace{1ex}
{\raggedright $^{(1)}$ The species  in red means their transitions are marginally
detected or highly blended. 
The emission of SiO and its isotopologues is not modeled. 
See Sect. \ref{sec_c2h5cn_vib} for the fitting of C$_2$H$_5$CN $\varv$13/$\varv$21.\\
$^{(2)}$ The emission source size and $T_{\rm ex}$ for each spectral component are fixed 
(Sect. \ref{sect_model_mol}).
The spectral modeling of this work is mainly used for line identification, 
and  particular caution should be taken if the modeled parameters (e.g. $N_{\rm tot}$) 
listed in Table \ref{model_pars_table} are 
used for comparison with results of other work.\\
$^{(3)}$ For species in vibrational state, to derive the column density, 
the partition function of the corresponding molecule accounting 
for all vibrational states is adopted. Thus  the column density should be interpreted as the
column density of the corresponding molecule. 
If another emission source size is adopted, the column densities could be 
recalculated through multiplying the values listed here by a factor of $\rm (size/size^{new})^2$. 
\par}
\vspace{0.5em}
{\centering \textbf{Table \thetable} {\it continued}}
\end{table*}

\begin{table*}
\centering
{\centering \textbf{Table \thetable} {\it (continued)}}\\
\vspace{-0.5em}
\begin{tabular}{llllllclllllll}
\hline
\multicolumn{1}{|l}{Species} & size & $T_{\rm ex}$ & $N_{\rm tot}$ & $\Delta V$ & $V_{\rm lsr}$ & &
Species & size & $T_{\rm ex}$ & $N_{\rm tot}$ & $\Delta V$ & \multicolumn{1}{l|}{$V_{\rm lsr}$} \\
\multicolumn{1}{|l}{}       & (\arcsec)& (K)          & (cm$^{-2}$)   & (km s$^{-1}$) &(km s$^{-1}$)& &
       & (\arcsec)& (K)          & (cm$^{-2}$)   & (km s$^{-1}$) & \multicolumn{1}{l|}{(km s$^{-1}$)}\\
\hline
\multicolumn{1}{|l}{}                    & 10       & 100      &      1.2e+14 & 6.0      & \multicolumn{1}{l|}{5.0     }  &    &
\multicolumn{1}{|l}{CH$_3$CHO}           & 15       & 50       &      2.4e+14 & 3.0      & \multicolumn{1}{l|}{8.0     }  \\
\multicolumn{1}{|l}{}                    & 5        & 200      &      1.1e+14 & 20.0     & \multicolumn{1}{l|}{3.0     }  &    &
\multicolumn{1}{|l}{}                    & 30       & 150      &      2.4e+14 & 25.0     & \multicolumn{1}{l|}{9.0     }  \\
\multicolumn{1}{|l}{}                    & 10       & 90       &      1.6e+14 & 20.0     & \multicolumn{1}{l|}{3.0     }  &    &
\multicolumn{1}{|l}{CH$_3$OCHO}          & 30       & 60       &      4.8e+14 & 4.0      & \multicolumn{1}{l|}{8.0     }  \\
\multicolumn{1}{|l}{C$_2$H$_5$CN}        & 5        & 275      &      1.9e+16 & 5.0      & \multicolumn{1}{l|}{5.5     }  &    &
\multicolumn{1}{|l}{}                    & 30       & 150      &      1.9e+15 & 25.0     & \multicolumn{1}{l|}{9.0     }  \\
\multicolumn{1}{|l}{}                    & 10       & 110      &      1.4e+15 & 13.0     & \multicolumn{1}{l|}{4.0     }  &    &
\multicolumn{1}{|l}{}                    & 15       & 110      &      1.7e+16 & 4.0      & \multicolumn{1}{l|}{7.5     }  \\
\multicolumn{1}{|l}{}                    & 25       & 65       &      3.0e+14 & 20.0     & \multicolumn{1}{l|}{4.0     }  &    &
\multicolumn{1}{|l}{}                    & 10       & 300      &      2.4e+16 & 4.0      & \multicolumn{1}{l|}{7.5     }  \\
\multicolumn{1}{|l}{\textcolor{red}{CH$_3$NH$_2$}} & 15       & 100      &      2.4e+14 & 4.0      & \multicolumn{1}{l|}{6.0     }  &    &
\multicolumn{1}{|l}{}                    & 10       & 250      &      7.7e+15 & 10.0     & \multicolumn{1}{l|}{5.5     }  \\
\multicolumn{1}{|l}{CH$_3$OH}            & 30       & 50       &      2.4e+16 & 4.0      & \multicolumn{1}{l|}{8.0     }  &    &
\multicolumn{1}{|l}{CH$_3$OCHO $\varv_t=1$} & 15       & 100      &      1.1e+16 & 3.0      & \multicolumn{1}{l|}{8.0     }  \\
\multicolumn{1}{|l}{}                    & 15       & 110      &      2.4e+17 & 4.0      & \multicolumn{1}{l|}{7.5     }  &    &
\multicolumn{1}{|l}{CH$_2$OCH$_2$}       & 15       & 50       &      6.0e+13 & 3.0      & \multicolumn{1}{l|}{7.5     }  \\
\multicolumn{1}{|l}{}                    & 30       & 150      &      2.4e+15 & 25.0     & \multicolumn{1}{l|}{9.0     }  &    &
\multicolumn{1}{|l}{}                    & 15       & 50       &      1.2e+13 & 1.5      & \multicolumn{1}{l|}{7.5     }  \\
\multicolumn{1}{|l}{$^{13}$CH$_3$OH}     & 30       & 50       &      2.4e+14 & 4.0      & \multicolumn{1}{l|}{8.0     }  &    &
\multicolumn{1}{|l}{CH$_3$OCH$_3$}       & 15       & 100      &      1.4e+16 & 3.0      & \multicolumn{1}{l|}{7.5     }  \\
\multicolumn{1}{|l}{}                    & 15       & 110      &      2.4e+15 & 4.0      & \multicolumn{1}{l|}{7.5     }  &    &
\multicolumn{1}{|l}{CH$_3$COCH$_3$}      & 10       & 100      &      1.8e+15 & 4.0      & \multicolumn{1}{l|}{5.5     }  \\
\multicolumn{1}{|l}{A-CH$_3$OH $\varv_t=1$} & 15       & 110      &      1.2e+17 & 4.0      & \multicolumn{1}{l|}{7.5     }  &    &
\multicolumn{1}{|l}{HNCO}                & 30       & 60       &      6.0e+13 & 4.0      & \multicolumn{1}{l|}{9.0     }  \\
\multicolumn{1}{|l}{E-CH$_3$OH $\varv_t=1$} & 15       & 110      &      1.2e+17 & 4.0      & \multicolumn{1}{l|}{7.0     }  &    &
\multicolumn{1}{|l}{}                    & 30       & 125      &      3.0e+14 & 25.0     & \multicolumn{1}{l|}{6.0     }  \\
\multicolumn{1}{|l}{C$_2$H$_5$OH}        & 15       & 60       &      7.2e+14 & 4.0      & \multicolumn{1}{l|}{8.0     }  &    &
\multicolumn{1}{|l}{}                    & 15       & 110      &      4.8e+14 & 4.0      & \multicolumn{1}{l|}{7.5     }  \\
\multicolumn{1}{|l}{HCOOH}               & 10       & 50       &      3.0e+14 & 3.0      & \multicolumn{1}{l|}{7.5     }  &    &
\multicolumn{1}{|l}{}                    & 10       & 225      &      8.4e+14 & 10.0     & \multicolumn{1}{l|}{5.5     }  \\
\multicolumn{1}{|l}{H$_2$CO}             & 30       & 50       &      6.0e+14 & 25.0     & \multicolumn{1}{l|}{6.0     }  &    &
\multicolumn{1}{|l}{}                    & 5        & 300      &      6.0e+15 & 5.0      & \multicolumn{1}{l|}{5.5     }  \\
\multicolumn{1}{|l}{}                    & 15       & 50       &      6.0e+15 & 3.0      & \multicolumn{1}{l|}{8.0     }  &    &
\multicolumn{1}{|l}{\textcolor{red}{NH$_2$CHO}} & 10       & 100      &      1.2e+14 & 3.0      & \multicolumn{1}{l|}{7.0     }  \\
\multicolumn{1}{|l}{}                    & 10       & 50       &      1.2e+15 & 10.0     & \multicolumn{1}{l|}{5.5     }  &    &
\multicolumn{1}{|l}{NH$_3$}              & 10       & 400      &      2.0e+16 & 30.0     & \multicolumn{1}{l|}{8.0     }  \\
\multicolumn{1}{|l}{H$_2^{13}$CO}        & 15       & 50       &      1.8e+14 & 3.5      & \multicolumn{1}{l|}{9.5     }  &    &
\multicolumn{1}{|l}{}                    & 10       & 300      &      1.2e+16 & 8.0      & \multicolumn{1}{l|}{6.0     }  \\
\multicolumn{1}{|l}{H$_2$CCO}            & 10       & 100      &      1.8e+15 & 3.0      & \multicolumn{1}{l|}{8.0     }  &    &
\multicolumn{1}{|l}{NH$_2$D}             & 10       & 100      &      6.0e+14 & 8.0      & \multicolumn{1}{l|}{6.0     }  \\
\hline
\end{tabular}\\
\end{table*}

\section{Line identification} \label{seclineid}
\subsection{Extract possible emission features} \label{sect_linefeature}
The spectrum was visually checked for a preliminary identification of  bad channels and channels
containing possible line features. The bad channels were masked out.
Gaussian fittings were then applied to the possible line features one by one. 
For strong lines with obvious non-Gaussian shapes or multiple blended emission components,
multiple Gaussian fittings  were applied to approach their line profiles.
In total, 597 Gaussian components were extracted as listed in Tables \ref{linelist} and \ref{linelist_rrls}
in Appendix. Table \ref{linelist} lists the molecular lines, and Table \ref{linelist_rrls} lists the
RRLs which are not blended with molecular lines.

\subsection{Identification of RRLs}
The radio recombination lines (RRLs) of hydrogen (H) and Helium (He) are from the H{\sc ii} regions,
with typical linewidths $>$10 km s$^{-1}$ \citep{2002ASSL..282.....G}.
The emission features with line widths larger than 10 km s$^{-1}$ were firstly labeled as
candidates of RRLs) of H and He.
Since the RRLs of carbon (C) mainly originate from the photon-dissociation regions (PDRs) between the M42 and the Orion Bar
\citep[e.g.,][]{2015A&A...575A..82C,2017A&A...605A..76R}, 
the constraint of line width was not applied to identifying RRLs of carbon.
Emission features close to helium RRLs were marked as carbon RRL candidates.
We then crossmatched those RRL candidates  with the rest frequencies of RRLs of H, He, and C, which can be calculated
through
\begin{equation}
\nu^{\rm RRL}_{\rm rest}(n+\Delta n,n) = \nu_0^{\rm RRL}\left(\frac{1}{n^2}-\frac{1}{(n+\Delta n)^2}\right)\ \ {\rm MHz},
\end{equation} 
with $\nu_0^{\rm RRL}$ adopted as $3.28805129\times10^9$, $3.28939118\times10^9$,
and $3.28969187\times10^9$ for H, He, and C, respectively \citep{2002ASSL..282.....G}.
For a RRL candidate which is not blended with strong lines of molecules (e.g.,
SiO, SO, HC$_3$N, CH$_3$OH, CH$_3$OCH$_3$, and C$_2$H$_5$CN), it was assigned to a specific RRL transition
if only one RRL transition can be found
within 20 km s$^{-1}$ of it. 
If it is blended, multiple Gaussian fitting was carefully
reconducted to separate the contributions of blended components. 
We marked an RRL candidate as an unresolved blended line if it 
is highly blended and inseparable by Gaussian decomposition.

In total, 177 recombination lines are matched, including 
126, 40 and 11 lines corresponding to H, He and, C, respectively.
Among them, 39 are blended lines which can not be well resolved and the other 138 can be spectrally resolved.
Basic parameters of the matched RRL lines, including the observed frequency ($f_{\rm obs}$), 
the name of the RRL transition, the rest frequency ($f_{\rm rest}$), and the Gaussian fitting results
(the integrated intensity $\int T_{\rm MB} dV$, the line width $\Delta V$ and the peak intensity $T^{\rm peak}_{\rm MB}$), 
are listed in Table \ref{linelist}.

\subsection{Identification of molecular lines} \label{Sect_idmol}
To identify the emission of molecular lines, 
we consulted
the frequencies of  molecular transitions
from the databases at the CDMS\footnote{\url{https://cdms.astro.uni-koeln.de/cdms/portal/}} \citep{2001A&A...370L..49M}, the 
JPL\footnote{\url{https://spec.jpl.nasa.gov/}} \citep{1998JQSRT..60..883P},
and the Splatalogue\footnote{\url{https://splatalogue.online/}}.
The transition parameters of C$_2$H$_5$CN $\varv 13/\varv 21$ are
adopted from \citet{2021JMoSp.37511392E}.

The identification of molecular lines started with  strong emission features. 
All strong emission features ($T^{\rm peak}_{\rm MB}>100$ mK) can be easily assigned with no ambiguities.  
To identify a weak emission feature,  
we first checked whether there is a probable transition of the already identified species.
If so, we assigned it to that matched species.
If not, we searched the databases to find candidate species which has transitions near the objective line  
and empirically has non-negligible abundances.
For a matched candidate species, we queried
the databases to get all its transitions with frequencies covered by this survey. 
If more than one strong transition (with large line strength and reasonable upper-level energy) of a candidate species has
corresponding emission feature in our spectrum, that candidate species was marked as an identified one. 
We repeated this process until all emission features have been checked.
We then modeled the emission of identified species (including both molecular emission and RRLs; 
see Sect. \ref{modelfit_sec}).
If the blended features can be accounted only partly by the identified species, 
we tried to assign the unmatched emission features to unidentified species. 
If there were strong transitions of a species that can not be well fitted, that species was removed from the set of identified species.
We iterated the above procedure of weak line identification until no further adjustment could be made.
The remaining unmatched features are labeled as unidentified (U) lines. 
Basic parameters of the matched molecular lines, including the observed frequency ($f_{\rm obs}$), 
the name of the species, the rest frequency ($f_{\rm rest}$), the transition labels, the
upper-level energy ($E_{\rm up}$), the Einstein coefficients ($A_{ij}$), and the Gaussian fitting results, 
are also listed in Table \ref{linelist}.

In total, 371 molecular transitions of 53 species were identified (Table \ref{allspeciestab}).
Here, isotopologues and molecules in different vibrational states  are treated as
different species.
The transition parameters of C$_2$H$_5$CN $\varv 13/\varv 21$ are not publicly available,
and we will further discuss about this species in Sect. \ref{sec_c2h5cn_vib}.

\begin{figure*}[!thb]
\centering
\includegraphics[width=0.95\linewidth]{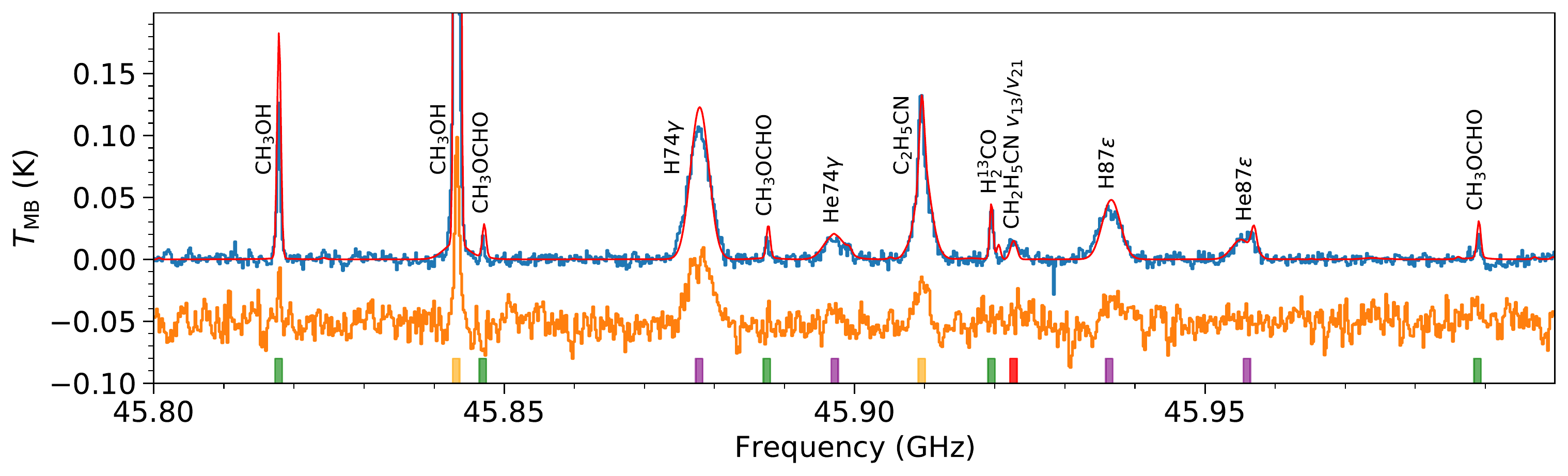}
\caption{An example of comparison  between the spectra of this survey in blue and of \citet{2017A&A...605A..76R} in orange. 
Spectrum  shown here has been smoothed to have the same spectral resolution of \citet{2017A&A...605A..76R}.
The red line represents the results of model fitting (Sect. \ref{modelfit_sec}). 
The bottom purple strips denote the detected RRLs. The yellow strips 
denote the molecular lines which have also been firmly
detected and resolved by \citet{2017A&A...605A..76R}. The green strips denote the 
molecular lines detected by TMRT 65 m but have not been detected by  \citet{2017A&A...605A..76R}.
The red strips denote the lines of C$_2$H$_5$CN $\varv$13/$\varv$21.  \label{fig_comparerizzo}
The $x$ axis is the frequency which has been converted to the rest frame of Orion KL with a $V_{\rm LSR}$ of 6 km s$^{-1}$ (Sect. \ref{sec_dr}).
}
\end{figure*}

\subsection{Unidentified lines} \label{sect_unline}
\begin{figure} 
\includegraphics[width=0.95\linewidth]{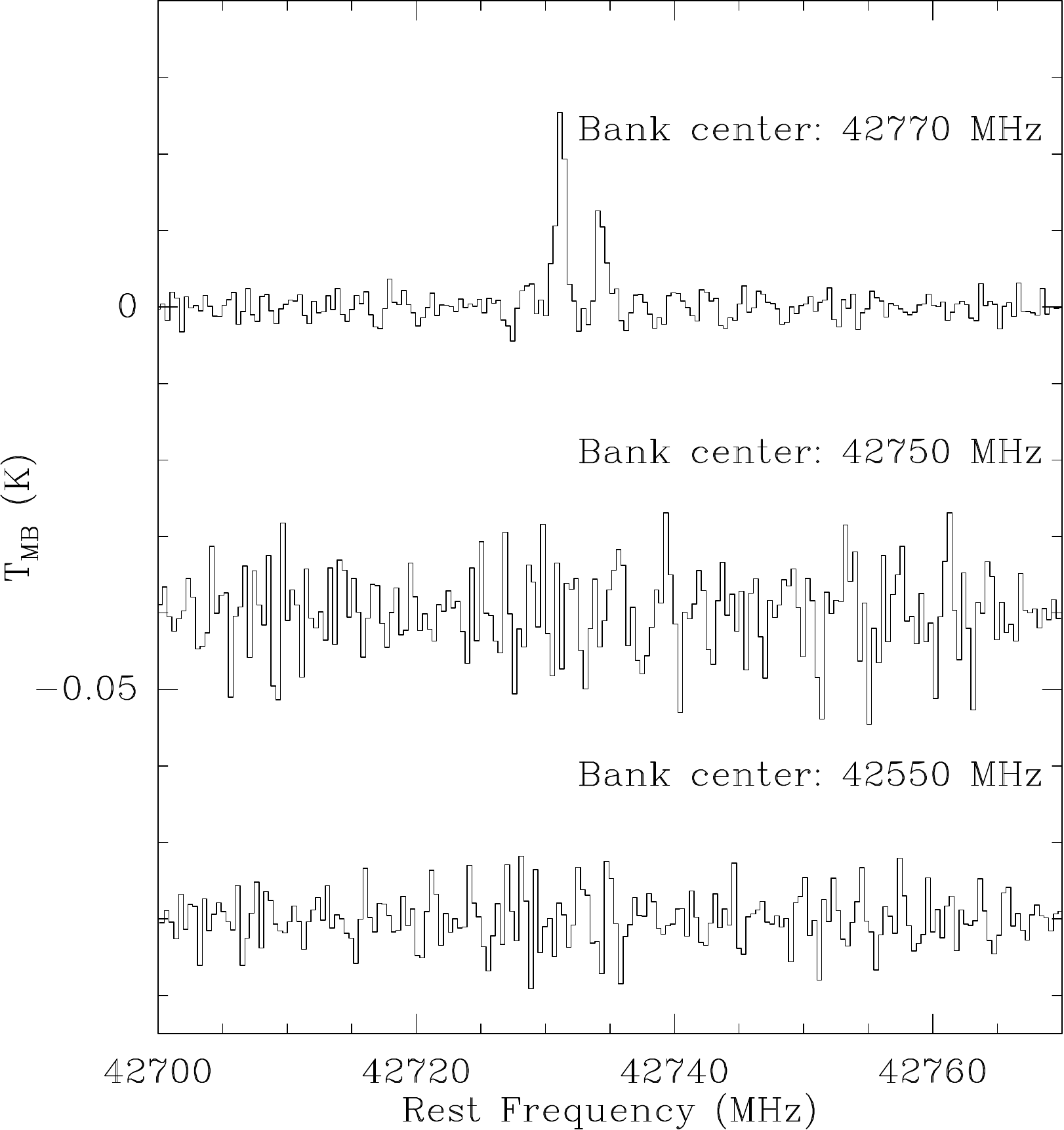}
\caption{The spectra of the 42732/42735 doublet observed in three different days.
The fake signals appeared  only once under unknown particular situation and no such phenomena happened during other 
observation setups (Sect. \ref{sect_unline}).
The fake doublets are kept in the spectrum shown in Fig. \ref{specset_example}. \label{fig_fake}}
\end{figure}

There are 39 emission features that have not been successfully associated with any RRLs and molecular transitions. 
They are also listed in Table \ref{linelist} labeled as `U'. Most of them are weak emission features or blended  with
strong lines (Fig. \ref{specset_example}).
Two doublets (42732/42735 MHz and 43482/43485 MHz) are exceptions.
Their strong intensities can not be assigned to any RRLs and molecular emission.
The nearby strong doublet of SiO maser (43122/43124 MHz) with peak intensity $>50$ K may saturate the system and
contribute to these fake doublets. 
However, this reason can be ruled out
since the fake doublets appeared only once, with central frequencies of 
42770 MHz and 44020 MHz for bank A and bank B, respectively. 
During the observations of another day with 
a slightly shifted LO frequency (the central frequencies were
42750 MHz and 44000 MHz for bank A and bank B, respectively), such fake doublets did not appear. 
We have also checked the observations with a larger LO frequency offset (the central frequencies were 
42550 MHz and 43750 MHz for bank A and bank B, respectively), and such fake doublets still did not appear (Fig. \ref{fig_fake}).
The relevant observations were conducted in three different days, and they all covered the SiO $J$=1-0 maser. 
Thus, we tend to assign the fake doublets to interference lines arising from the telescope system
under very particular situation. 
No such phenomena happened during other 
observation setups.

\section{Model fitting} \label{modelfit_sec}
To reproduce the spectrum of Orion KL observed by the TMRT and re-affirm the line identification,  we fit it
with a simple radiative transfer model. 
The emission of Orion KL (especially for strong lines) 
is complex, with contributions from several physical components including the
foreground {H}{\sc ii} region  M42 \citep{1997A&A...327.1177W}, the PDR
between M42 and the molecular cloud \citep[e.g.,][]{1994ApJ...428..209N}, the 
externally heated ``compact ridge'' \cite[e.g.,][]{1993ApJS...89..123M,2011A&A...527A..95W,2016ApJ...832...12T},
the hot cores \citep[e.g.,][]{1993ApJS...89..123M,2021A&A...647A..42J},
the extended ridge and plateau \citep[e.g.,][]{1989ARA&A..27...41G,2021ApJ...906...55B} 
as well as some other millimeter continuum sources
\citep[e.g.,][]{2014ApJ...791..123W}.
Since only single-point data were available in our observations, 
it is difficult to assign the detected emission to
specific physical structures within the beam, especially for molecular emission. 
Most of the detected lines are optically thin even if a beam filling factor of 0.1 is adopted.
Thus, the inferred column densities from the model fitting are highly coupled with the source size 
and the beam filling factor. 
We note that the spectral modeling of this work is mainly used for line identification, 
and  particular caution should be taken if the modeled parameters (e.g. $N_{\rm tot}$) listed in Table \ref{model_pars_table} are 
used for comparison with results from other authors.
The fitted column density of the molecule
should be multiplied by a factor accordingly if a different  size of the emission region is 
adopted. The fitted emission measures ($EM$s) of 
{H}{\sc ii} region and PDR region (Sect. \ref{sect_fitresult_RRL}) are more robust since the emission of RRL 
are more extended than the observing beam.  

\begin{figure} 
\includegraphics[width=0.95\linewidth]{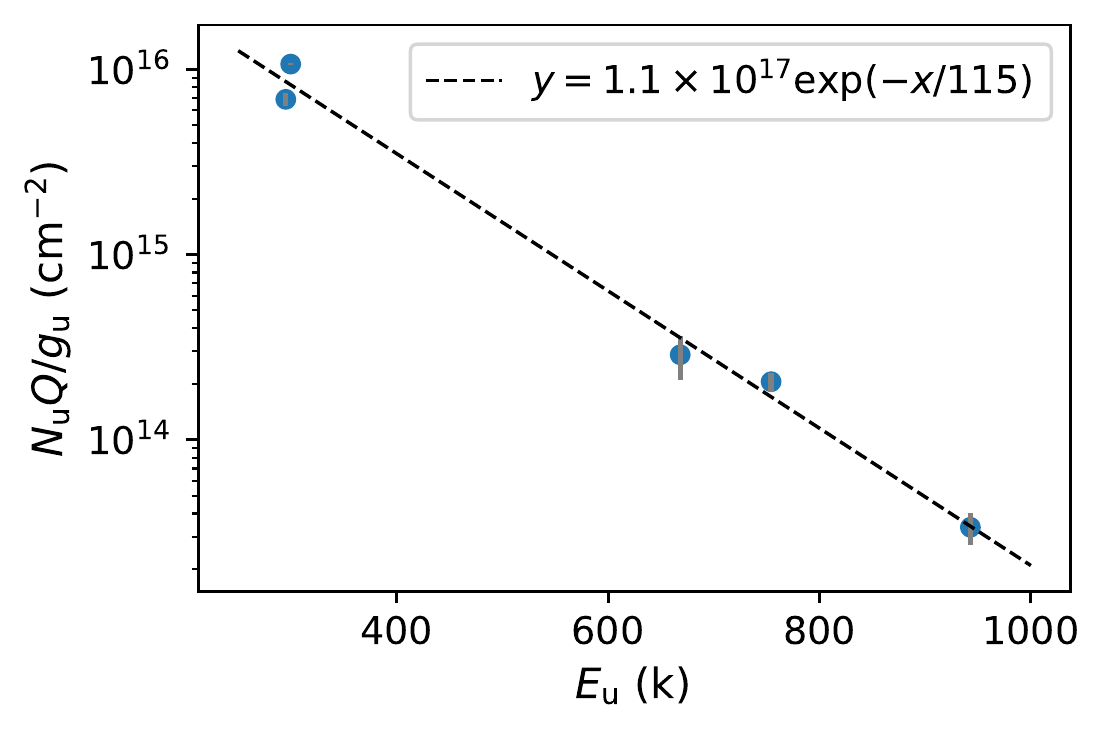}
\caption{The rotational diagram of E-CH$_3$OH $\varv_t=1$.\label{fig_rot}}
\end{figure}

\begin{figure*}[!thb]
\centering
\includegraphics[width=0.95\linewidth]{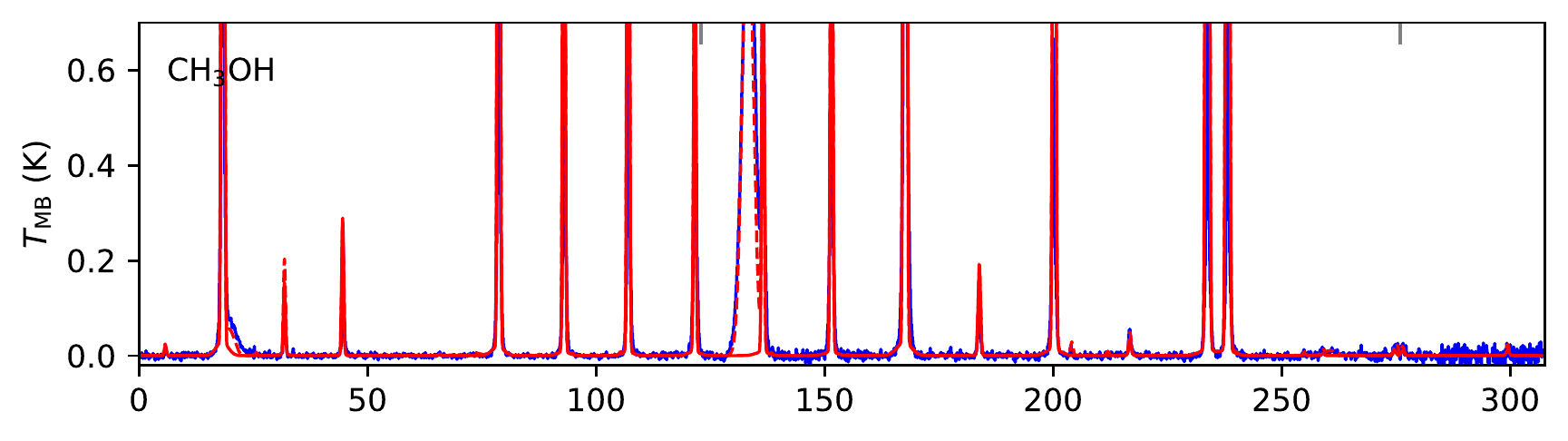}
\includegraphics[width=0.95\linewidth]{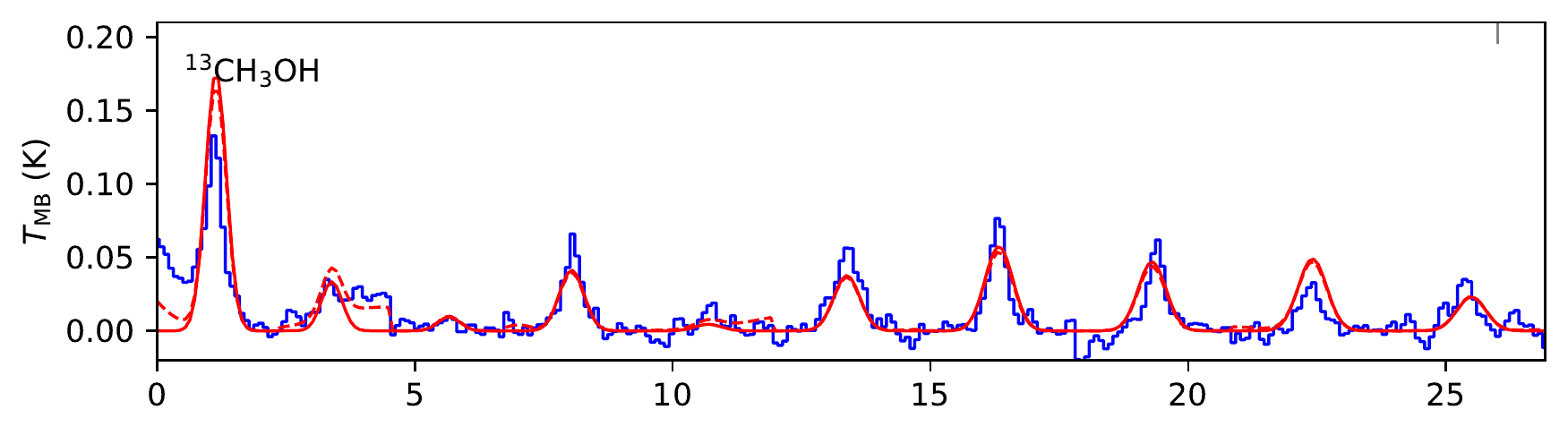}
\caption{Model fitting for CH$_3$OH (upper panel) and $^{13}$CH$_3$OH (lower panel) as an example. Spectra at different transitions are spliced together
with segments marked by the gray ticks on the top axis. 
The x axis is in unit of MHz with a channel width of 91.553 kHz (no smoothing applied).
The solid red line represents the model fitting of the corresponding species, while the 
dashed line includes the contributions of all the modeled molecular species (Table \ref{model_pars_table}) and RRLs
(see Sect. \ref{modelfit_sec} for details). 
See Fig. \ref{continued_fit_spec} for other species.\label{example_fit_spec}}
\end{figure*}

\begin{figure}[!htb]
\centering
\includegraphics[width=0.9\linewidth]{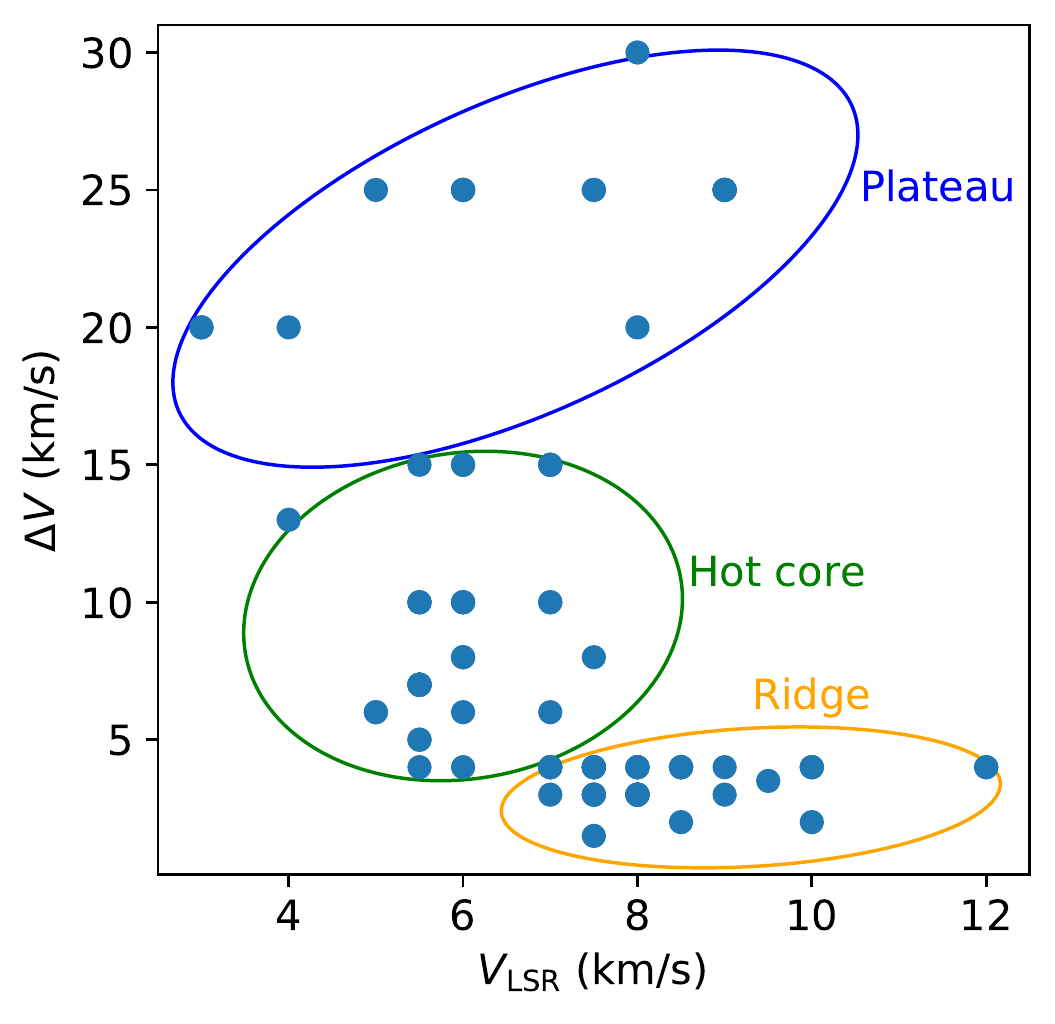}
\includegraphics[width=0.9\linewidth]{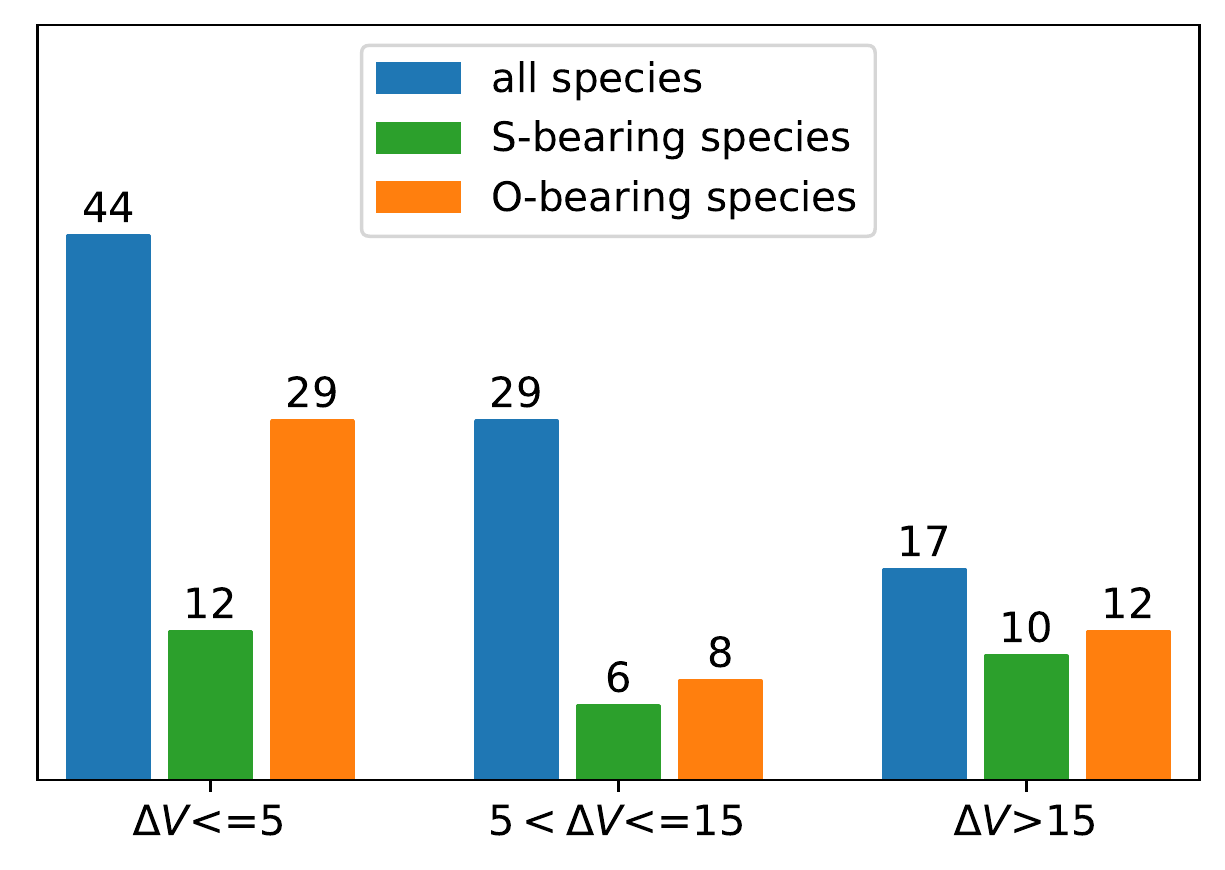}
\caption{Upper: Groups of spectral components (Table \ref{model_pars_table}) by K-means (Sect. \ref{sec_gps}).
Lower: The number of spectral components with $\Delta V$ (in units of km s$^{-1}$) in three different ranges.
\label{v_dv_fig}}
\end{figure}

\subsection{Model fitting of RRLs}
\subsubsection{Emission model of RRLs}
The emission of RRLs  was modeled through adding up the spectra of all the RRL transitions of
H, He, and C within the frequency
range of this survey.  For species X (denoting H and He), its RRL spectrum 
can be calculated using   \citep{2002ASSL..282.....G}
\begin{equation}\label{kappa_RRL}
\tau_{n_1,n_2} = 3.867\times 10^{-12} \frac{b_{n_2}}{\Delta \nu}  \frac{\Delta n}{n_1} f_{n_1,n_2}
\left(1-\frac{3\Delta n}{2n_1}\right) \frac{EM}{T_e^{5/2}} , 
\end{equation}
and
\begin{equation}
T_{\rm X} = \sum_{n_1,n_2} T_e\tau_{n_1,n_2}\frac{n_X}{n_e}\exp\left( -\frac{ [\nu-\nu_{\rm rest}^{\rm RRL}(n_1,n_2)
(1-\frac{\varv}{c})]^2}
{2\Delta\nu^2/(8\ln(2))} \right)
\end{equation}
Here, the parameters are adopted as the values in cgs units,
$EM$ is the emission measure, $n_{\rm X}$ is the volume density of X, $\varv$ is the velocity, 
$b_{n_2}$ is the upper-level departure coefficient, and $T$ is the excitation temperature.
The optically thin limit ($\tau_{n_1,n_2}\ll 1$) is satisfied for all the detected RRLs.
The term of $\exp(E_{n_2}/kT)$  is omitted in Eq. (\ref{kappa_RRL}) since $E_{n_2}<<kT$ for the detected RRLs.
The oscillator strength $f_{n_1,n_2}$ can be approximated as \citep{1968Natur.218..756M} 
\begin{equation}
f_{n_1,n_2} \sim n_1M_{\Delta n} \left( 1+1.5\frac{\Delta n}{n_1} \right) \label{eq_tau_fnn}
\end{equation}
for high-level transitions of hydrogen-like atoms.
We empirically fitted the values of $M_{\Delta n}$ tabulated in \citet{1968Natur.218..756M} as
\begin{equation}
M_{\Delta n} \sim 0.1905 \left(1/\Delta n\right)^{2.887} \label{eq_Mdn}
\end{equation}
The oscillator strength can be calculated more accurately
from the method described in \citet{1971MNRAS.153..471B} and \citet{Regemorter_1979}.
The accurate values have been used in recent models of  
\citet{2018MNRAS.478.2766P} and
\citet{2019ApJ...881...14Z,2022arXiv220609545Z}. 
The modeled spectrum in this work will not 
change obviously when the more accurate values of $f_{n_1,n_2}$
are adopted.

The electron temperature ($T_e$) of H{\sc ii} regions
are assumed as 8000 K \citep[e.g.,][]{1967ApJ...149L..61Z,1997A&A...327.1177W}.
If the RRL emitters are in LTE state, the departure coefficients should be unity for all levels. 
For $n_e=10^4$ cm$^{-3}$ and $T=8000$ K, we calculated the 
population of hydrogen using the method by \citet{2019ApJ...881...14Z,2022arXiv220609545Z}.
The departure coefficients $b_n$ increase from 0.85 to 0.99 as 
$n$ increasing from 50 to 100. 
It  means that the departure coefficients in Eq. (\ref{eq_tau_fnn}) 
can not be omitted  to precisely reproducing the observed RRLs and calculating the $EM$.
The RRLs of H and He from H{\sc ii} regions 
under strong ultraviolet fields originate from similar emission regions.
The states of helium atoms considered in this work are 
approximately hydrogenic in nature with $n\gg 1$.
\citet{1995MNRAS.272...41S} suggested that the departure coefficients for 
hydrogen atoms can also be applied to these states of helium atoms.  

At large $n$, the carbon atom is physically similar to H and He, thus the above equations for H and He
are also valid for deriving the emission of  carbon RRLs \citep{2017ApJ...837..141S}.
For carbon in photon-dissociation regions, an electron temperature of
2000 K is adopted considering $(E_{\rm ionize}^{\rm H}-E_{\rm ionize}^{\rm C})/k< 3000$ K. 
We also try to fit the carbon lines using a $T_e$ of 300 K (Sect. \ref{sect_fitresult_RRL}).
We assume that the carbon atoms are  in LTE state with the departure coefficients of carbon adopted as unity,  
although this is 
a somehow uncertain.

To determine the values of line widths and velocities of the modeled RRLs, we have referred to 
the Gaussian fitting results of the detected RRLs (see Table \ref{linelist}
and Sect. \ref{analyse_RRLs}) and the values of previous studies \citep{2009ApJ...691.1254G,
2015A&A...581A..48G,2017A&A...605A..76R}.
The $\Delta \nu$ is the FWHM line width of the modeled spectrum in frequency scale, which can be calculated through
\begin{equation}
\Delta \nu=f_{n_1,n_2}^{\rm RRL} \frac{\Delta \varv}{c}.
\end{equation}
Here, the $\Delta \varv$ is fixed as 20 km s$^{-1}$ for H/He RRLs and 5 km s$^{-1}$ for C RRLs.
The velocity (in $V_{\rm LSR}$) of H/He and C are fixed as -3 km s$^{-1}$ and 9 km s$^{-1}$, 
respectively (see Sect. \ref{sec_HHeratio} for further
discussion).

\subsubsection{Fitting results of RRLs} \label{sect_fitresult_RRL}
The beam filling factor of H/He RRLs is adopted as unity, since they 
mainly originate from the extended M42 H{\sc ii} region. 
The emission measure is fitted as 3.9$\times$10$^{6}$ cm$^{-6}$ pc. 
The abundance ratio between  He and H (He/H) is fitted to  8.5\%. 
We note that this value is derived with an assumption that the 
H and He have identical ionized regions. We will further discuss the 
He/H ratio in Sect. \ref{sec_HHeratio}.
The fitted intensities of unblended H/He RRLs are all consistent with the observed values within 20 percent. 
The fitted emission measure of H{\sc ii} region ($EM^{\mbox{\footnotesize H\sc ii}}$) 
is compatible with the value of 7.5$\times$10$^6$ cm$^{-6}$ pc derived by
\citet{2009ApJ...705..226D} using the 21.5 GHz data of the Green Bank Telescope with a beam size of 33.5\arcsec.

The carbon RRLs mainly originate from PDR between the M42 and molecular clouds. 
Since the main body of the Orion bar is larger than the beam of this survey,
a beam filling factor of unity is also adopted for carbon RRLs.
If the ionized carbons contribute to most of the electrons within the PDR region,
the emission measure of PDR can be fitted to
$EM^{\rm PDR}=3.9\times 10^3\ {\rm cm^{-6}\ pc}$.
If an electron temperature of 300 K (instead of 2000 K) is adopted,  
the fitting result gives 
$EM^{\rm PDR}= 1.8\times 10^2\ {\rm cm^{-6}\ pc}$.

\subsection{Model fitting of molecular lines} \label{sect_model_mol}
\subsubsection{Emission model of molecular lines}
The molecular emission was modeled by adding up a set of 
spectral components contributed by different species. 
A species (denoted as $X$) may contribute multiple spectral components with different velocities.
A spectral component may consist of more than one line features contributed by different transitions of its corresponding species.
Each spectral component is assumed to be in LTE state with 
an identical  emission source size ($D$) and the same excitation temperature ($T_{ex}$)
for all transitions of the corresponding species.  
For each spectral component (denoted as $s$), its spectrum ($T_s$ in main beam temperature scale) 
could be modeled through adding up 
a set of emission lines of the transitions of the corresponding species. 
Those lines have Gaussian-shape optical depths with the same line width and $T_{\rm ex}$. Specifically, a spectral 
component can be calculated following
\citep[e.g.,][]{2015PASP..127..266M}
\begin{equation}
\tau_{ij}^s =  \frac{A_{ij}  c^2}{8\pi(\nu_{ij})^2}\left[
\exp\left(\frac{h\nu_{ij}}{kT_{\rm ex}}\right) - 1
\right] N^s_{\rm tot} \frac{g_u}{Q} \exp\left( \frac{-E_u}{kT_{ex}}\right) \frac{1.06}{\Delta \nu_s}
\end{equation}
and 
\begin{equation}
T_{\rm s} = \sum_{i,j}f T_{ex} \left\{ 1- \exp\left[ -\tau_{i,j}^s\exp\left( -\frac{ [\nu-\nu_{i,j}
(1-\frac{\varv_s}{c})]^2}
{2\Delta\nu_s^2/(8\ln(2))} \right) \right]\right\}  \label{eq_model_mol}
\end{equation}
Here, $\tau_{ij}^s$ is the peak optical depth, 
$N_{\rm tot}^s$ is the column density of the corresponding species, $\Delta v_s$ is the line width in
frequency scale ($\Delta \nu= \nu_{i,j} \Delta V_s/c$), $f$ is the beam filling factor $(D/\Theta_{\nu_{i,j}})^2$,
$\Theta_{\nu_{i,j}}$ is the beam size at ${\nu_{i,j}}$, $E_u$ is the upper-level energy,
$g_u$ is the degree of degeneracy of the upper level, and $Q$ is the partition function.
The Q is a function of $T_{ex}$ and it could be interpolated from the tabulated values quoted from the on-line databases
(Sect. \ref{Sect_idmol}). The modeled spectrum of a species  is
\begin{equation}
T_X = \sum_s T_s.
\end{equation}
For each species, we fit $T_X$ to the observed spectrum.
Since we only tried to build a radiative transfer model  with the goal to reproduce our observed
spectrum, source size and $T_{\rm ex}$ for each spectral component are fixed. 
The values of line widths ($\Delta V_s$), velocity ($\varv_s$) and column density ($N^s_{\rm tot}$)
are fitted. 

The parameters for different species were basically independent.
However, to reproduce the profiles of emission lines of isotopologues or species with similar chemical characteristics, 
we tend to adopt similar fixed values for their $D$ and $T_{ex}$,
and use similar initial guess for their fitted parameters.
For most species whose transitions have similar upper-level energies, the fitting  result is not sensitive to  $T_{ex}$.
For those species, $T_{ex}$ is manually adopted as 50 K, 100 K, 150 K, 200 K, or 300 K. 
We have also referred to literature \citep{2010A&A...517A..96T,2017A&A...605A..76R} to  guess the 
initial values of the fitted parameters.
We try to model the spectrum with as few spectral components as possible.
After each iteration of model fitting, we manually checked the fitting result to see whether there are  line peaks, skewed line shoulders or line wings
that are obviously not well fitted. If so, we added a spectral component to fit them.

The spectra of different species are fitted separately.
The species with strong lines are first fitted.
For each of the rest species, the fitting procedure is conducted on the 
residual spectrum (the difference between the original spectrum 
and the modeled spectrum of all other fitted species).
This procedure is iterated until all the species are well fitted.

\subsubsection{Fitting results of molecular lines}
All the identified species except SiO and its isotopologues were modeled.
The transition parameters of C$_2$H$_5$CN $\varv$13/$\varv$21
are not yet publicly available from the databases of JPL and CDMS,
and we obtained those parameters from \citet{2021JMoSp.37511392E} through private communication.
We will discuss the fitting of C$_2$H$_5$CN $\varv$13/$\varv$21 in 
Sect. \ref{sec_c2h5cn_vib}. 

The spectrum of E-CH$_3$OH $\varv_t=1$ can not be well fitted with a $T_{ex}$ of 100 K.
Adopting $T_{ex}$  as 110 K, the temperature of compact ridge \citep{2010A&A...517A..96T},
the spectrum of E-CH$_3$OH $\varv_t=1$ can be better reproduced.
It is consistent with the result of its rotational 
diagram (Fig. \ref{fig_rot}), which yields a $T_{ex}$ of 115$\pm$5 K for E-CH$_3$OH $\varv_t=1$.
We  adopted  $T_{ex}$ as 110 K to fit the corresponding spectral components of E-CH$_3$OH $\varv_t=1$,
A-CH$_3$OH $\varv_t=1$, CH$_3$OH  and $^{13}$CH$_3$OH. 
The spectra of CH$_3$OCHO, HNCO, C$_2$H$_3$CN and C$_2$H$_5$CN  can be well fitted with the spectral components quoted
from \citet{2017A&A...605A..76R} with $T_{ex}$ unaltered. For CH$_3$OCHO and HNCO, the deviation between the fitted column densities of this survey
and those of \citet{2017A&A...605A..76R} are smaller than 50 percent. 
However, for C$_2$H$_3$CN and C$_2$H$_5$CN, the fitted column densities of this survey are 3 to 10 times lower than the values of
\citet{2017A&A...605A..76R}.
Only one line of CH$_3$CN is covered, and it is strong with a highly non-Gaussian shape. 
The fitting results of CH$_3$CN should be treated as rough estimations with large uncertainties. 

The fitting results of the modeled parameters are listed in Table \ref{model_pars_table}.
For each species, we extracted the spectral segments that may contain emission of 
that species. Those spectral segments were then spliced together (referred to as the 
`spliced spectrum'). The spliced spectra for all fitted species are shown in
Figs. \ref{example_fit_spec} and \ref{continued_fit_spec}.
For the transitions of most species, the deviations between the modeled  intensities and
the observed values are smaller than 10 percent.
The spliced spectrum of NH$_2$D is an exception.
NH$_2$D has two detected transitions (43042 MHz and 49962 MHz), and
those two lines can not be simultaneously well reproduced.

\begin{figure}[!htb]
\centering
\includegraphics[width=0.495\linewidth]{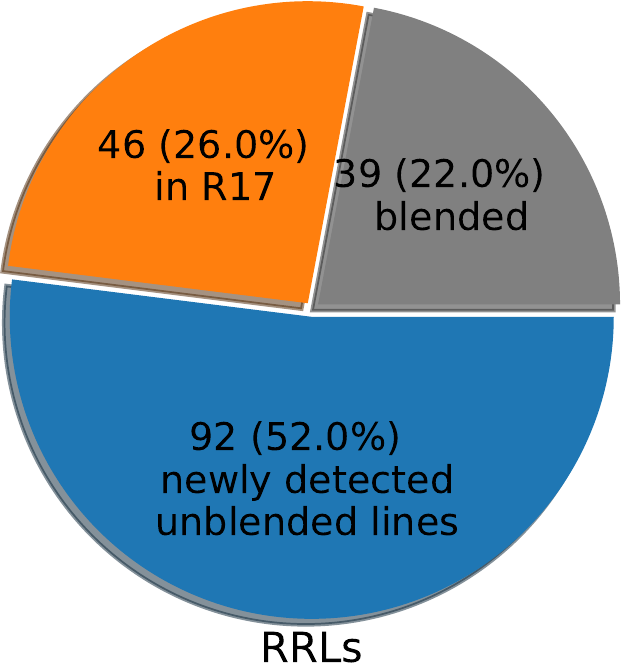}
\includegraphics[width=0.495\linewidth]{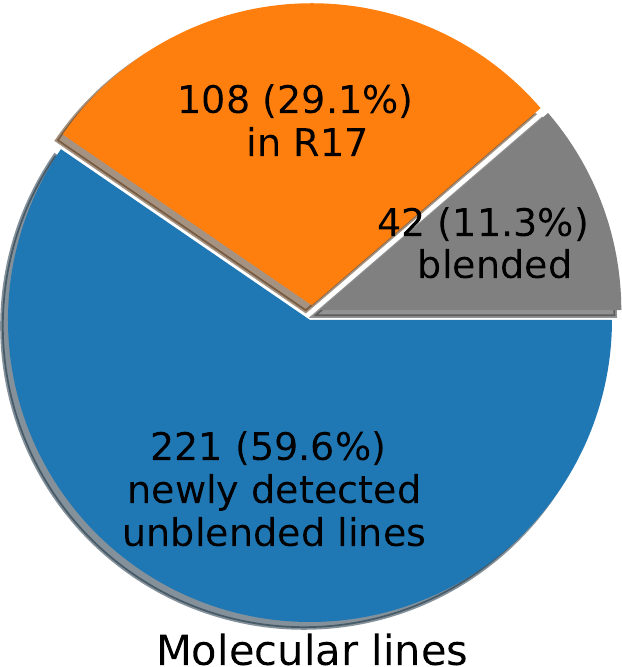}
\caption{Pie diagrams to compare the number of molecular lines and RRLs of this survey with and without
detection by R17 \citep{2017A&A...605A..76R}.
\label{fig_venn_compare}}
\end{figure}

\subsubsection{Groups of spectral components} \label{sec_gps}
The spectral components are divided into three groups
in the velocity and line width ($V_{\rm LSR}$--$\Delta V$) 
space using the K-means algorithm \citep{Lloyd82} implemented in SciPy\footnote{\url{https://pypi.org/project/scipy/}},
as shown in Fig. \ref{v_dv_fig}.
The three groups are associated with three gas components of Orion KL, the plateau, 
the hot core and the extended/compact ridge \citep{2010A&A...517A..96T,2017A&A...605A..76R}.
The median $V_{LSR}$ of the three groups are 7 km s$^{-1}$, 6 km s$^{-1}$, and 8.5 km s$^{-1}$,
respectively.
The median $\Delta V$ of the three groups are 25 km s$^{-1}$, 8 km s$^{-1}$, and 3 km s$^{-1}$,
respectively.
Roughly, the line widths of spectral components in groups of  extended/compact ridge, hot core and plateau 
are in range of 1--5 km s$^{-1}$, 5--15 km s$^{-1}$, 15--30 km s$^{-1}$, respectively. 
The kinetic  temperature of the compact ridge, $\sim$100 K, is higher than the value of the 
extended ridge, $\sim$60 K \citep{2010A&A...517A..96T}.
However,  most of the $T_{\rm ex}$ listed in Table \ref{model_pars_table}  have quite large uncertainties, and thus
it is difficult to  explicitly assign a spectral component with $\Delta V<$ 5 km s$^{-1}$  to the compact ridge or the extended ridge.

For oxygen-bearing species, most of their spectral components  (29/49) have $\Delta V<=$ 5 km s$^{-1}$ (Fig. \ref{v_dv_fig}).
It is expected since most of the oxygen-bearing species have spectral components originated from the extended/compact ridge. 
Thirty-eight percent (10/28) of the spectral components of  sulfur-bearing species 
have $\Delta V>$ 15 km s$^{-1}$. In contrast, for oxygen-bearing species,  only twenty-four percent (12/49) of the spectral components
have $\Delta V>$ 15 km s$^{-1}$.
Aldehyde-containing species and species with both sulfur and oxygen elements tend to have spectral components 
originated from the plateau (Table \ref{model_pars_table}), and this supports a possible enhancement of these species by shocks
\citep[e.g.,][]{2015A&A...581A..71F,2017MNRAS.469L..73L,2021ApJ...912..148L}.

\begin{figure}[!ht]
\centering
\includegraphics[width=0.995\linewidth]{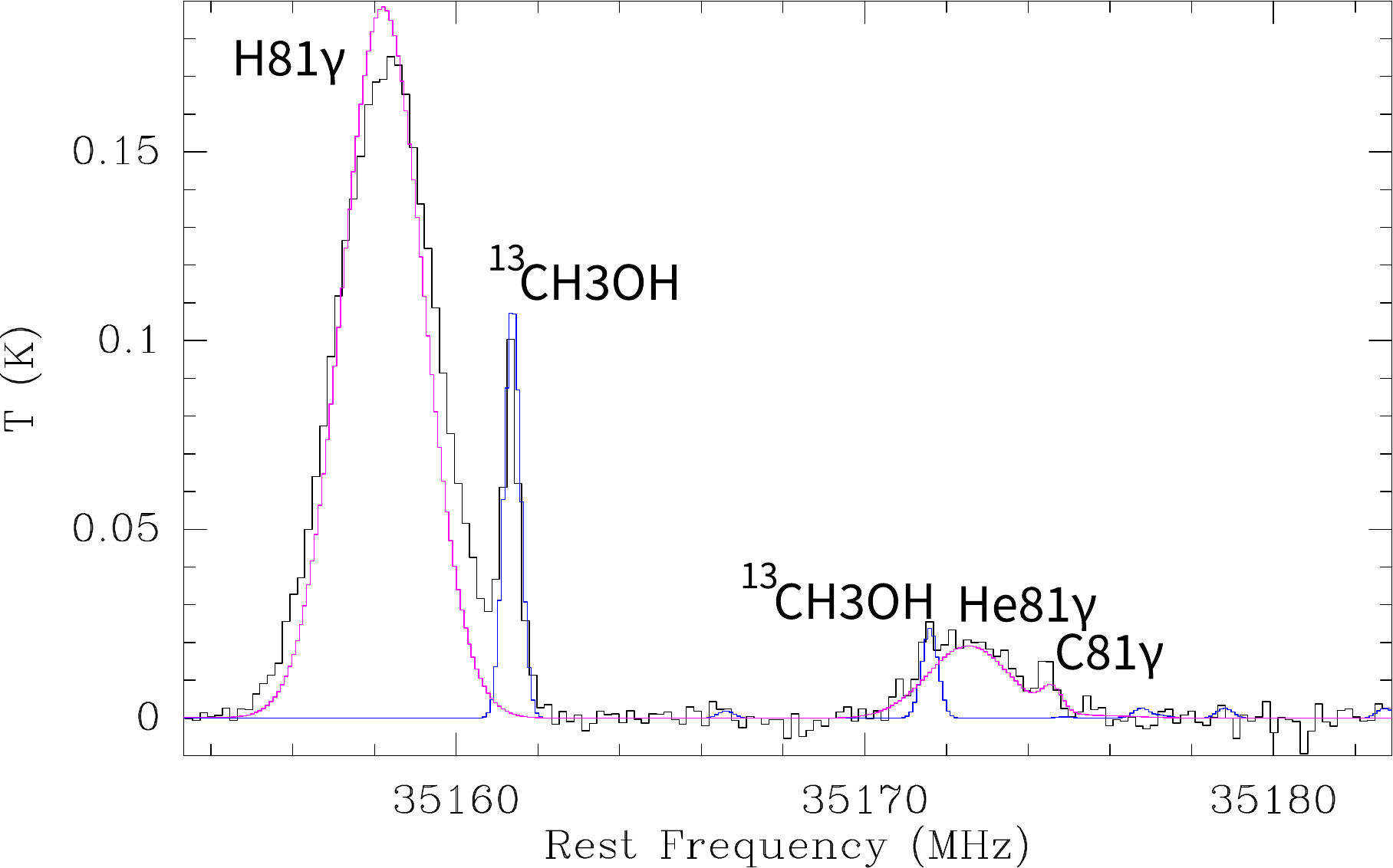}
\includegraphics[width=0.995\linewidth]{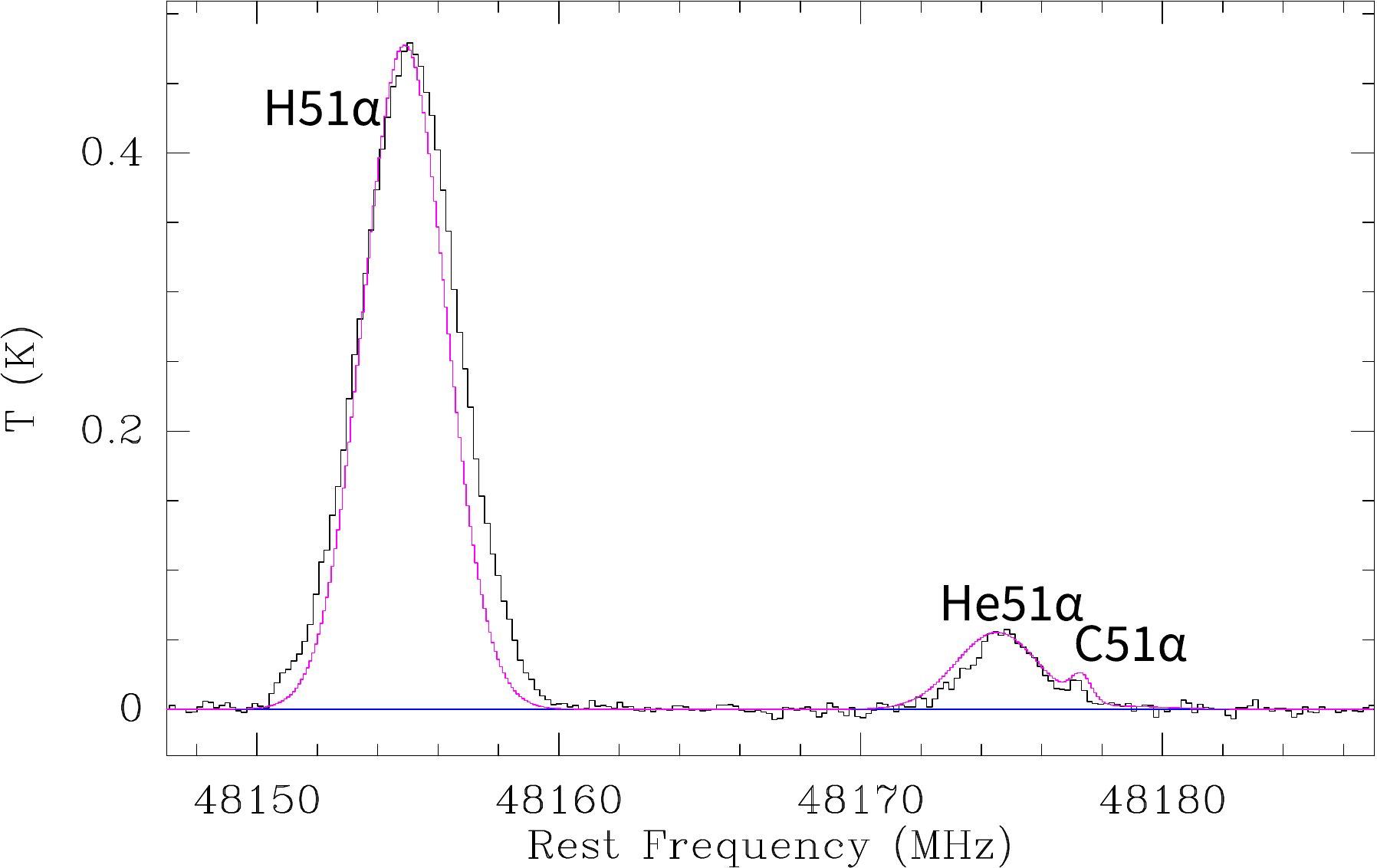}
\caption{Upper: The black line is the observed  81$\gamma$ lines of H, He and C. 
It has been smoothed to a channel width of 183 kHz (1.56 km s$^{-1}$). The
purple line is the modeled RRLs, and the blue line is the modeled molecular lines (Sect. \ref{modelfit_sec}).
Lower: same as the upper panel but for the 51$\alpha$ lines. 
\label{figC81gamma} }
\end{figure}

\begin{figure}[!thb]
\includegraphics[width=0.95\linewidth]{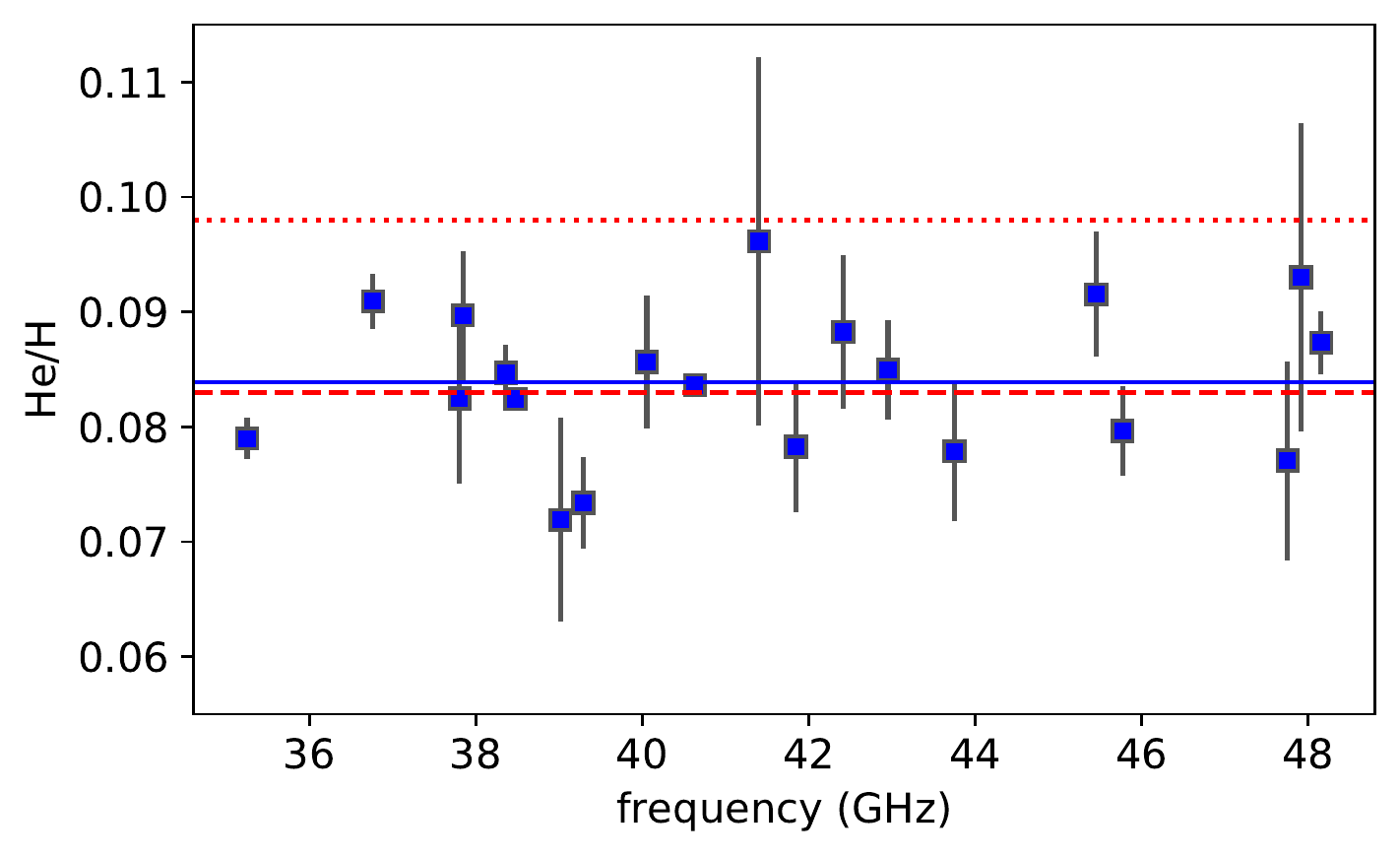}\\
\includegraphics[width=0.95\linewidth]{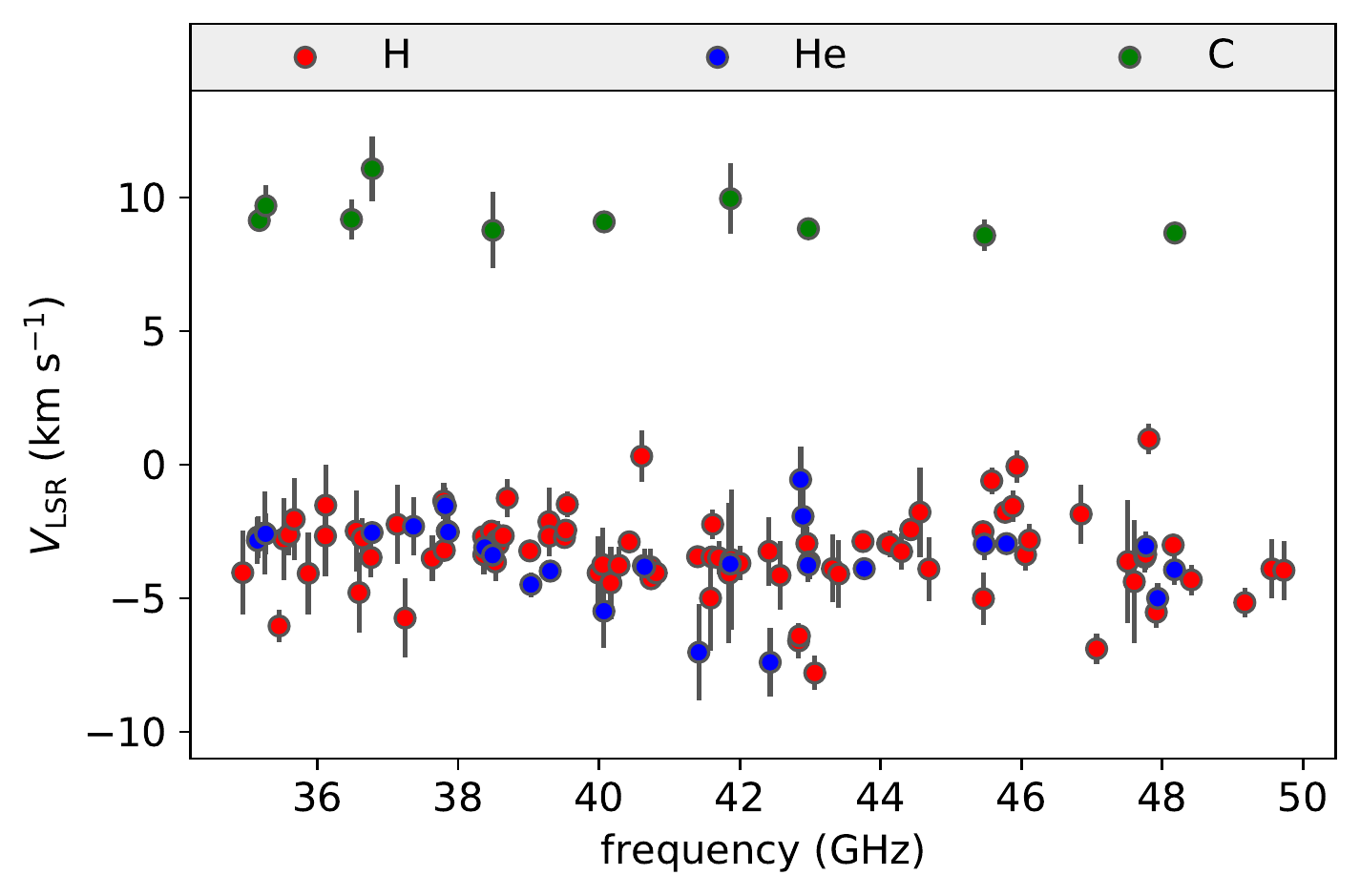}
\caption{Upper: The intensity ratios between RRLs of He and H. The red dotted and dashed lines
represent the He/H abundance ratio of 8.3\% due to Big Bang nucleosynthesis \citep{2004ApJ...617...29O}
and the solar value $\sim$9.8\% \citep{1994ARA&A..32..191W}, respectively.
The blue solid line is the mean value of the data.
Lower: the $V_{\rm LSR}$ of unblended RRLs of H, He and C derived by Gaussian fittings.
\label{fig_HHeC}}
\end{figure} 

\section{Analysis and Discussion} \label{secdis}
\subsection{RRLs} \label{analyse_RRLs}
Fig. \ref{fig_venn_compare} shows the comparison between the numbers of  
emission lines of this survey with and without detection by \citet{2017A&A...605A..76R}.
Forty-six of those RRLs were detected and spectrally resolved by
\citet{2017A&A...605A..76R}.  
\subsubsection{The maximum  $\Delta n$ of detected RRLs} \label{sect_largestdn}
Among the unblended hydrogen RRLs detected in this survey, 
H135$\pi$ has the maximum $\Delta n$ of 16.
The maximum $\Delta n$ of this survey is larger than the values of the Q-band survey of \citet{2017A&A...605A..76R} and the
radio K-band (1.3 cm) survey of
\citet{2015A&A...581A..48G}. The maximum  $\Delta n$  detected by
\citet{2017A&A...605A..76R} and \citet{2015A&A...581A..48G} 
was 11 for both.
For spectrally resolved helium RRLs in this survey, the maximum $\Delta n$ is 7, and it 
is also higher than the values of 4 detected
by \citet{2017A&A...605A..76R} and \citet{2015A&A...581A..48G}.
The carbon RRLs detected by \citet{2015A&A...581A..48G} are all carbon alpha lines highly blended with
helium RRLs. Only two carbon RRLs (C52$\alpha$ and C53$\alpha$) were firmly 
detected and resolved by \citet{2017A&A...605A..76R}. 
On the contrary, 10 RRLs of carbon are resolved in this survey, 
including the C81$\gamma$ with an $\Delta n$ of 3
(Fig. \ref{figC81gamma}).
Thanks to the higher sensitivity and wider frequency coverage of this survey,
this work has more than doubled  the number of detected RRLs,
particularly those with larger $\Delta n$, in the Q band compared with \citet{2017A&A...605A..76R}.

\subsubsection{Intensity ratios and dynamics of RRLs} \label{sec_HHeratio}
Benefiting from a large number of RRLs detected in this survey, we can estimate
the abundance ratio of He/H via the intensity ratio between their RRLs with the 
same $n$ and $\Delta n$:
\begin{equation}
y = \frac{N({\rm He})}{N({\rm H})} \sim \frac{1}{R^{1/3}}\times \frac{\int T_\nu({\rm He}) {\rm d}\nu }
{\int T_\nu({\rm H}) {\rm d}\nu}.
\end{equation}
Here,  $\int T_\nu({\rm X}){\rm d}\nu$ is the integrated intensity of the RRL of X, 
and the $R$ is the volume ratio between the  ionized regions of He and H.
If the $R$ value is adopted as unity considering the similarity between the line widths of hydrogen and helium RRLs (Table \ref{linelist}), 
the inferred mean value of $y$ is 8.4\% with a standard deviation of 0.2\%  (Fig. \ref{fig_HHeC}).

The $y$ value derived here is consistent with the result of global fitting described in Sect. \ref{sect_fitresult_RRL}.
It is also consistent with the values of (8.7$\pm$0.7)\% derived by \citet{2015A&A...581A..48G}, 
(8.3$\pm$1.2)\% by \citet{2017A&A...605A..76R}, and (8.8$\pm$0.6)\% by \citet{1991ApJ...374..580B} derived 
based on optical
observations. 
The value of $y$ seems to be compatible with the value of  8.3\% due to Big Bang nucleosynthesis \citep{2004ApJ...617...29O}
and lower than the solar value, 9.8\% \citep{1994ARA&A..32..191W}.
However, if an $R$ value of 0.6 is adopted \citep{1983Ap&SS..91..381C}, the derived value of $y$ is 10\%,
much closer to the solar value.

The velocities of unblended hydrogen, helium, and carbon RRLs are shown in the lower panel of
Fig. \ref{fig_HHeC}. The velocities of hydrogen and helium RRLs at different frequencies
are consistent with a value of $-$3$\pm$1.5 km s$^{-1}$. The velocities of carbon RRLs
are $\sim+$9$\pm$1 km s$^{-1}$.  
The computed velocities are compatible with previous results
that the hydrogen and helium RRLs arise from the M42, 
while the carbon RRLs originate from the photon-dominated region (PDR) 
between M42 and the associated molecular clouds 
\citep[the Orion Bar, see ][]{2015A&A...575A..82C,2015A&A...581A..48G,2017A&A...605A..76R}.

\begin{figure}[!t]
\centering
\includegraphics[width=0.95\linewidth]{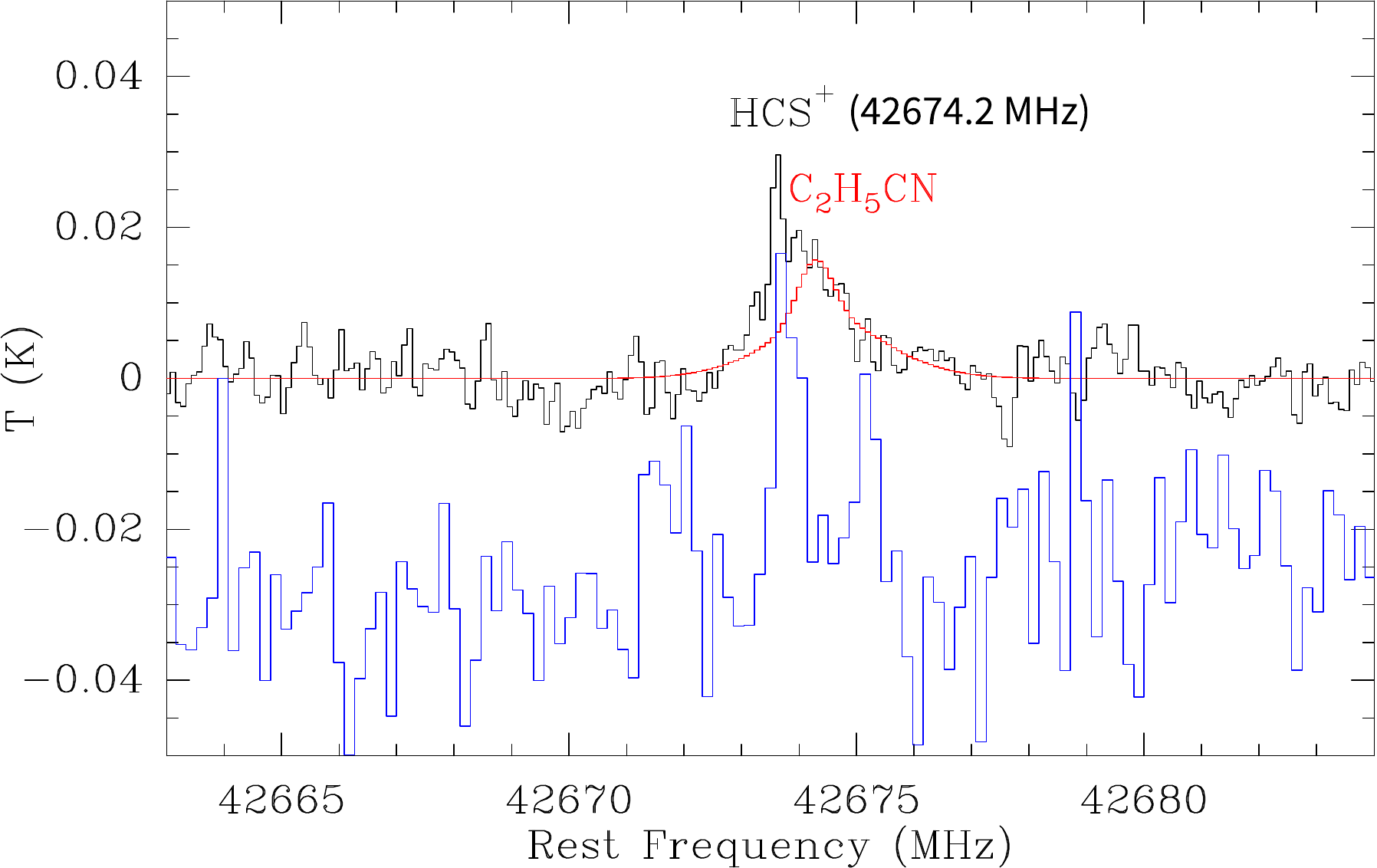}
\includegraphics[width=0.95\linewidth]{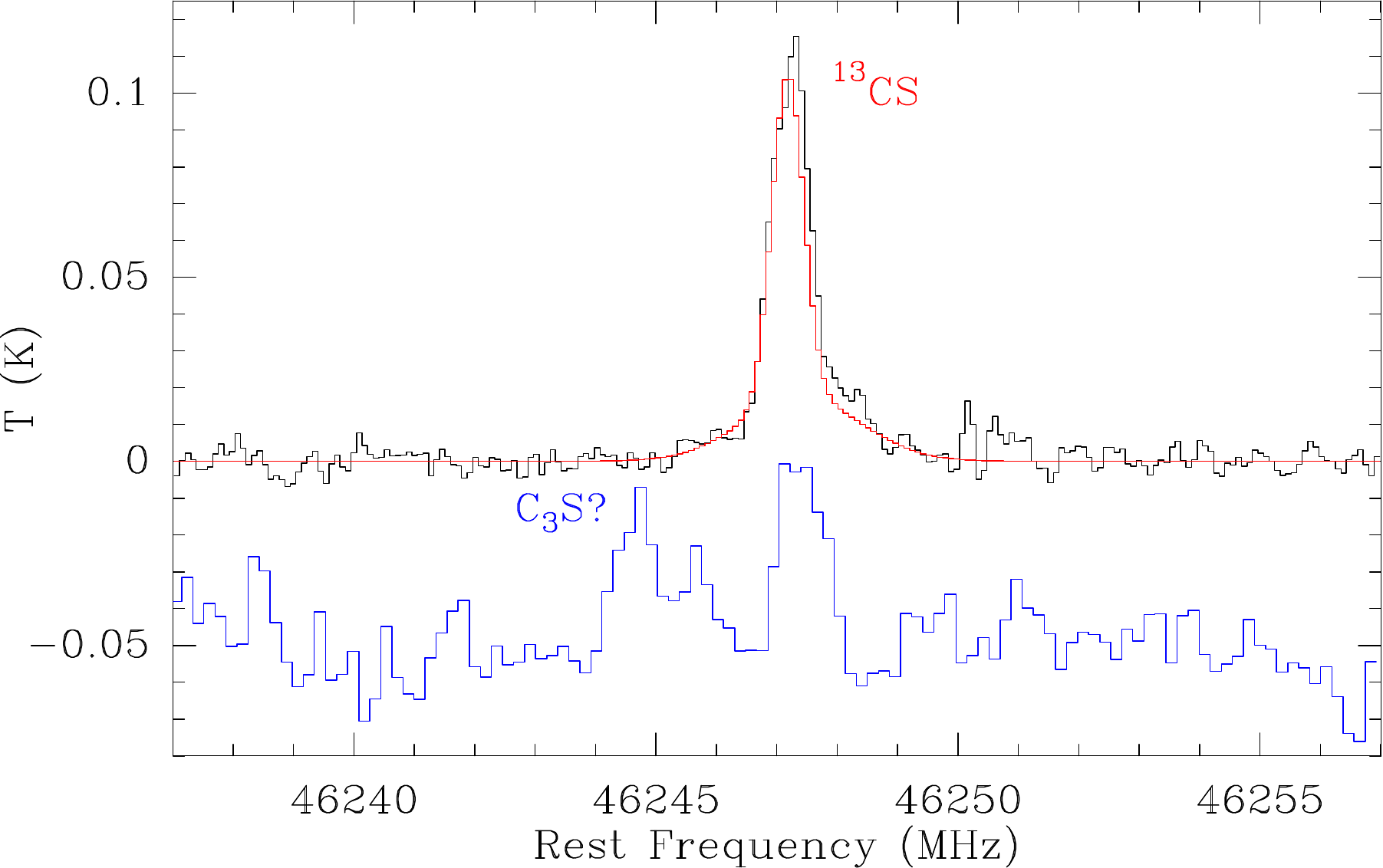}
\caption{In both panels, the black lines are the spectra detected in our survey,  
and the blue 
lines are the spectra of \citet{2017A&A...605A..76R}.
In the upper panel, the red line is the model fitting of C$_2$H$_5$CN.
In the lower panel, the red line is the model fitting of $^{13}$CS.
The small peak near 46245 MHz (blue line) was identified as 
C$_3$S $J = 8-7$ by \citet{2017A&A...605A..76R}. See Sect. \ref{sec_sulfur} for the 
discussion about the emission features of C$_3$S. 
\label{fig_hcsp}}
\end{figure}

\begin{figure*}[!t]
\centering
\includegraphics[width=0.9\linewidth]{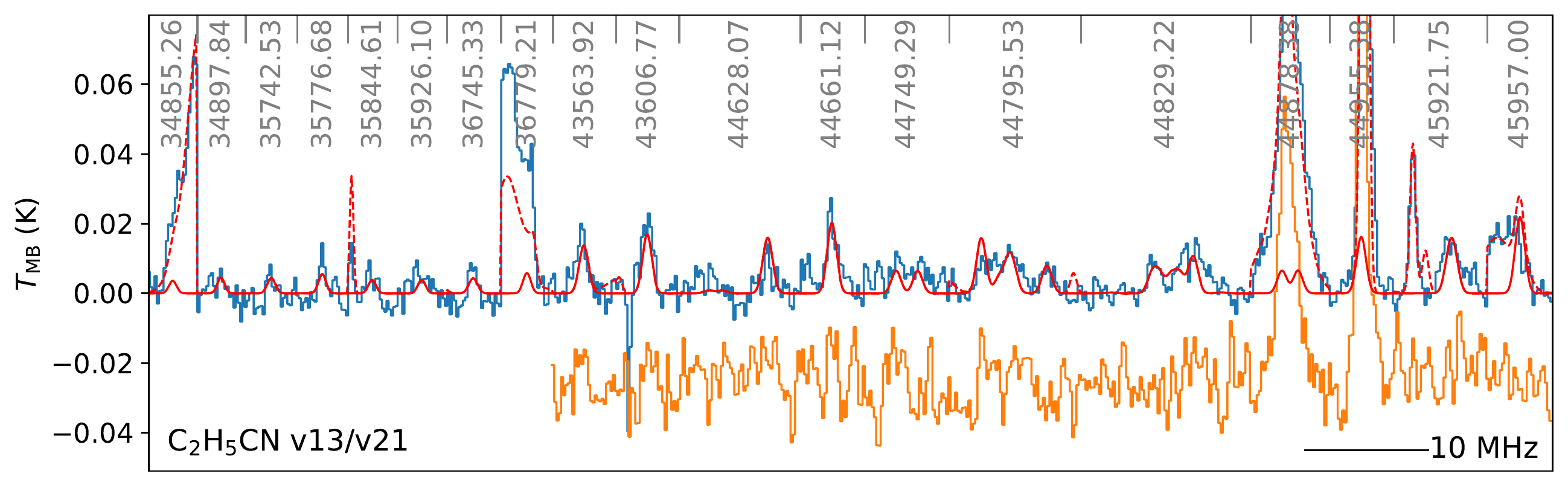}
\includegraphics[width=0.9\linewidth]{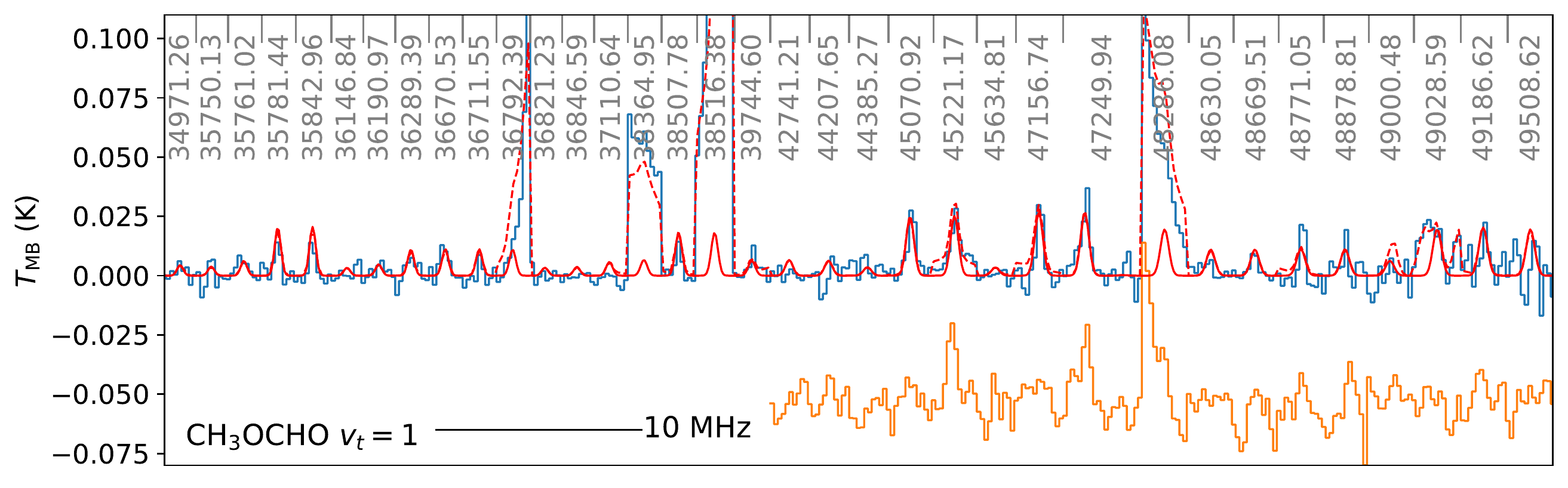}
\caption{The ``spliced'' spectra of C$_2$H$_5$CN $\varv$13/$\varv$21 (upper panel) and CH$_3$OCHO $\varv_t=1$ (lower panel).
The blue and orange lines are spectra from this survey and \citet{2017A&A...605A..76R}, respectively.
The spectra have been smoothed to have a spectral resolution of 183 kHz (twice the spectral resolution of this survey).
The center frequencies of each segments are shown on the top axis. The red line is 
the result of model fitting. 
The  red dashed line includes the contributions of all the fitted molecular species (Table \ref{model_pars_table}) and RRLs
(see Sect. \ref{modelfit_sec} for details).
The unsmoothed version for CH$_3$OCHO $\varv_t=1$ can be found in Fig. \ref{continued_fit_spec}. \label{fig_c2h5cn} }
\end{figure*}

\subsection{Molecular species}
As shown in Fig. \ref{fig_venn_compare}, 
among the 371 molecular transitions detected and identified by this survey, 329 are unblended and only 108 of 
them have been firmly detected and identified by 
\citet{2017A&A...605A..76R}.
Those molecular transitions belong to 53 species, most of them
are sulfur-bearing species, oxygen-bearing organic molecules, nitrogen-bearing species (cyanopolyynes, N-bearing COMs and NH$_3$), 
and their isotopologues .
Among the 53 detected species, 32 were firmly detected by \citet{2017A&A...605A..76R}.
The other 21 species were not firmly  detected by \citet{2017A&A...605A..76R},
including the tentatively detected NH$_2$CHO.

\subsubsection{Sulfur-bearing species} \label{sec_sulfur}
Sulfur-bearing molecules (e.g., H$_2$S, SO, SO$_2$, CS, OCS, and
H$_2$CS) have been detected in various star-forming environments \citep[e.g.,][]{1973ApJ...186..123T,
2019ApJ...885...82L,2019MNRAS.488..495W}.
Thirteen kinds of isotopologues of sulfur-bearing molecules are detected in this survey.
Among them, five species were not confirmed  by \citet{2017A&A...605A..76R}.

The comparison between the HCS$^+$ spectra of this survey and that of \citet{2017A&A...605A..76R} 
is shown in Fig. \ref{fig_hcsp}.
The narrow emission line of HCS$^+$ is overlapped with the broad emission of C$_2$H$_5$CN. 
Although \citet{2017A&A...605A..76R} marked this line as a blended line of   HCS$^+$ and C$_2$H$_5$CN,
the broad emission of  C$_2$H$_5$CN can not be seen in their spectrum.
This further reduces the degree of confidence of the possible HCS$^+$ emission feature with a
limited signal-to-noise ratio. 
In the spectrum of this survey, 
the emission of HCS$^+$  and C$_2$H$_5$CN can be spectrally resolved.
Our survey has higher sensitivity and smaller beam size, and this may make our
survey more sensitive to emission of compact sources such as
the emission regions of complex organic molecules (COMs).

Two transitions of H$_2$CS were detected in this survey.
The line width of H$_2$CS  is consistent with the value of HCS$^+$ (see Fig. \ref{continued_fit_spec} and Table
\ref{model_pars_table}).
This implies that H$_2$CS and HCS$^+$ may  originate from 
similar emission regions and have tight chemical correlation, through reactions such as \citep{2013A&A...550A..36M}
\begin{equation} 
\begin{aligned}
&\rm  CH_2^++OCS \to H_2CS^++CO\\
&\rm  CH_2^++OCS\to HCS^++HCO\\
\end{aligned}
\end{equation}

We note that the emission feature identified as the $J=8-7$ of C$_3$S by  \citet{2017A&A...605A..76R} can not be seen 
in the spectrum of this survey (Fig. \ref{fig_hcsp}). 
There is also no emission feature 
at C$_3$S $J=7-6$ (40465 MHz). Since $J=8-7$ is the only  line of C$_3$S detected by  \citet{2017A&A...605A..76R}
with limited S/N,
the contribution of noise can not be ignored.
The column density of C$_3$S may be much lower than the value estimated by \citet{2017A&A...605A..76R}.
However, since our observation has a smaller beam size, 
it can not be excluded that the C$_3$S emission
detected by \citet{2017A&A...605A..76R} originates from regions uncovered by the beam of this survey.
The CCS emission identified by \citet{2017A&A...605A..76R} is also not detected or only marginally detected in
this survey (Fig. \ref{continued_fit_spec}). 

One transition of O$^{13}$CS is observed by \citet{2017A&A...605A..76R} and two by this survey.
The emission of O$^{13}$CS is marginally detected in this survey and \citet{2017A&A...605A..76R}.
For our survey,  since there are weak emission features at both of its two transitions, 
O$^{13}$CS is marked as a detected species.

\subsubsection{Oxygen-bearing organic molecules}
In total, 15 kinds of oxygen-bearing organic species are detected in this survey.
We use three velocity components to fit the emission of CH$_3$OH,
including two narrow component with $\Delta V=4$ km s$^{-1}$ and a broad component
with $\Delta V=25$ km s$^{-1}$ (Table \ref{model_pars_table}).
The wide wings of $^{13}$CH$_3$OH have not been obviously
detected, thus the broad component ($\Delta V=30$ km s$^{-1}$) 
has not been reused from CH$_3$OH to fit the spectrum of $^{13}$CH$_3$OH.
The spectrum of $^{13}$CH$_3$OH can be well modeled when adopting the
column densities of $^{13}$CH$_3$OH as 0.01 times the values of the two narrow components of CH$_3$OH
(see Figs. \ref{example_fit_spec}). 
If the CH$_3$OH and $^{13}$CH$_3$OH have identical emission regions, it implies that
the abundance ratio between $^{13}$CH$_3$OH  and CH$_3$OH can be estimated as 0.01.
For E-CH$_3$OH $\varv_t=1$ and A-CH$_3$OH $\varv_t=1$,
the model parameter of CH$_3$OH   can also be applied to them
 without any modification to reproduce their spectra, assuming that the
A-type and E-type of CH$_3$OH have equal abundances.
C$_2$H$_5$OH is also detected in this survey. The model fitting gives an abundance ratio between C$_2$H$_5$OH
and CH$_3$OH of 0.003.

Apart from H$_2$CO, CH$_3$OCHO which have been detected by \citet{2017A&A...605A..76R}, 
the molecules containing an aldehyde group detected in this survey include H$_2^{13}$CO,
H$_2$CCO, CH$_3$CHO. The abundance ratio between H$_2^{13}$CO and
H$_2$CO is estimated as 0.02, and the value between  CH$_3$CHO and H$_2$CO is 0.06. 
The first vibrationally excited state of CH$_3$OCHO ($\varv_t=1$)  
was detected  by \citet{2017A&A...605A..76R} through stacking faint lines, most of them are near the detection limit.
In our survey, more than 10 lines of  CH$_3$OCHO $\varv_t=1$ have been detected with significant S/Ns
(Figs. \ref{fig_c2h5cn} and \ref{continued_fit_spec}).

Three lines of HCOOH (formic acid) are detected in this survey, with a velocity of $\sim$7.5 km s$^{-1}$.
\citet{2002ApJ...576..255L} observed the HCOOH lines at 225 GHz and 262 GHz using the 
Berkeley-Illinois-Maryland Association (BIMA) array,
and the line emission was found to peak in the velocity range of 6.9 and 8.4 km s$^{-1}$.
Near the compact ridge, the  HCOOH emission was spatially resolved by
\citet{2002ApJ...576..255L}, showing a partial shell
morphology. 
If an emission size of 10\arcsec~and an excitation temperature of 60 K is adopted,
the fitting of Q-band lines of this survey gives a HCOOH column densities of 3.0$\times$10$^{14}$ cm$^{-2}$
(Table \ref{model_pars_table}).
It is consistent with the value of 4.8$\times$10$^{14}$ cm$^{-2}$ derived by \citet{2002ApJ...576..255L} 
adopting an excitation temperature of 62 K.

Interstellar acetone (CH$_3$COCH$_3$) is a complex organic
molecule that has been previously detected in several hot cores and low-mass protostars \citep[e.g.,][]{1987A&A...180L..13C,
2011A&A...534A.100J,2012ApJS..201...17F,2017ApJ...849..139Z}.
Towards the Orion KL, 
the transitions of CH$_3$COCH$_3$ were not detected or only marginally detected by \citet{2017A&A...605A..76R}.
Six CH$_3$COCH$_3$ lines were identified by \citet{2009ApJ...691.1254G}. 
In this survey, more than ten weak emission features 
were assigned to transitions of CH$_3$COCH$_3$ (see Table \ref{linelist}).
However, most of the them are marginally detected.
We stacked the lines of CH$_3$COCH$_3$ to justify  the detection of 
CH$_3$COCH$_3$ in Sect. \ref{sec_stack}.


\subsubsection{Cyanopolyynes}
Cyanopolyynes (HC$_{2n+1}$N)  are chemically young species, and they are usually detected in cold cores,
cold/warm ambient gas of star-forming regions and the envelops of late-type carbon stars 
\citep[e.g.,][]{2001ApJ...558..693D,2013A&A...559A..51E,2017A&A...601A...4A,2021ApJ...912..148L}.
The vibrational levels of HC$_3$N are mainly excited by mid-IR radiation \citep{2000A&A...361.1058D},
and it makes HC$_3$N also a good tracer of hot cores \citep{1971ApJ...163L..35T,2020MNRAS.496.2790L}.

For  the isotopologues of HC$_3$N (including H$^{13}$CCCN, HC$^{13}$CCN, and HCC$^{13}$CN),
the observations of \citet{2017A&A...605A..76R} only covered their
$J=5-4$ transitions. 
This survey covers the transitions of both $J=4-3$ and $J=5-4$.
Thus, we can better constrain the emission model (Sect. \ref{sect_model_mol}) 
of HC$_3$N and its isotopologues. The modeled emission of H$^{13}$CCCN, HC$^{13}$CCN and HCC$^{13}$CN
with identical column densities and excitations approximately reproduces the observations for all their detected transitions
(Fig. \ref{continued_fit_spec}).
We speculate that the ion-molecule reactions are preferred to produce HC$_3$N in Orion KL, since they may
not lead to significant divergence of abundance among different isotopologues  
of HC$_3$N as compared with the neutral-neutral reactions \citep{2016ApJ...830..106T}.

Only two vibrationally excited HC$_3$N transitions were firmly detected in \citet{2017A&A...605A..76R}.
In this survey, at least two unblended transitions are detected for each of the three  vibrational levels 
($\varv_6=1$,  $\varv_7=1$, $\varv_7=2$) of  HC$_3$N.
Five transitions of HC$_5$N are also detected in this survey.

\begin{figure}[!t]
\centering
\includegraphics[width=0.8\linewidth]{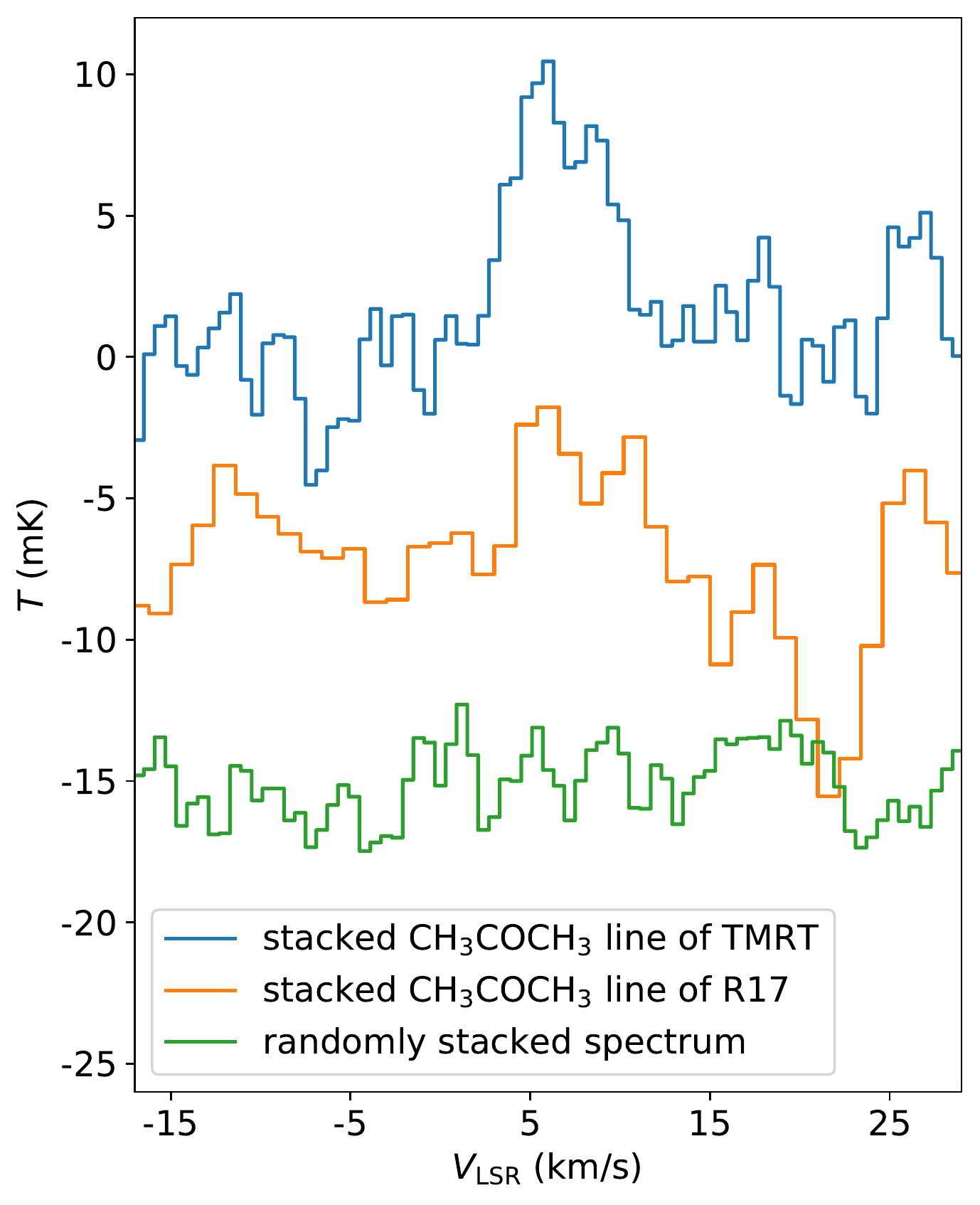}
\caption{The blue and orange lines are the stacked lines of CH$_3$COCH$_3$ of our survey and 
\citet{2017A&A...605A..76R}, respectively . 
The green line is stacked  from randomly chosen segments of our spectrum. 
See Sect. \ref{sec_stack} for details.
\label{fig_stacking}}
\end{figure} 

\begin{figure}[!t]
\centering
\includegraphics[width=0.95\linewidth]{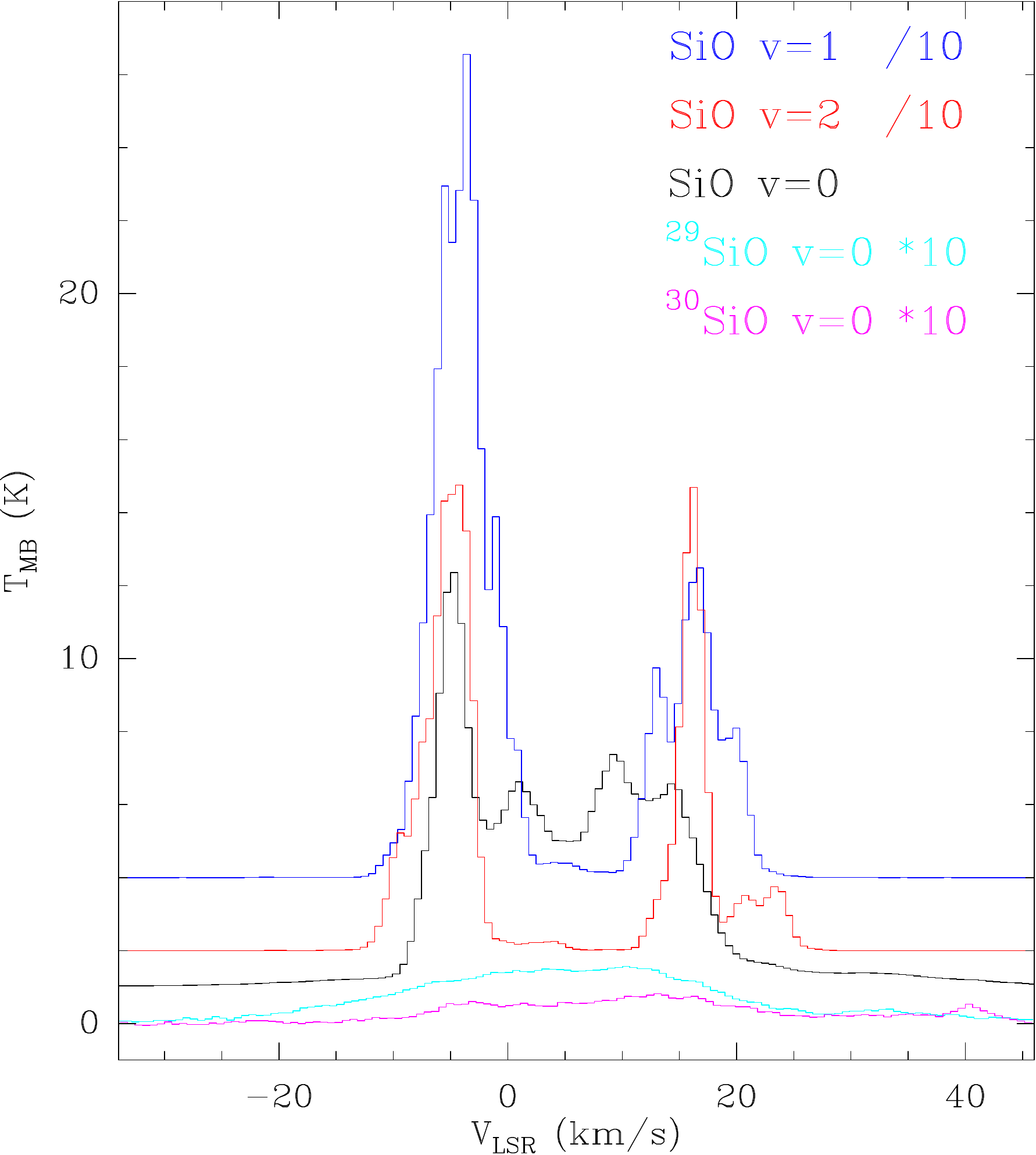}
\includegraphics[width=0.95\linewidth]{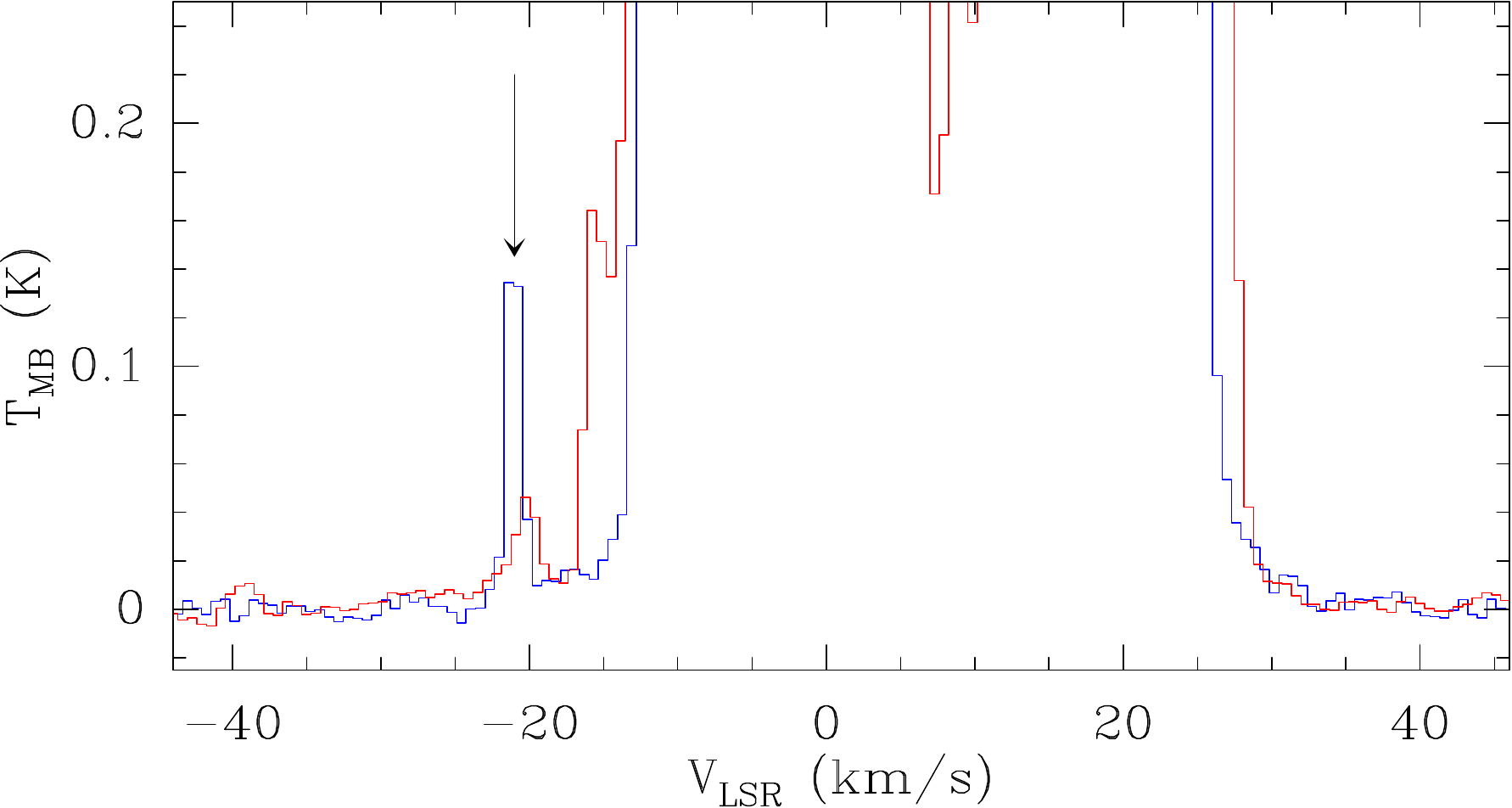}
\caption{Upper: the spectra of SiO and its isotopologues. Lower: Zoom in the spectra of SiO masers.
The spectra of SiO $\varv=1$ and $\varv=1$ in the lower panel has not been divided by 10 as
in the upper panel.
The black arrow marks the weak and narrow feature beside the main features of SiO $\varv=1$ (blue)
and $\varv=2$ (red). \label{fig_sio}}
\end{figure} 

\subsubsection{C$_2$H$_5$CN $\varv$13/$\varv$21} \label{sec_c2h5cn_vib}
Ten lines of C$_2$H$_5$CN $\varv 13/\varv 21$ are detected in this survey. Among those lines, nine are spectrally
resolvable and they are marked by red strips in Figs. \ref{fig_comparerizzo} and \ref{specset_example}.
Those lines are  all undetected or  only marginally detected (with S/N$<$3)  by \citet{2017A&A...605A..76R}.
To identify those lines, we privately communicated with the authors of \citet{2021JMoSp.37511392E} 
to obtain the transition parameters (rest frequency, $E_{\rm u}$, $A_{i,j}$, and g$_u$) 
of transitions of  C$_2$H$_5$CN $\varv$13/$\varv$21 within 34.8--50 GHz. 
Emission lines corresponding to those transitions are clearly detected in our survey (Fig. \ref{fig_c2h5cn}).
We modeled them using the method 
described in Sect. \ref{sect_model_mol}. 
The modeled spectrum is consistent with that observed in this survey,
as shown in the upper panel of Fig. \ref{fig_c2h5cn}.
A $T_{\rm ex}$ of 50 K is assumed to model the spectrum of C$_2$H$_5$CN $\varv$13/$\varv$21,
although $T_{\rm ex}$ can not be well constrained since all the detected transitions
have similar $E_{\rm u}$. The line width is fitted as  6.5 km s$^{-1}$. 
The column density of C$_2$H$_5$CN $\varv$13/$\varv$21 is
undecided since we do not know the partition function.

The  in-plane bending vibration $\varv 13 = 1$ and  torsional state $\varv 21=1$ of 
C$_2$H$_5$CN \citep{2013ApJ...768...81D} have been detected towards Sgr B2(N-LMH) by \citet{2004ApJ...608..306M}
with rest frequencies ranging from 100 to 270 GHz.
C$_2$H$_5$CN in vibrationally excited states  has also been identified in Orion KL 
by \citet{2013ApJ...768...81D} from the data of an
IRAM millimeter survey \citep{2011A&A...528A..26T}.
\citet{2017A&A...605A..76R}
tried to search for signals of those transitions in the Q-band spectrum of their survey,
but they failed to find signals with intensities higher than 3$\sigma$. 
The vibrationally excited states ($\varv 13$/$\varv 21$) of  
C$_2$H$_5$CN  in the Q band are for the first time  firmly detected
in this survey.

\subsubsection{HCN $\varv_2=1$ and NH$_3$} \label{NH3_sec}
HCN in the vibrational state ($\varv_2=1$) is detected in their direct $l$-type lines in this survey (Fig. \ref{continued_fit_spec}). 
HCN $\varv_2=1$ is a vibrationally bending mode with a 
ground state energy of  1024.4 K \citep{1970JRNBS..74..791R,2011A&A...529A..76R,2015A&A...575A..94B}.
If HCN in the vibrational state is thermally excited, an HCN column density of 1.8$\times$10$^{17}$ cm$^{-2}$ is derived 
with an assumed excitation temperature of 200 K and an emission size of 10\arcsec.
It is consistent with near-infrared observations towards the Orion KL using
the Short Wavelength Spectrometer of Infrared Space Observatory \citep{2003A&A...399.1047B},
where an HCN column density of $\sim 5\times 10^{16}$ cm$^{-2}$ was derived from the vibrational emission band
of HCN at 14.05 $\mu$m, adopting a
Doppler line width of 5 km s$^{-1}$  and an excitation temperature of 275 K.

Three transitions of NH$_3$ are detected, including the inversion lines of (14,14), (15,15) and (16,16).
The (14,14) is highly blended with the hydrogen RRLs and the emission of CH$_3$OCH$_3$.
The (15,15) is lightly blended with the emission of E-CH$_3$OCHO, which has a much narrower line width and
contributes no more than 10 percent intensities.
The (16,16) is close to a line feature of HC$_5$N, and it is easy to separate them through Gaussian fitting.
The emission of NH$_3$ lines is modeled by two components with line widths of 8 and 30 km s$^{-1}$.
The emission of NH$_3$ is necessary to reproduce the observed spectrum at the frequencies of the three transitions.
Thus, we consider NH$_3$ as a detected species.
The narrower component has a line width consistent with the value for NH$_2$D (8 km s$^{-1}$).
It is also consistent with the result of \citet{1993A&A...276L..29W}, who detected
inversion lines of NH$_3$ up to (14,14) with a line width of $\sim$10 km s$^{-1}$.
The detection of NH$_3$ (15,15) and (16,16) in this survey pushes  the upper-level energy of 
NH$_3$ emission lines detected towards Orion KL upwards to $>$2000 K.
 
\subsubsection{stacking lines of CH$_3$COCH$_3$}\label{sec_stack} 
We stacked the emission of CH$_3$COCH$_3$ to improve the S/N and to
consolidate the detection of  CH$_3$COCH$_3$ of this survey.
To avoid bias, all the unblended transitions of CH$_3$COCH$_3$ (including
both the detected ones and undetected ones) were stacked.
We first calculate the relative intensity ($r_i$) of the $i_{\rm th}$ transition through
Eq. \ref{eq_model_mol} assuming a $T_{\rm ex}$ of 100 K.
The $r_i$ is then normalized to have a maximum value of 1.
Then, the spectrum of each segments was rescaled through multiplying a factor of 1/$r_i$ to make them have
equal expected intensities. Obviously, the noise has also been amplified by  a factor of 1/$r_i$, simultaneously.  
The rescaled spectra are then averaged  to get the stacked spectrum, with a weight of $r_i^2$ for each. 
These procedures are equivalent to directly stacking up the original spectra weighted by $r_i/\sum_1^N r_i^2$. 
Here, $N$ is the number of stacked transitions.
We use $\sum_i^N r_i^2$ to represent the effective number of stacked transitions ($N_{\rm eff}$).
If $N_{\rm eff}$ is small, bias arising from possible noise spikes at
a few transitions may be non-ignorable.
The $N_{eff}$ is 11 and 9.5 for CH$_3$COCH$_3$ of this survey and of \citet{2017A&A...605A..76R}, respectively.
Thus, the stacking procedure is valid for CH$_3$COCH$_3$.
The stacked line of CH$_3$COCH$_3$ of our survey 
shows an obvious emission feature (Fig. \ref{fig_stacking}).
In the contrast, the stacked line of CH$_3$COCH$_3$ of \citet{2017A&A...605A..76R}
only has a weak emission feature lower than $3\sigma$.
We also stack a group of  spectral segments, with each of the $N$ segments
having a rest frequency randomly chosen within the unblended frequency ranges.
The randomly stacked spectrum show no emission feature as expected.
Thus, the weak emission features we assign to CH$_3$COCH$_3$
should be real, and CH$_3$COCH$_3$ is considered  a detected species.

\subsection{SiO emission}
\subsubsection{Overall description of SiO emission}
The upper panel of Fig. \ref{fig_sio} shows the $J=1-0$ spectra of SiO and its isotopologues
(SiO $\varv=0$, SiO $\varv=1$, SiO $\varv=2$, $^{29}$SiO $\varv=0$ and $^{30}$SiO $\varv=0$)  detected in this survey.
Overall, the line shapes agree with the single-dish results presented by \citet{2009ApJ...691.1254G} and \citet{2017A&A...605A..76R}. 
The $\varv=0$ lines of $^{29}$SiO  and $^{30}$SiO are broad and smooth, while the 
spectrum of SiO $\varv=0$ shows narrow features (Fig. \ref{fig_sio}).
This is consistent with previous results that the emission of  SiO $\varv=0$ is part maser
and part thermal \citep{1995ApJ...455L..67C,2009ApJ...691.1254G}.

SiO $\varv= 1,2$ transitions are known to have been inverted
since their discovery \citep{1974ApJ...192L..33T,1974ApJ...192L..97B}. 
The  SiO $\varv= 1$ of this survey and  \citet{2009ApJ...691.1254G} have a nearly identical shape. 
This implies that the shape of SiO $\varv= 1$ may have not obviously changed during the past ten years under the single-dish
view. However, the bright and very narrow feature in the $\varv=2$ line at $\sim-1.4$ km s$^{-1}$ 
reported by \citet{2017A&A...605A..76R} can not be seen in the spectra of \citet{2009ApJ...691.1254G} and this survey.
The velocity component of SiO $\varv= 2$ with a $V_{\rm LSR}$ of $\sim 22$ km s$^{-1}$ has a much lower intensity compared with
\citet{2009ApJ...691.1254G}. The SiO $\varv= 2$ seems to have experienced a more significant change
than SiO $\varv= 1$ during the past ten years.

\subsubsection{A very narrow and weak SiO maser component}
There is a very  narrow (spectrally not well resolved) and weak ($\sim 0.1$ K) emission feature 
at $V_{\rm LSR}\sim-22$ km s$^{-1}$ in the 
SiO $\varv=1$ spectrum (Fig. \ref{fig_sio}).
We have conducted observations covering this line on three different days adopting different
frequencies of the local oscillator (LO), and this narrow and weak feature was detected in
all sets. In the spectrum of SiO $\varv=2$, a broader feature was detected
at a similar velocity (Fig. \ref{fig_sio}). Thus, we identify the narrow and weak
emission feature in the  SiO $\varv=1$ spectrum as a narrow velocity component of SiO $\varv=1$. 


\subsection{Future aspects}
We took full use of the Q-band receiver of  TMRT to search for
emission lines of Orion KL under a sensitivity on the order of mK.
This survey proves that the TMRT is sensitive and stable enough for
conducting deep line surveys.
In this work, we have only displayed some preliminary results of this survey.
More detailed scientific analysis and modeling will be conducted in future work.
Further, we plan to make deeper Q-band integrations towards Orion KL using the
TMRT in the near future, aiming to improve the S/Ns of newly detected transitions, to
confirm the weak or blended emission features detected in this
survey and to search for the emission of new species.
We also plan to extend our survey to other sources including SgrB2 and G010.47+0.03,
and to other frequency bands available by TMRT, especially the Ka band,
which could fill the gap between the K band and Q band.

\section{Summary} \label{secsummary}
We have conducted a line survey towards the Orion KL using the 
Q-band receiver of Tianma 65 m telescope, covering a frequency range of 34.8--50 GHz. So far, 
this survey is  the most sensitive wide-band Q-band survey towards the Orion KL.
Compared with the survey of \citet{2017A&A...605A..76R}, this survey extends the frequency 
coverage from  41.5--50 GHz to 34.8--50 GHz with sensitivities and spectral resolution twice better in average. 
The main results of this survey include:
\begin{itemize}
\item[1.] In total, 597 emission features were extracted from the Q-band spectrum of Orion KL. 
Gaussian fitting was applied to those emission features. 
Among them, 177 radio recombination lines (RRLs) are identified, including 
126, 40, and 11 RRLs corresponding to hydrogen, helium, and carbon, respectively.
Further, 371 molecular transitions are identified, containing 330 unblended ones.
The detected  molecular transitions come from 53 species (including 
isotopologues and molecules in different vibrational levels). 
The thermal lines were then fitted with a radiative transfer model to
reproduce the observed spectrum and re-affirm the line identification.

\item[2.] For RRLs, the emission measure ($EM$) of the {H}{\sc ii} region is fitted to be 3.9$\times$10$^6$ cm$^{-3}$ pc
with an intensity ratio between He/H of 8.4\%. 
The maximum $\Delta n$ of unblended RRLs is 16 for hydrogen,
7 for helium, and 3 for carbon. The carbon RRLs are confirmed to originate from PDR regions
with a $V_{\rm LSR}\sim 9$  km s$^{-1}$.
The $EM$ of PDR region is estimated to be $3.9\times 10^3$ (1.8$\times$10$^2$) cm$^{-6}$ pc, if
an electron temperature of 2000 (300) K is assumed.

\item[3.] The line shape of SiO $\varv= 1$ maser may have not obviously changed since \citet{2009ApJ...691.1254G},
while  the $\varv= 2$ maser tend to be more time-varying in the past ten years. 
A very  narrow (spectrally not well resolved) and weak ($\sim 0.1$ K, more than 1000 times lower than the
main peak) emission feature 
of SiO $\varv=1$ is detected at $V_{\rm LSR}\sim-22$ km s$^{-1}$.

\item[4.] The detected molecular species are mainly cyanopolyynes, 
sulfur-bearing species and oxygen-bearing organic molecules. Twenty-one
of the 53 detected species have not been firmly detected by \citet{2017A&A...605A..76R}, including
species such as
H$_2$CS, HCOOH, C$_2$H$_5$OH, H$_2^{13}$CO, 
H$_2$CCO, CH$_3$CHO, CH$_2$OCH$_2$, HCN $\varv_2=1$,  CH$_3$OCHO $\varv_t=1$ and C$_2$H$_5$CN $\varv 13$/$\varv$21.
The transitions of CH$_3$COCH$_3$ are stacked to confirm the detection of CH$_3$COCH$_3$.
The vibrationally excited states of  C$_2$H$_5$CN  in the Q band are for the first time  firmly detected
in this survey.
The detection of NH$_3$ (15,15) and (16,16) in this survey pushes  the upper-level energy of 
NH$_3$ emission lines detected towards Orion KL upwards to $>$2000 K.
\end{itemize}

This is the first systematic line survey of the TMRT, using its Q-band receiver and targeting on Orion KL.
In this work, we have only displayed some preliminary results of this survey.
More detailed scientific analysis, modeling and further follow-up surveys
will be conducted in future work.   

\begin{acknowledgments} \small
We wish to thank the staff of the Tianma 65 m for their
help during the observations.
Xunchuan Liu acknowledges the supports by NSFC No. 12033005.
Tie Liu acknowledges the supports by National Natural Science Foundation of China (NSFC) through grants No.12073061 and No.12122307, the international partnership program of Chinese Academy of Sciences through grant No.114231KYSB20200009, Shanghai Pujiang Program 20PJ1415500 and the science research grants from the
China Manned Space Project with no. CMS-CSST-2021-B06.
K.T. was supported by JSPS KAKENHI (Grant Number 20H05645). 
\end{acknowledgments}

\facilities{Tianma 65 m (TMRT)}
\software{GILDAS/CLASS \citep{2000ASPC..217..299G}, astropy \citep{2013A&A...558A..33A}}

\bibliography{ms}
\bibliographystyle{aasjournal}

\clearpage   
\appendix 
\section{Line list of molecular lines}
Table \ref{linelist} shows the results of identification and Gaussian fitting  of 
molecular lines, including RRLs which are blended with molecular lines.
The unblended RRLs are listed in Table \ref{linelist_rrls}. \\
{
\footnotesize
\textbf{Table Notes}:
\begin{itemize}\setlength{\itemsep}{0pt} \setlength{\parskip}{0pt}
\item[(1)] Doppler correction has been applied to $f_{\rm obs}$ assuming a 
source velocity  of 6 km s$^{-1}$ in LSR (Sect. \ref{sec_dr}). \\
The first-column item in blue means the corresponding 
transition has been firmly detected and resolved by \citet{2017A&A...605A..76R}.\\
Rows with empty $f_{\rm obs}$ correspond to blended lines.
\item[(2)] Rows with a same species name, $f_{\rm rest}$ and transition label correspond
to different emission components of a same transition.\\
The transition labels for HCN $\varv_t=1$ are $J^p$ with $p$ the parity.\\
The transition labels for CH$_3$OCH$_3$ are  $J_K^s$ with $s$ the symmetry substate.\\
The transition labels are $(J,K)$ for NH$_3$ convention lines.\\
The transition labels for CH$_3$CN $\varv_t=1$ are  $J_K^p$ with $p$ the parity.
\item[(3)] The numbers in brackets in the 7th and 8th columns 
represent the uncertainties of the \textbf{last digital} of corresponding parameters.\\
\end{itemize}
}

\startlongtable
\begin{deluxetable}{lllllllll}\label{linelist}
\tablecaption{Linelist}
\tablehead{
\colhead{$f_{\rm obs}$} & \colhead{Species} & \colhead{$f_{\rm rest}$} & \colhead{Transition} & \colhead{$E_u$} & \colhead{$A_{ij}$} & 
\colhead{$\int {T_{\rm MB}}{\rm d}V$} & \colhead{$\Delta$V} & \colhead{$T_{\rm MB}^{\rm peak}$} \\
\colhead{(MHz)} & \colhead{}  & \colhead{(MHz)} & \colhead{$J_{K_a,K_c}$} & \colhead{(K)} 
& \colhead{(10$^{-6}$ s$^{-1}$)} & \colhead{(K km s$^{-1}$)} & \colhead{(km s$^{-1}$)} & \colhead{K}
}
\startdata
34824.24                      & C$_2$H$_5$CN                    & 34824.070   & $4_{1,4}-3_{1,3}$                       & 5.3     & 3.04        & 0.94(5)             & 13(2)               & 7.04E$-$02\\
34831.44                      & U                               &             &                                         &         &             & 0.08(1)             & 3.7(9)              & 2.02E$-$02\\
34857.29                      & $^{34}$SO                       & 34857.543   & $N_J=3_{2}-2_{2}$                       & 20.9    & 0.2         & 1.71(5)             & 22(2)               & 7.20E$-$02\\
34954.41                      & He89$\delta$                    & 34954.331   &                                         &         &             & 0.48(5)             & 27(2)               & 1.66E$-$02\\
                              & HCN $\varv_2=1$                 & 34953.760   & $12^{-}-12^{+}$                         & 1357.7  & 0.03        &                     &                     &           \\
35135.15                      & H106$\eta$                      & 35132.748   &                                         &         &             & 2.68(2)             & 35.1(4)             & 7.19E$-$02\\
                              & H101$\zeta$                     & 35135.235   &                                         &         &             &                     &                     &           \\
                              & CH$_3$OCH$_3$                   & 35135.469   & $11_{5,6}-12_{4,9}$                     & 95.2    & 0.02        &                     &                     &           \\
\enddata
\end{deluxetable}
\section{Line list of RRLs unblended with molecular lines}
Table \ref{linelist_rrls} shows the results of identification and Gaussian fitting of
RRLs. The RRLs blended with molecular lines are not included (see Table \ref{linelist}).\\
{
\footnotesize
\textbf{Table Notes}:
\begin{itemize}\setlength{\itemsep}{0pt} \setlength{\parskip}{0pt}
\item[(1)] Doppler correction has been applied to $f_{\rm obs}$ assuming a 
source velocity  of 6 km s$^{-1}$ in LSR (Sect. \ref{sec_dr}). \\
The first-column item in blue means the corresponding 
transition has been firmly detected and resolved by \citet{2017A&A...605A..76R}.\\
Rows with empty $f_{\rm obs}$ correspond to blended lines.
\item[(2)] The numbers in brackets in the 4th and 5th columns 
represent the uncertainties of the \textbf{last digital} of corresponding parameters.\\
\end{itemize}
}

\startlongtable
\begin{deluxetable}{lllllllll}\label{linelist_rrls}
\tablecaption{Linelist}
\tablehead{
\colhead{$f_{\rm obs}$} & \colhead{Species} & \colhead{$f_{\rm rest}$}  & 
\colhead{$\int {T_{\rm MB}}{\rm d}V$} & \colhead{$\Delta$V} & \colhead{$T_{\rm MB}^{\rm peak}$} \\
\colhead{(MHz)} & \colhead{}  & \colhead{(MHz)} & \colhead{(K km s$^{-1}$)} & \colhead{(km s$^{-1}$)} & \colhead{K}
}
\startdata
34941.26                      & H89$\delta$                     & 34940.093    & 1.98(5)             & 19(2)               & 1.00E$-$01\\
35150.59                      & He101$\zeta$                    & 35149.553    & 0.10(1)             & 13(2)               & 7.41E$-$03\\
35158.29                      & H81$\gamma$                     & 35157.272    & 5.44(7)             & 23.6(7)             & 2.17E$-$01\\
35174.45                      & C81$\gamma$                     & 35174.814    & 0.05(1)             & 2.8(6)              & 1.67E$-$02\\
35251.78                      & H71$\beta$                      & 35250.774    & 10.63(5)            & 24(2)               & 4.19E$-$01\\
35266.15                      & He71$\beta$                     & 35265.138    & 0.84(1)             & 14.3(7)             & 5.52E$-$02\\
35267.93                      & C71$\beta$                      & 35268.362    & 0.12(1)             & 5.1(7)              & 2.15E$-$02\\
35457.42                      & H118$\kappa$                    & 35456.000    & 0.24(1)             & 14(1)               & 1.61E$-$02\\
\enddata
\end{deluxetable}

\clearpage
\begin{figure*}[!htb] 
\centering
\caption{Model fitting for different species. Continued from Fig. \ref{example_fit_spec}.
    \label{continued_fit_spec} 
}
\includegraphics[width=0.485\linewidth]{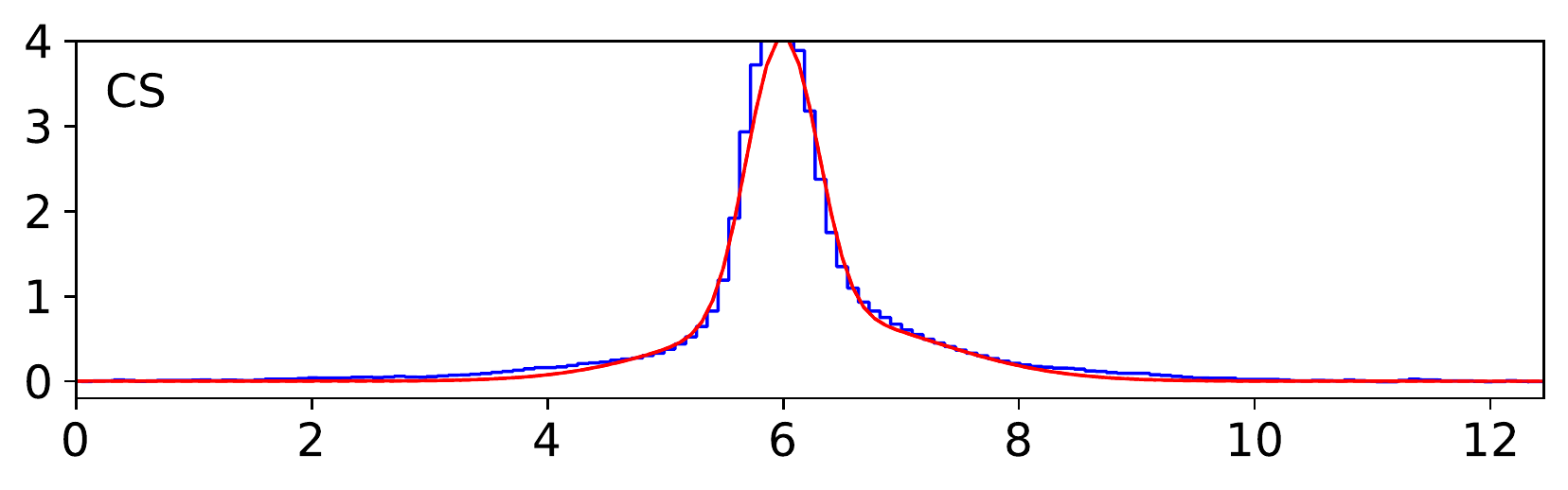}
\includegraphics[width=0.485\linewidth]{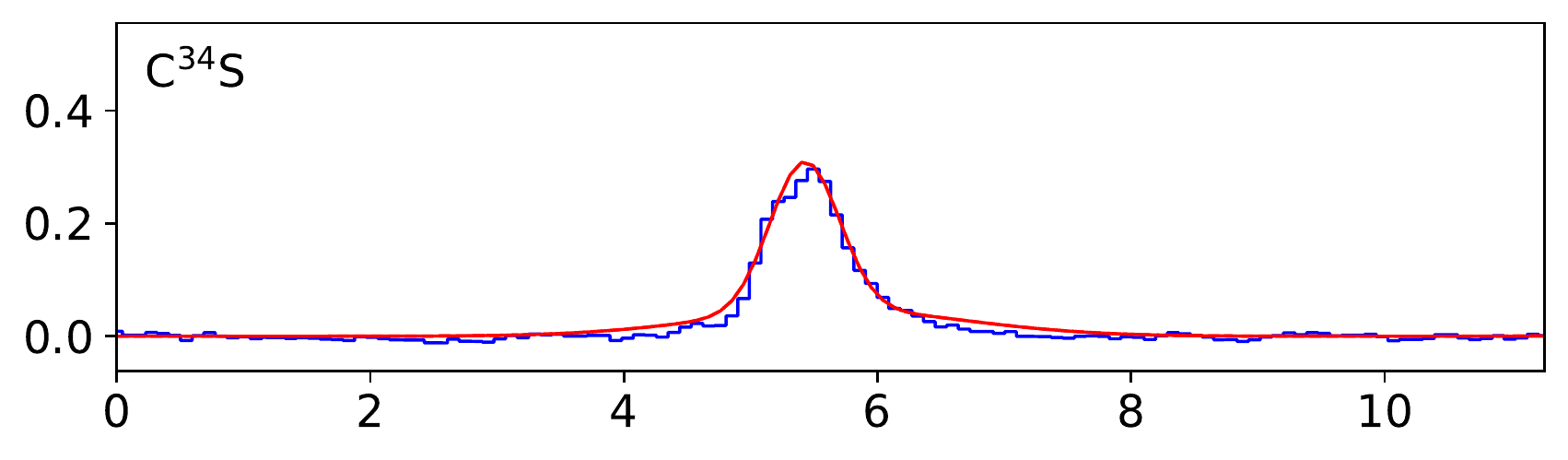}
\\
\includegraphics[width=0.485\linewidth]{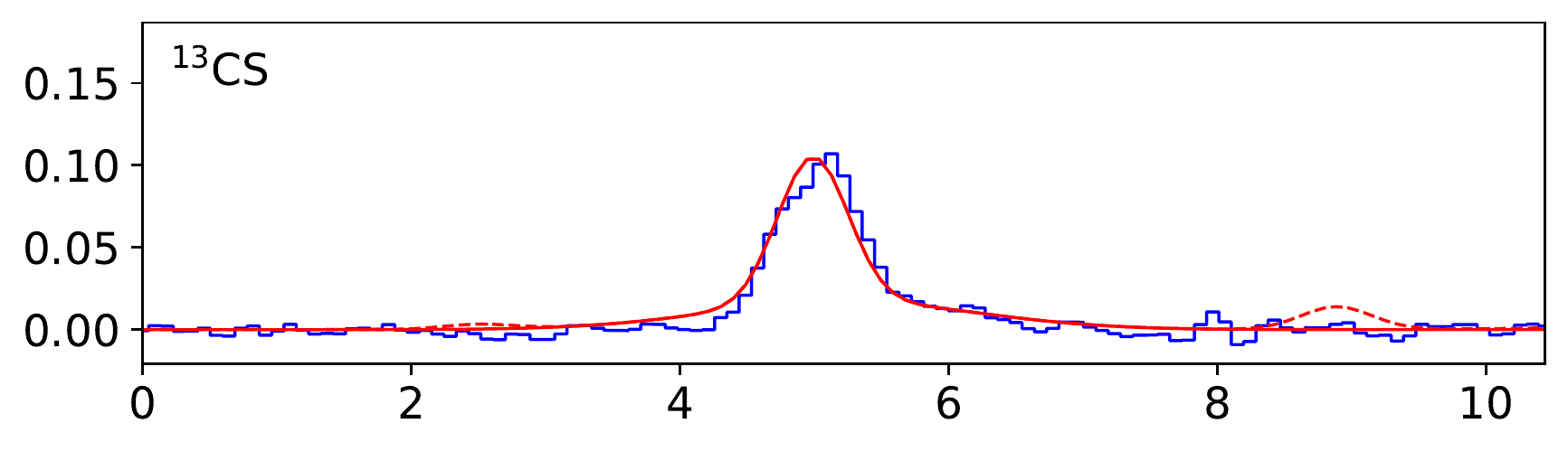}
\includegraphics[width=0.485\linewidth]{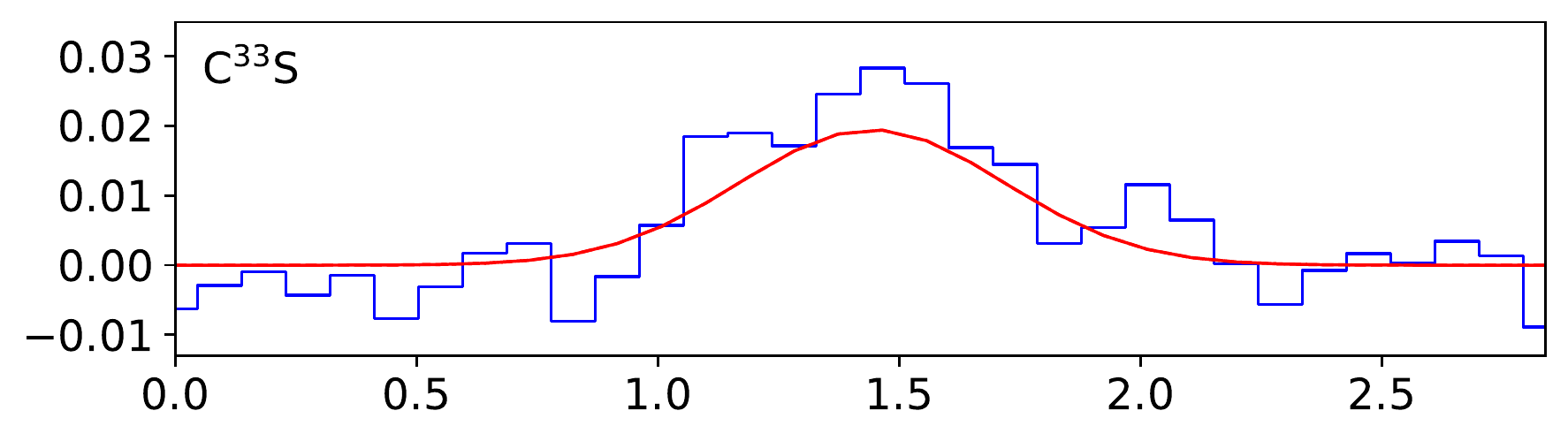}
\\
\includegraphics[width=0.485\linewidth]{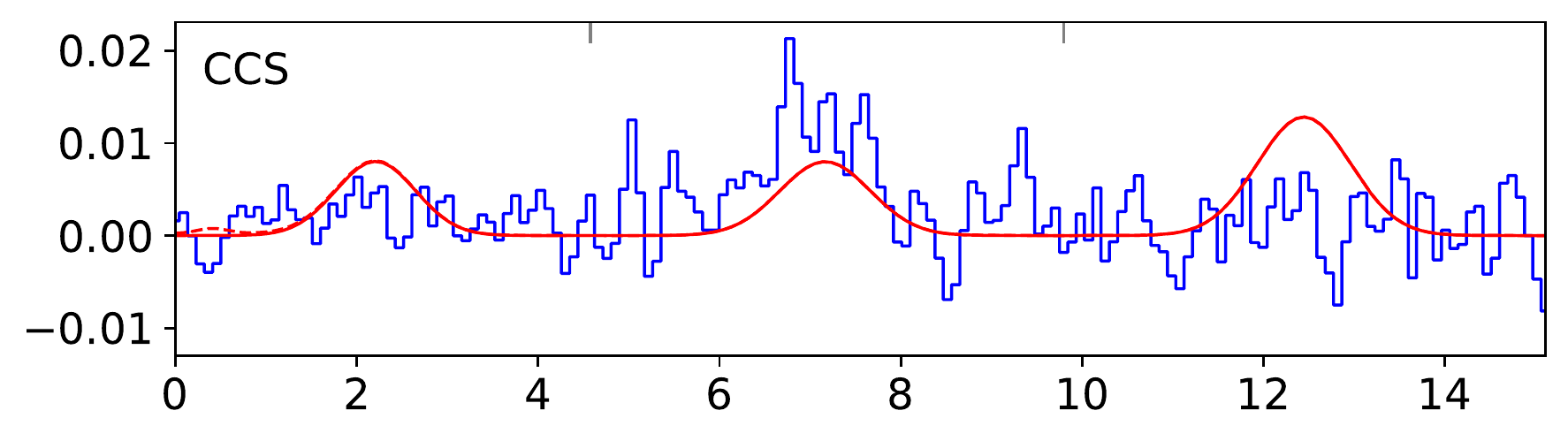}
\includegraphics[width=0.485\linewidth]{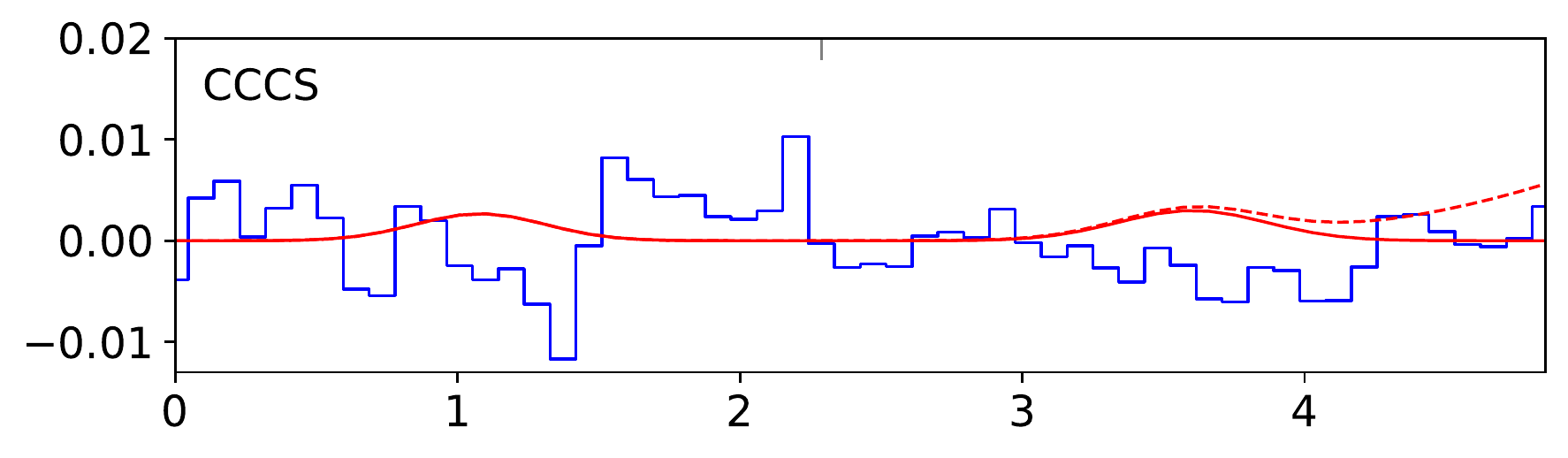}
\\
\includegraphics[width=0.485\linewidth]{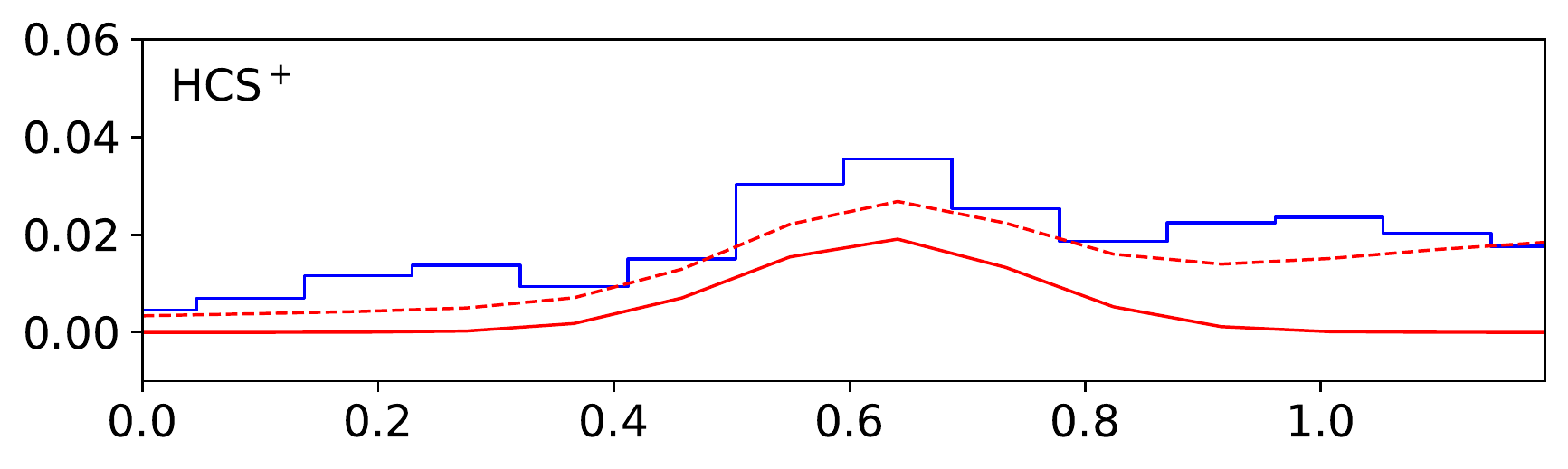}
\includegraphics[width=0.485\linewidth]{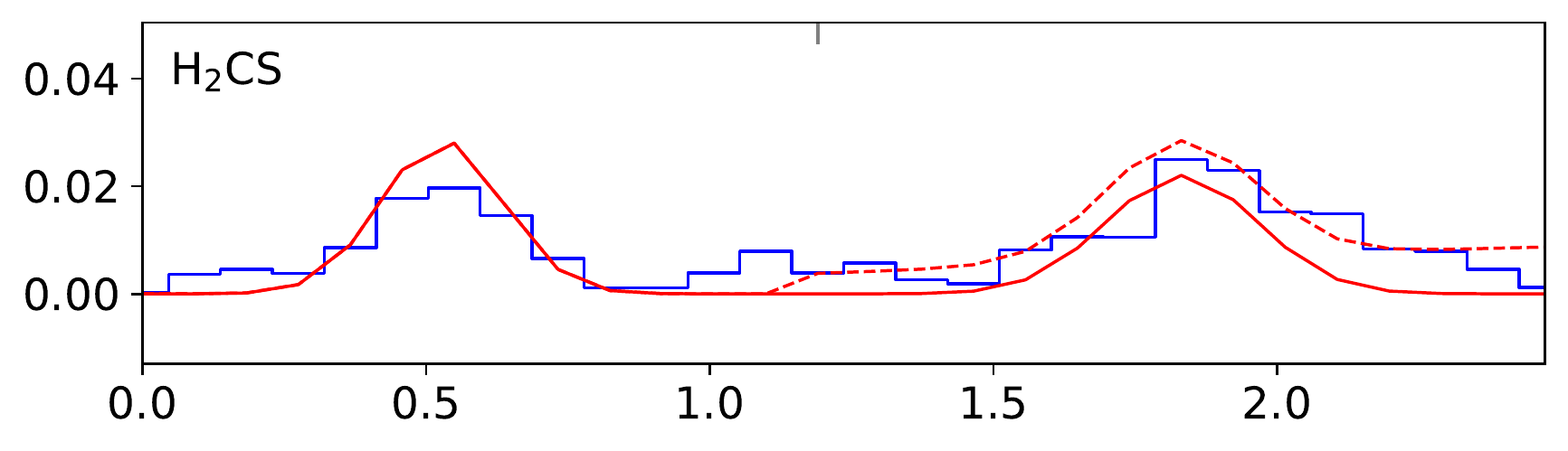}
\\
\includegraphics[width=0.485\linewidth]{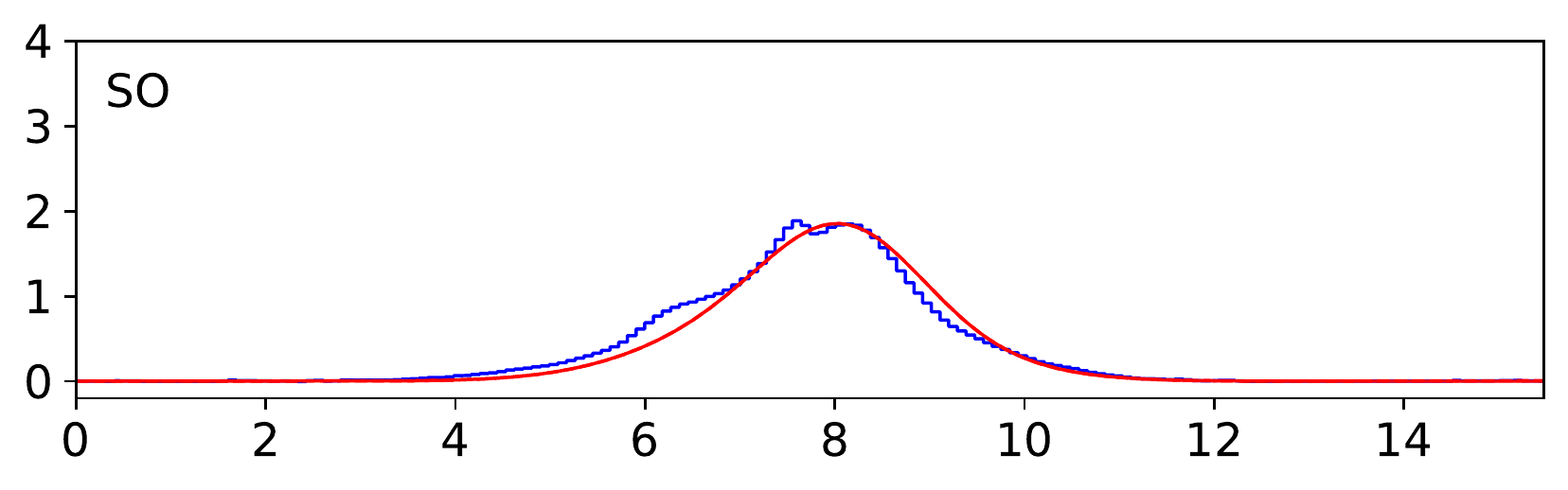}
\includegraphics[width=0.485\linewidth]{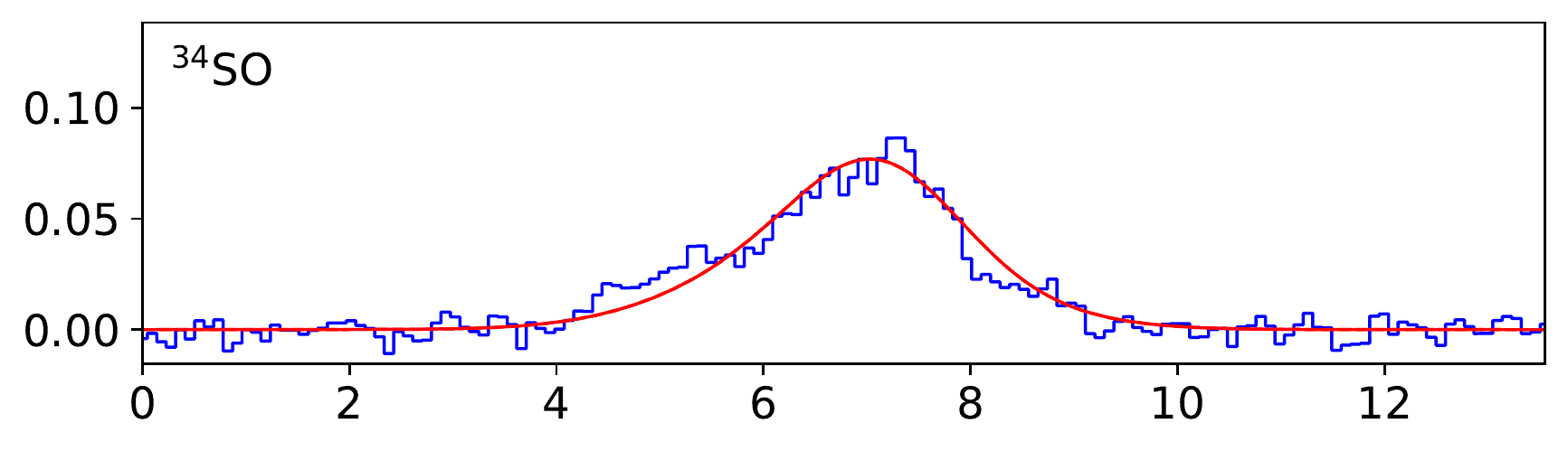}
\\
\includegraphics[width=0.485\linewidth]{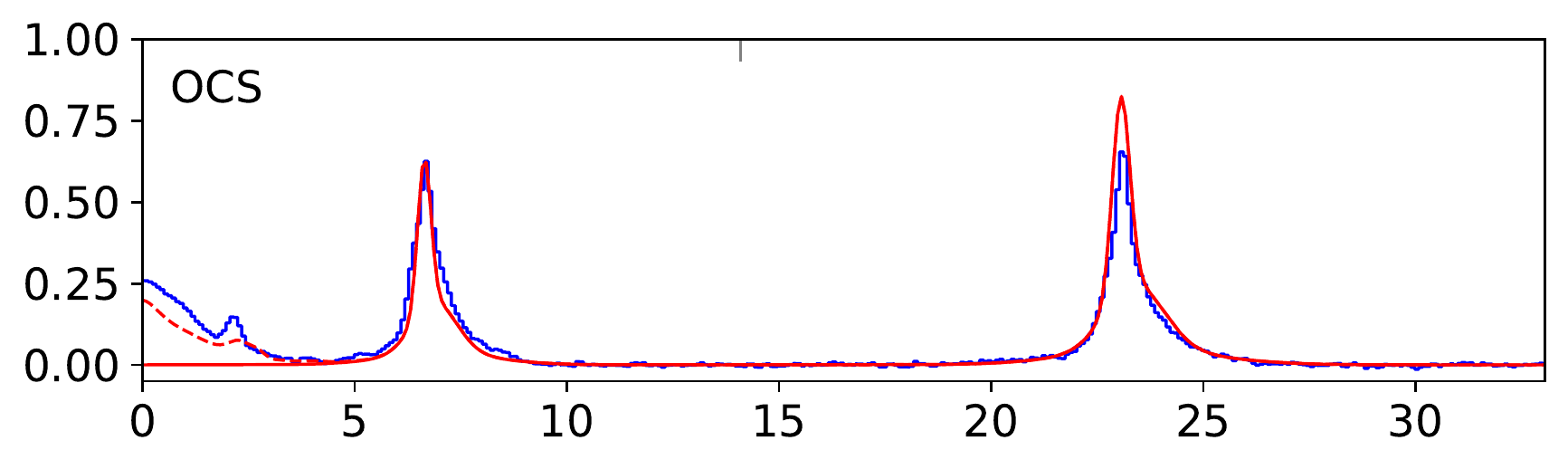}
\includegraphics[width=0.485\linewidth]{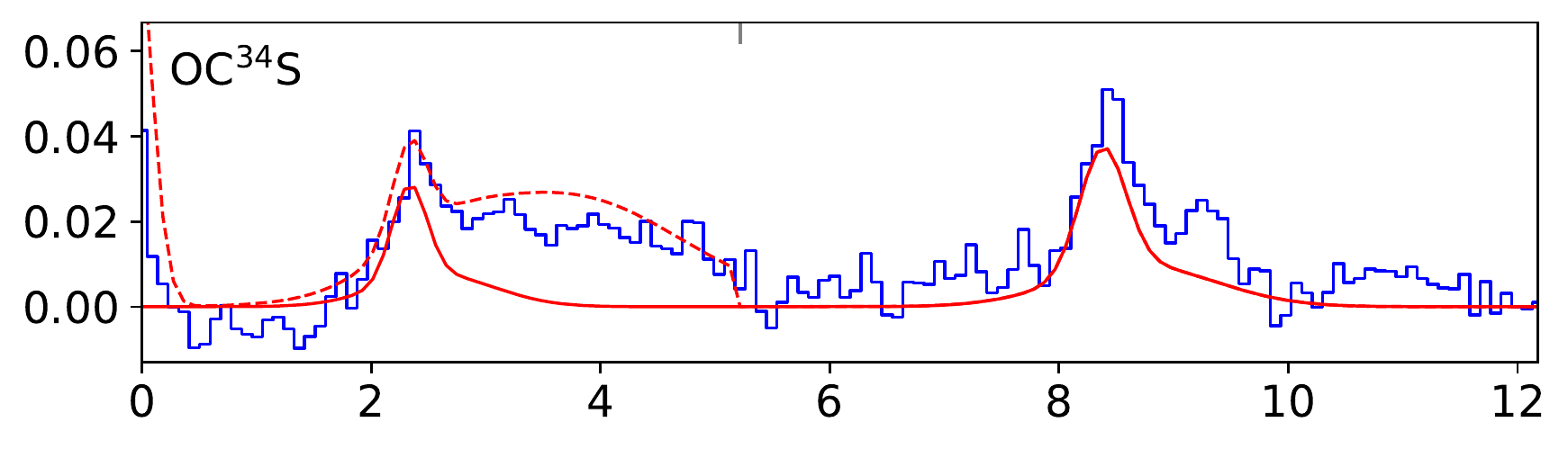}
\\
\includegraphics[width=0.485\linewidth]{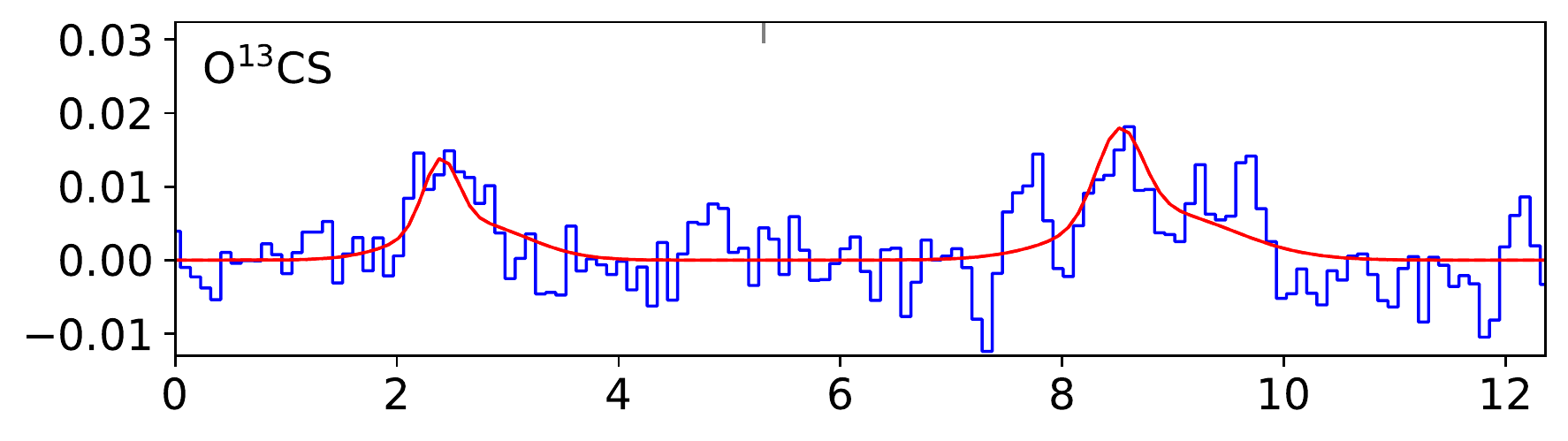}
\includegraphics[width=0.485\linewidth]{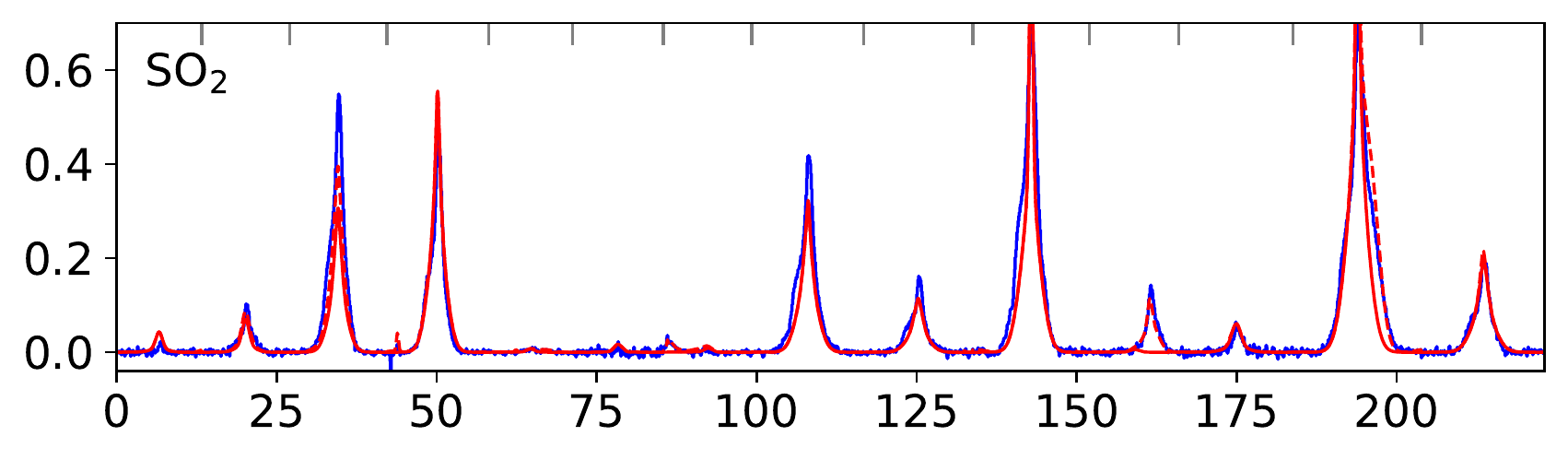}
\\
\vspace{1ex}
{\raggedright \footnotesize Note: In each panel, the spectrum between two neighbouring upper ticks are independent frequency segment
containing transitions of the corresponding species. 
The solid red line represents the model fitting of the corresponding species, while the 
dashed line includes the contributions of all the modeled molecular species (Table \ref{model_pars_table}) and RRLs
(see Sect. \ref{modelfit_sec} for details).
The $x$ axis is in unit of MHz. 
Each spectrum has been smoothed to a spectral resolution of 183 kHz.  \par}
{\raggedright \centering  \textbf{Figure \thefigure} {\it continued} \par}
\end{figure*}
        
\begin{figure*}
\centering
{\raggedright \centering  \textbf{Figure \thefigure} {\it (continued)} \par}
\includegraphics[width=0.485\linewidth]{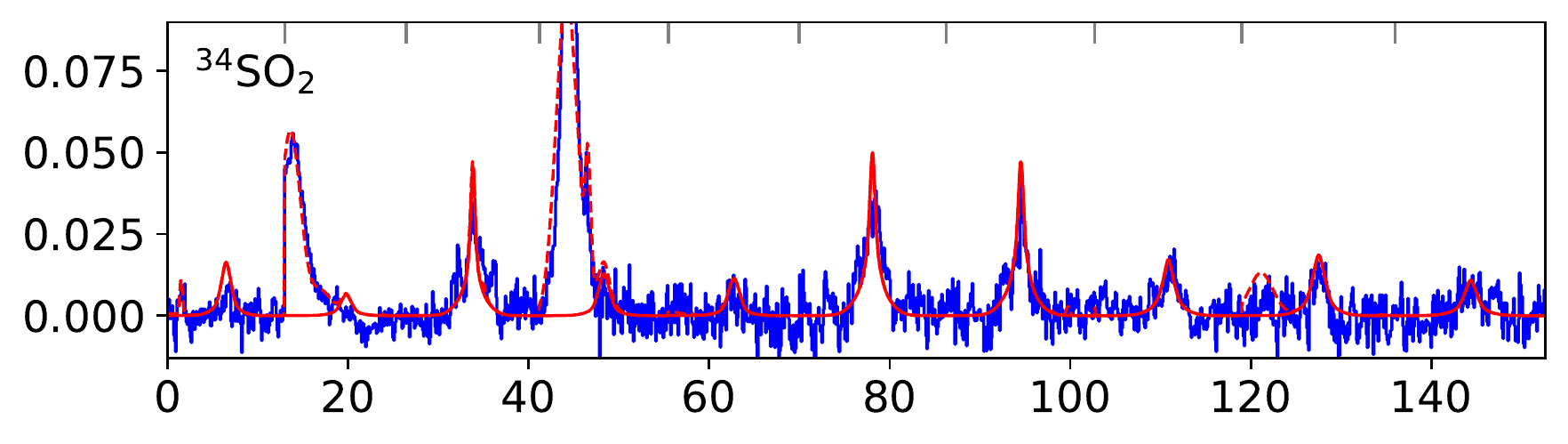}
\includegraphics[width=0.485\linewidth]{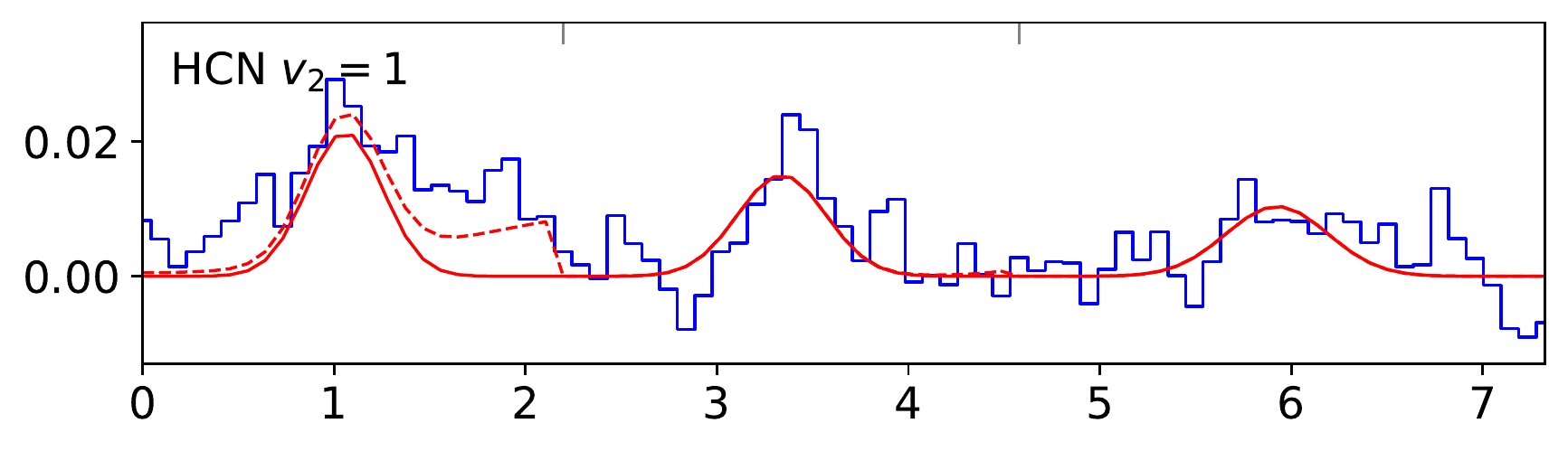}
\\
\includegraphics[width=0.485\linewidth]{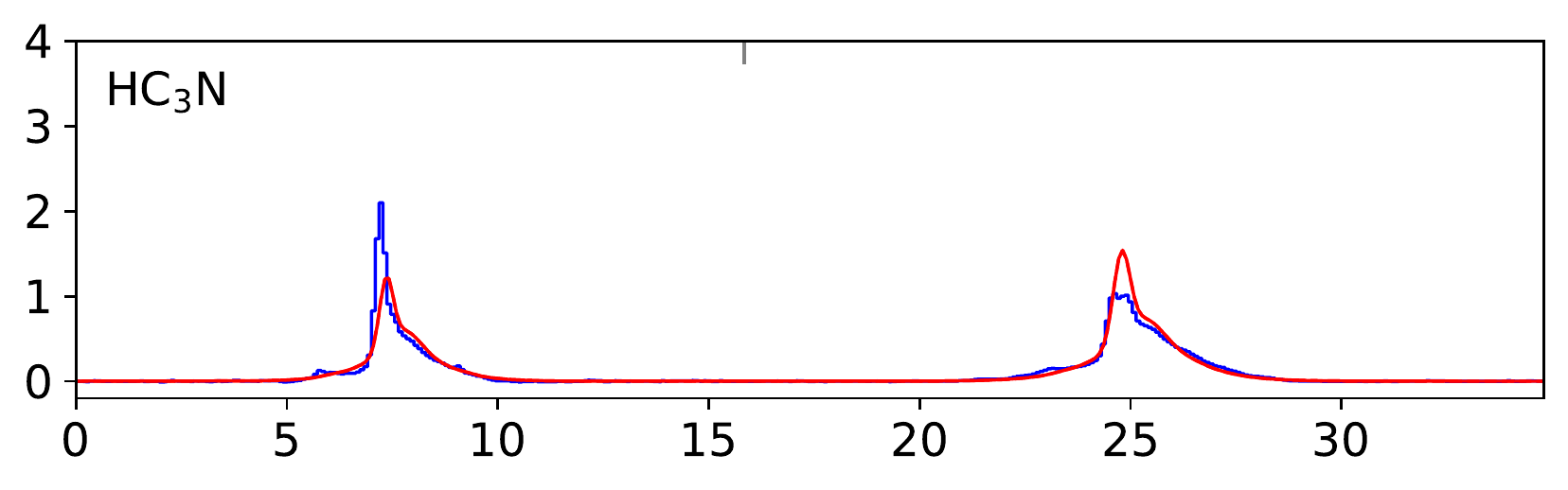}
\includegraphics[width=0.485\linewidth]{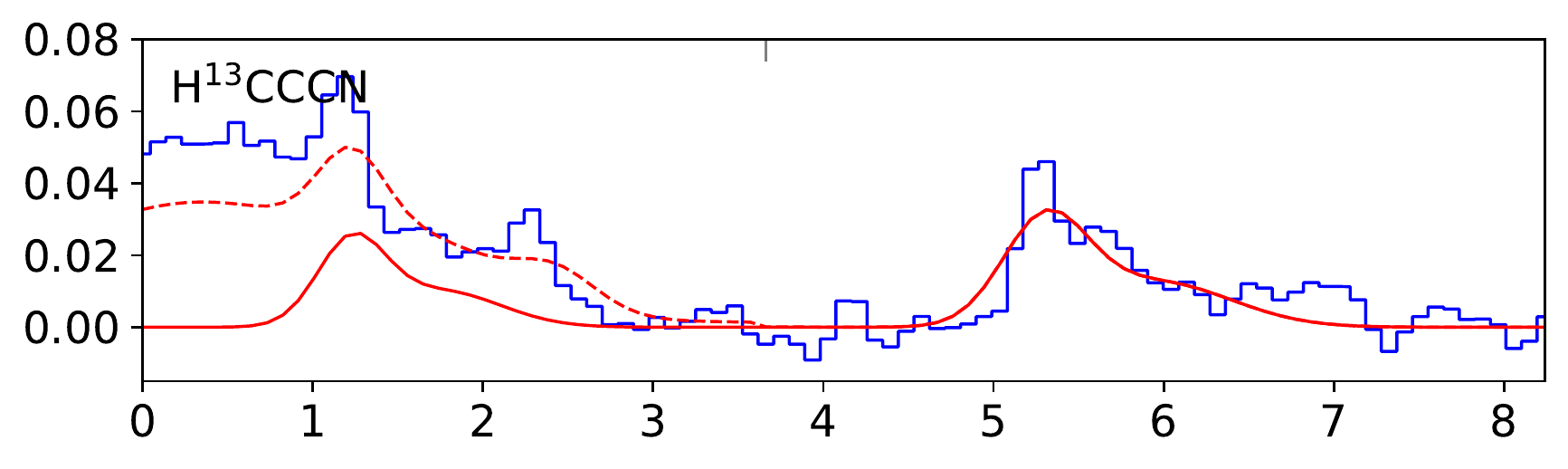}
\\
\includegraphics[width=0.485\linewidth]{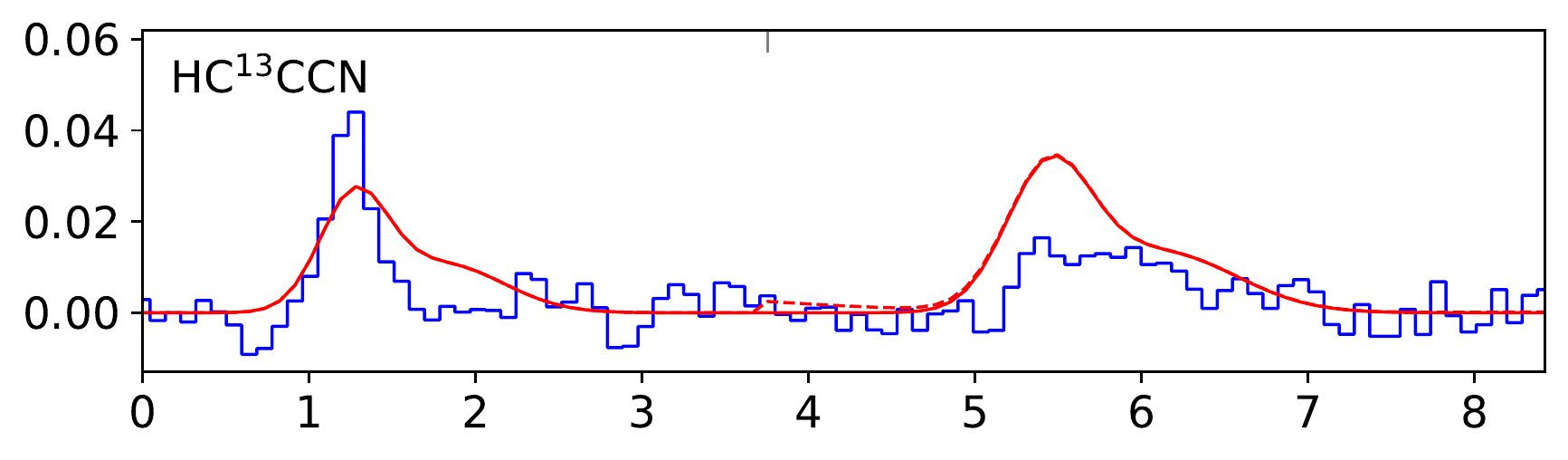}
\includegraphics[width=0.485\linewidth]{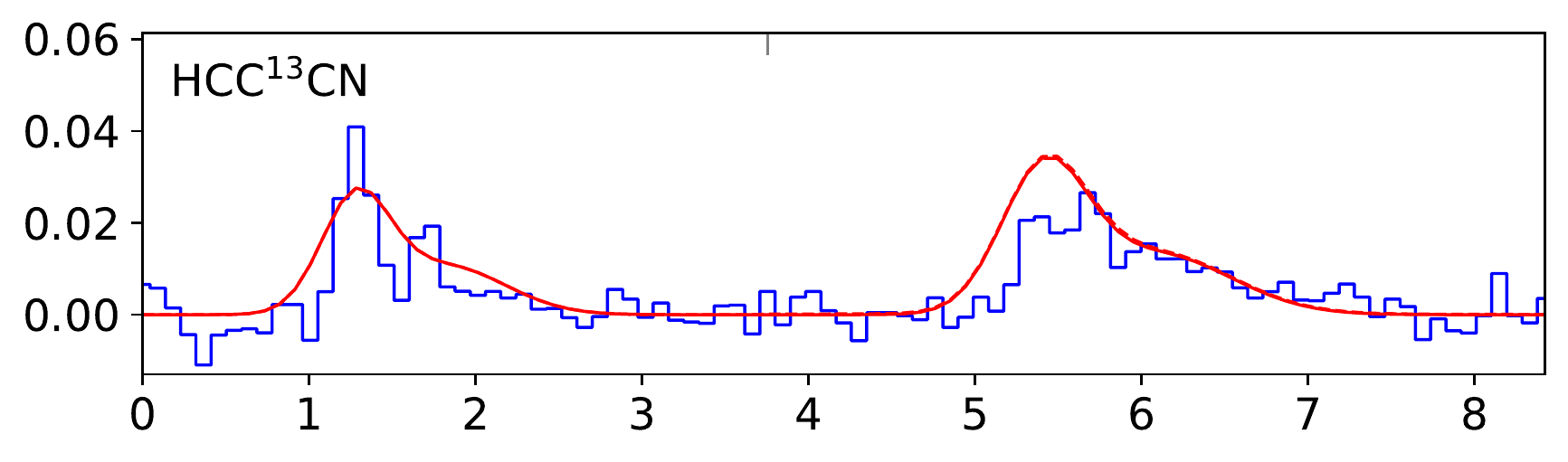}
\\
\includegraphics[width=0.485\linewidth]{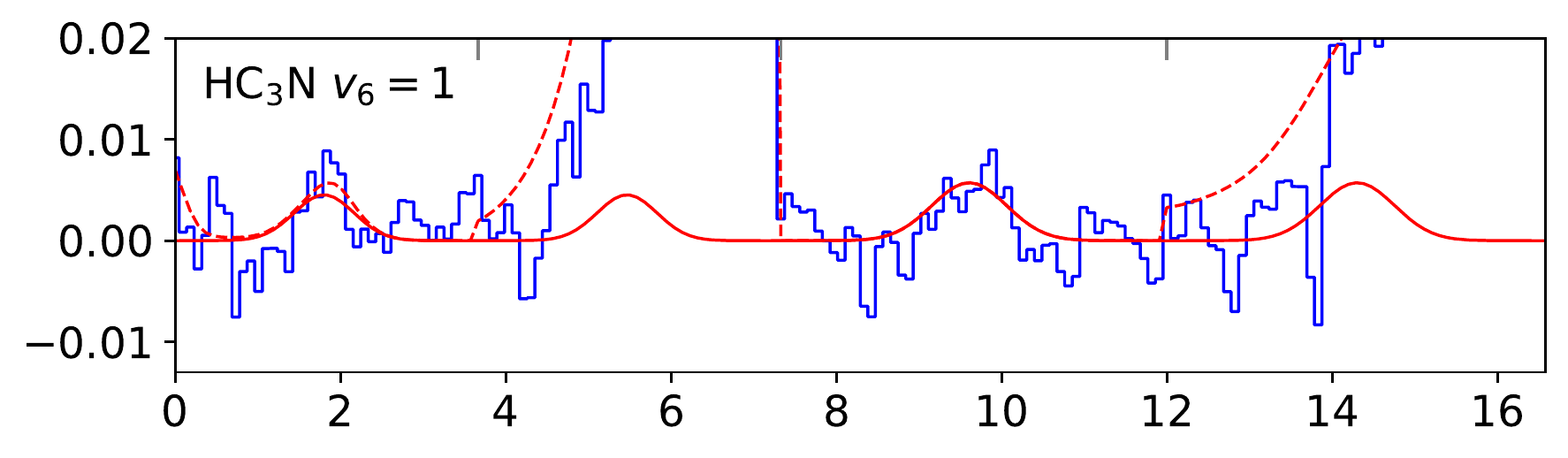}
\includegraphics[width=0.485\linewidth]{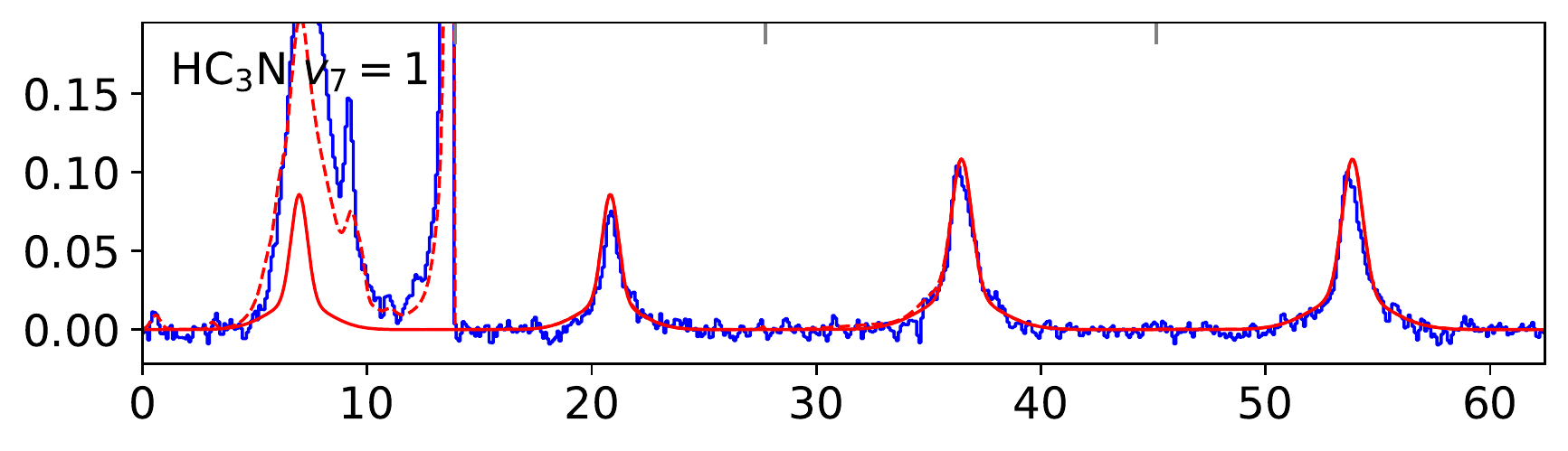}
\\
\includegraphics[width=0.485\linewidth]{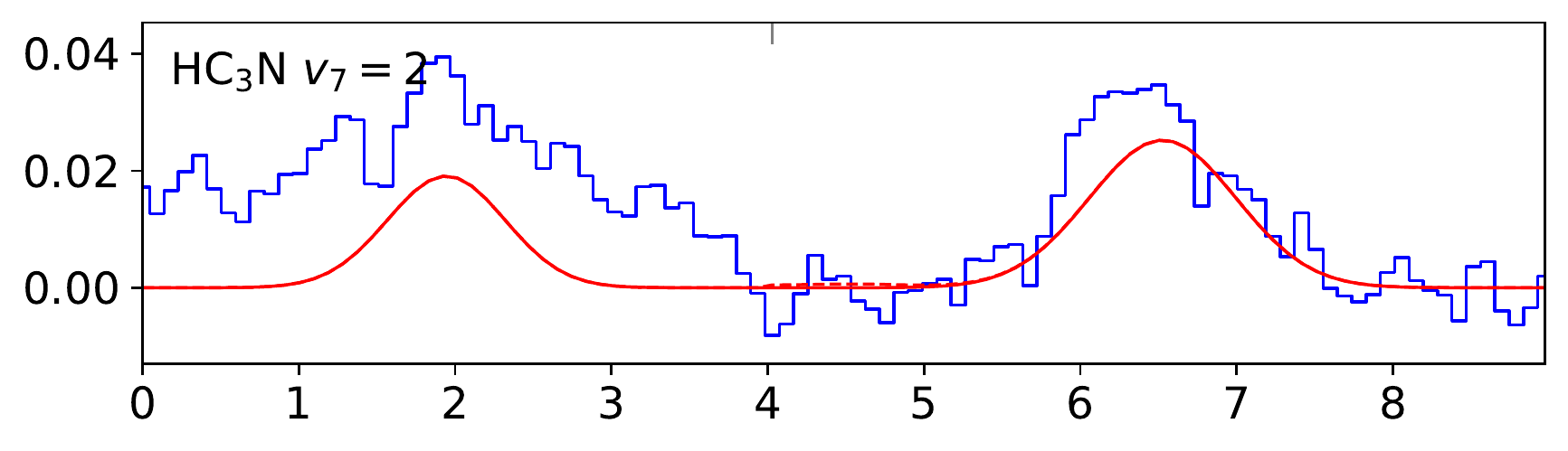}
\includegraphics[width=0.485\linewidth]{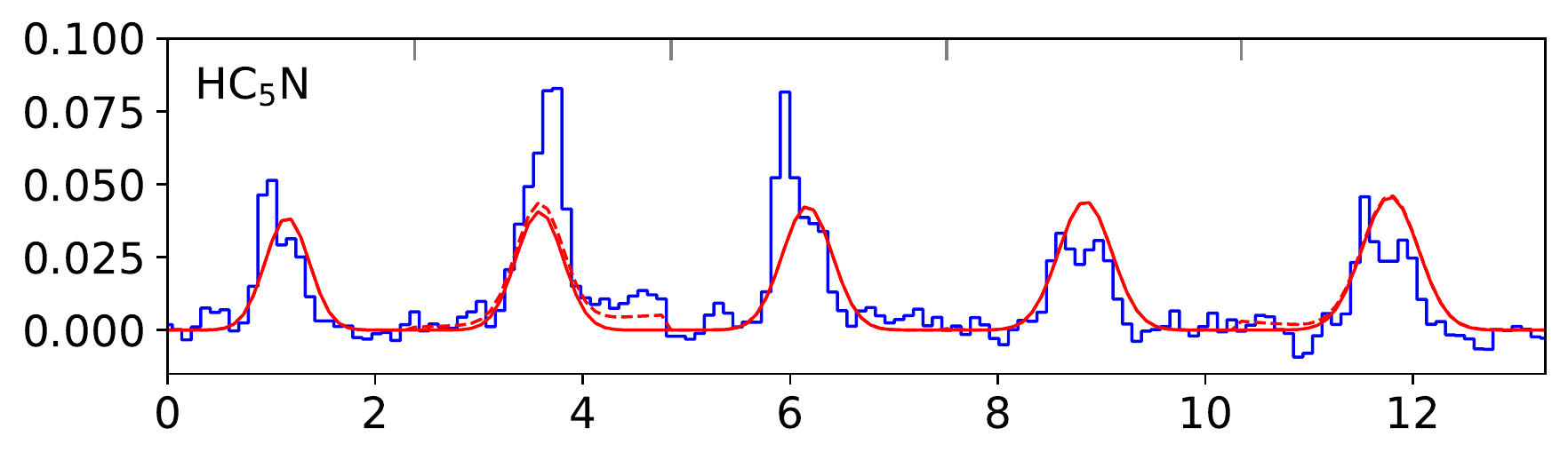}
\\
\includegraphics[width=0.485\linewidth]{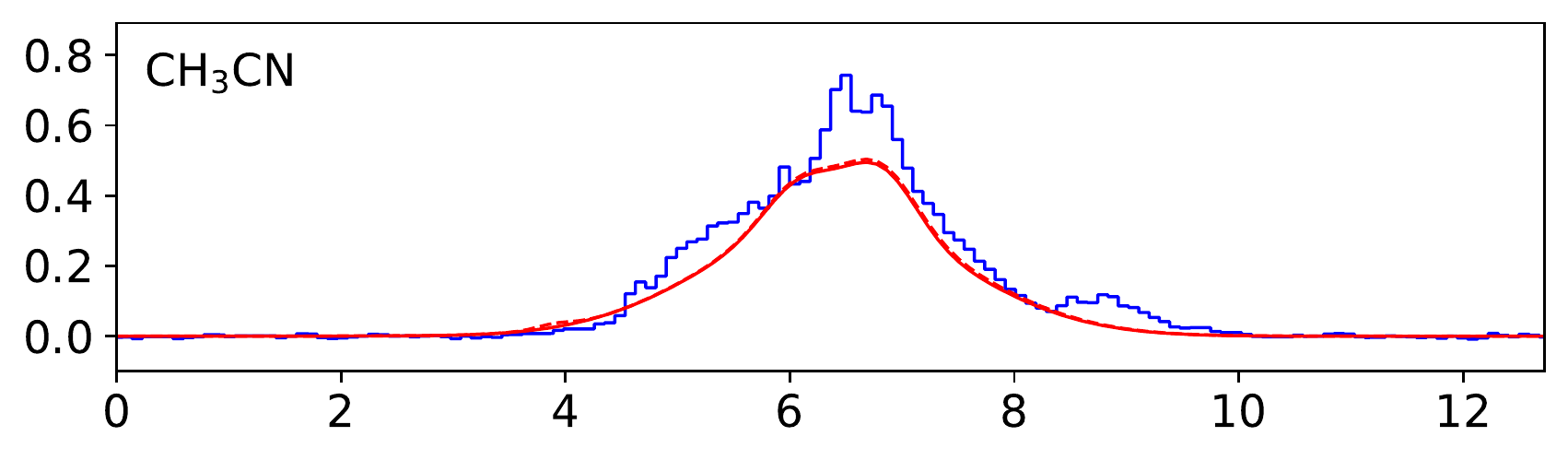}
\includegraphics[width=0.485\linewidth]{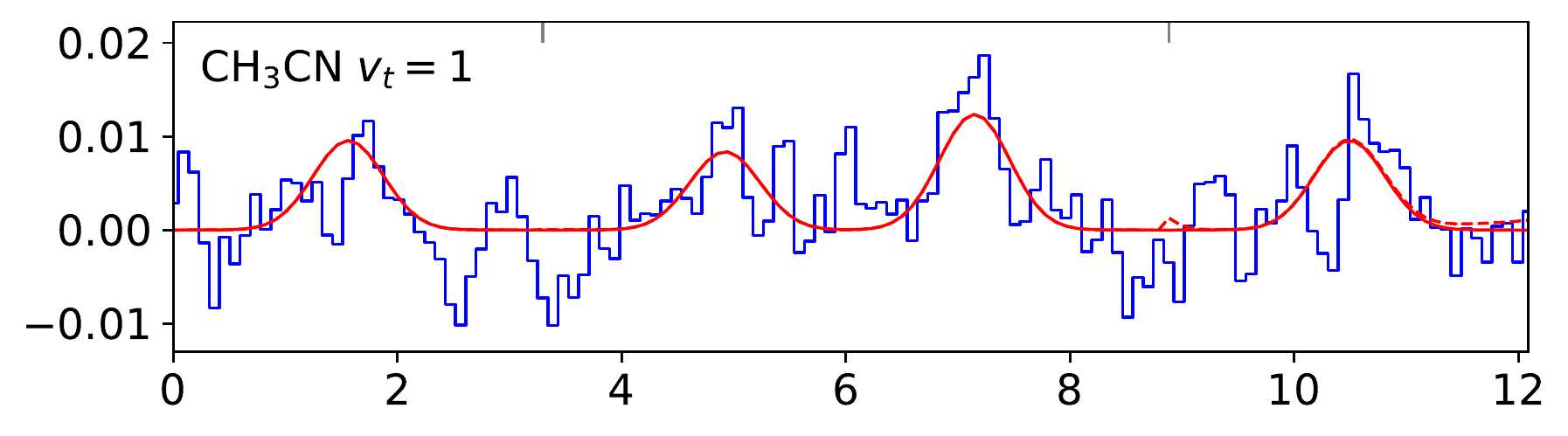}
\\
\includegraphics[width=0.485\linewidth]{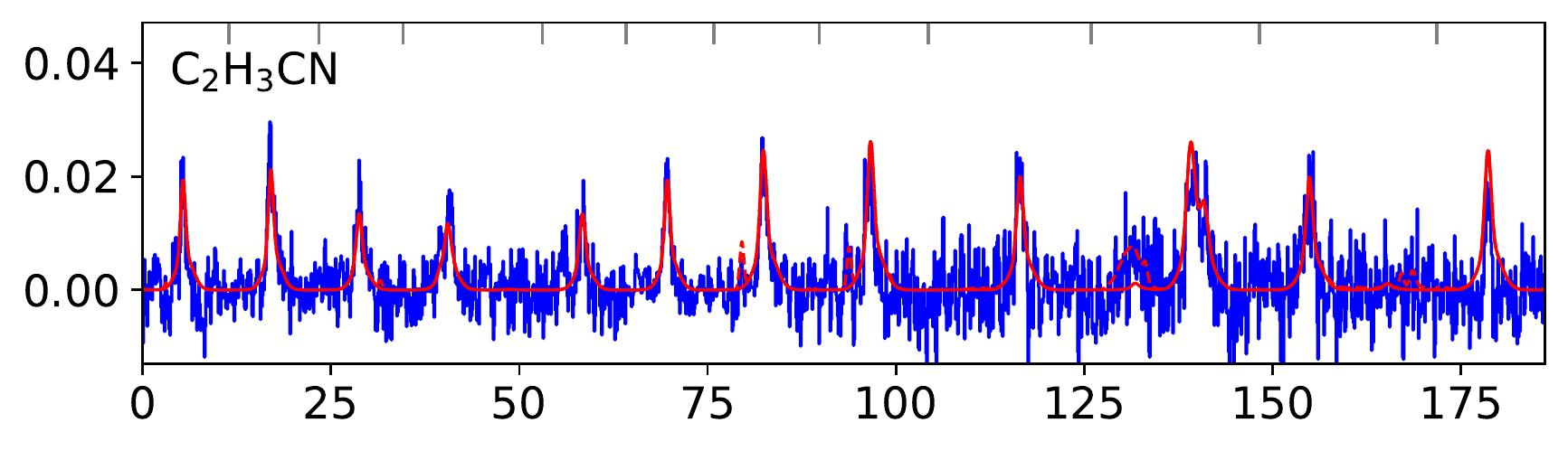}
\includegraphics[width=0.485\linewidth]{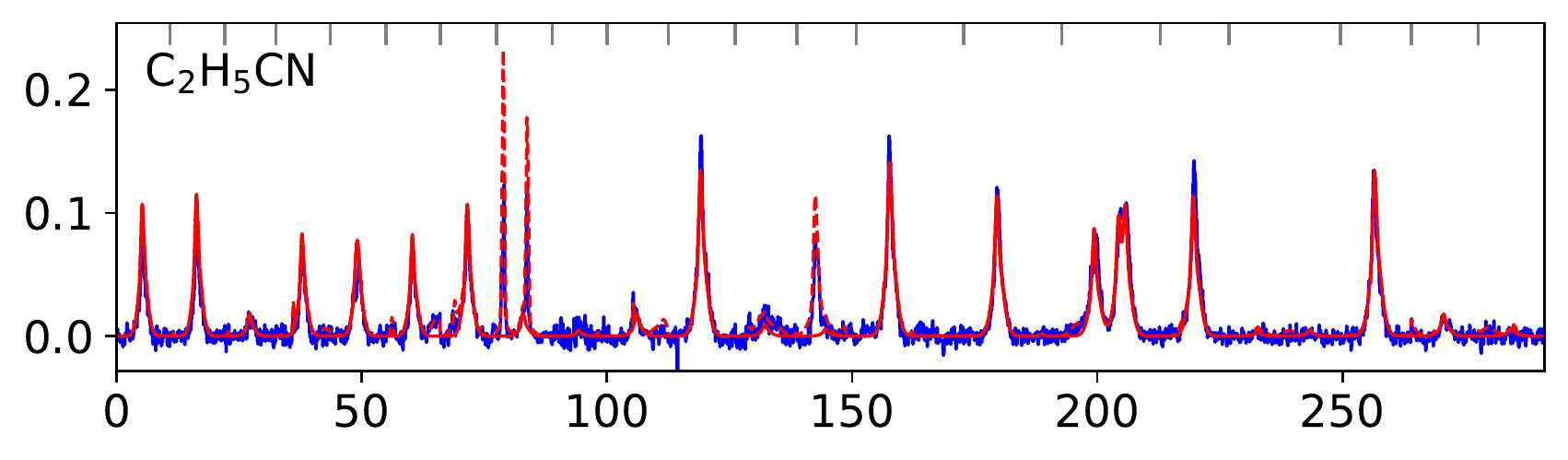}
\\
\includegraphics[width=0.485\linewidth]{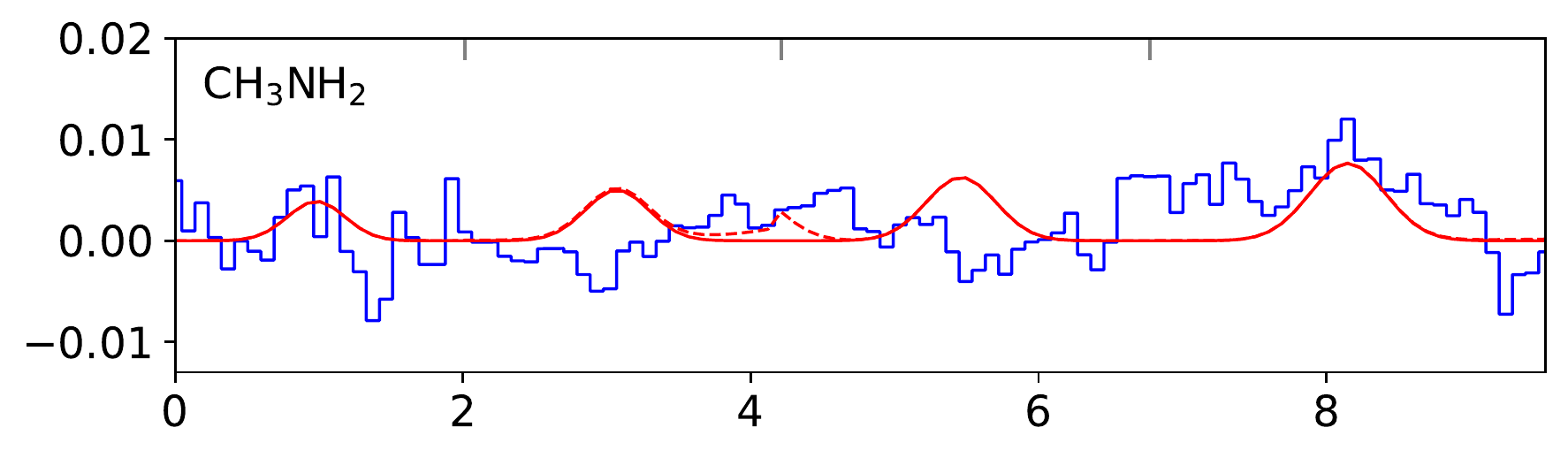}
\includegraphics[width=0.485\linewidth]{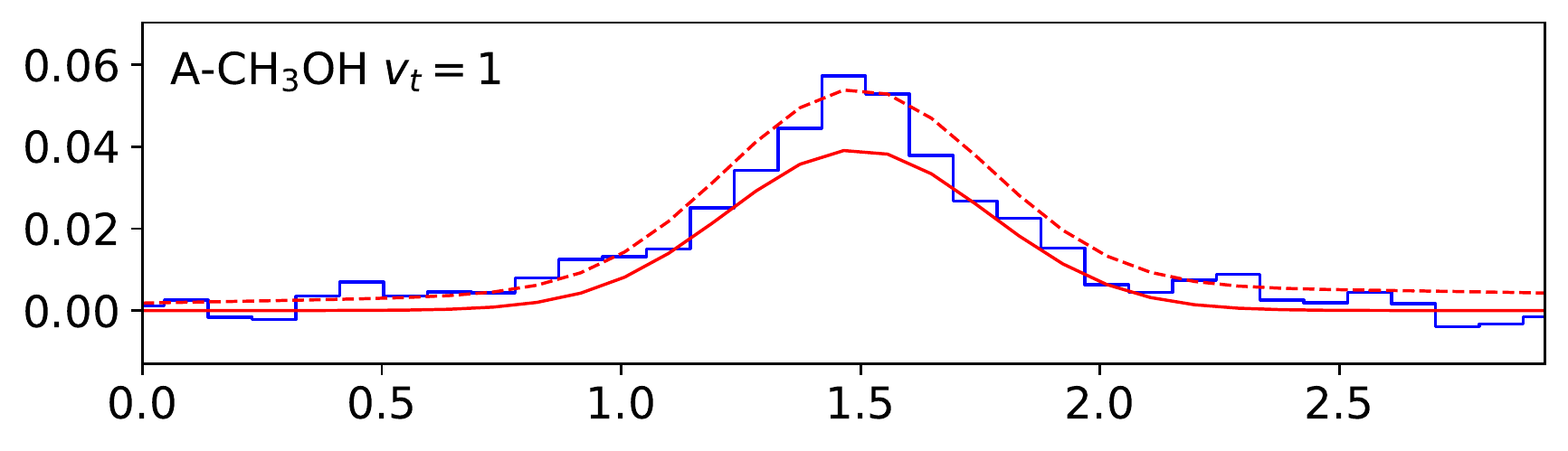}
\\
{\raggedright \centering  \textbf{Figure \thefigure} {\it continued} \par}
\end{figure*}
        
\begin{figure*}
\centering
{\raggedright \centering  \textbf{Figure \thefigure} {\it (continued)} \par}
\includegraphics[width=0.485\linewidth]{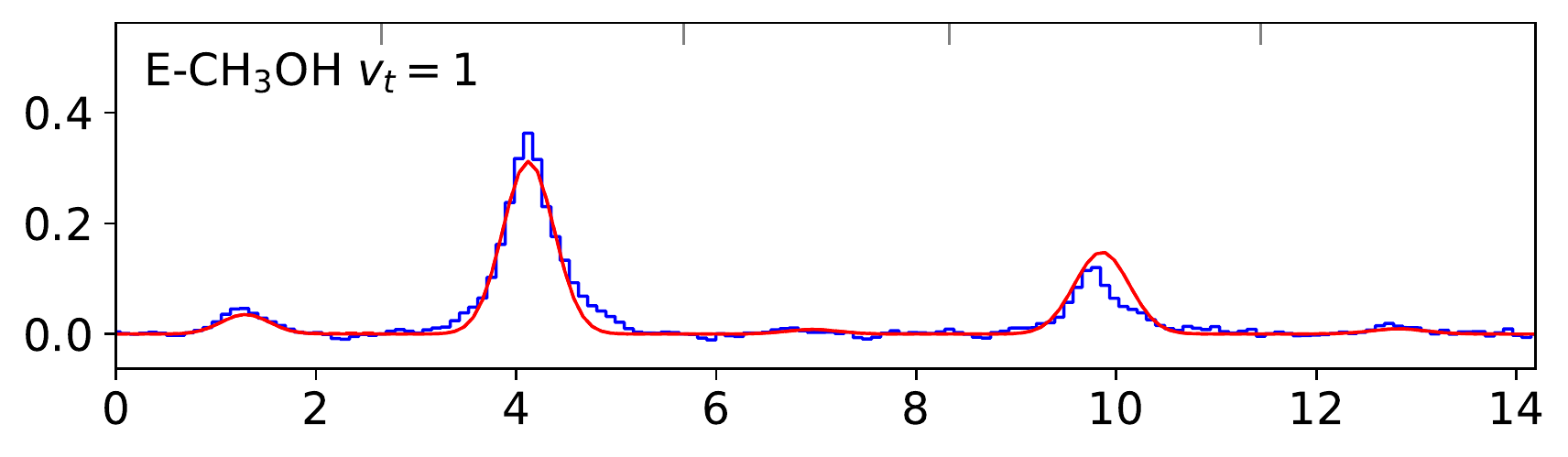}
\includegraphics[width=0.485\linewidth]{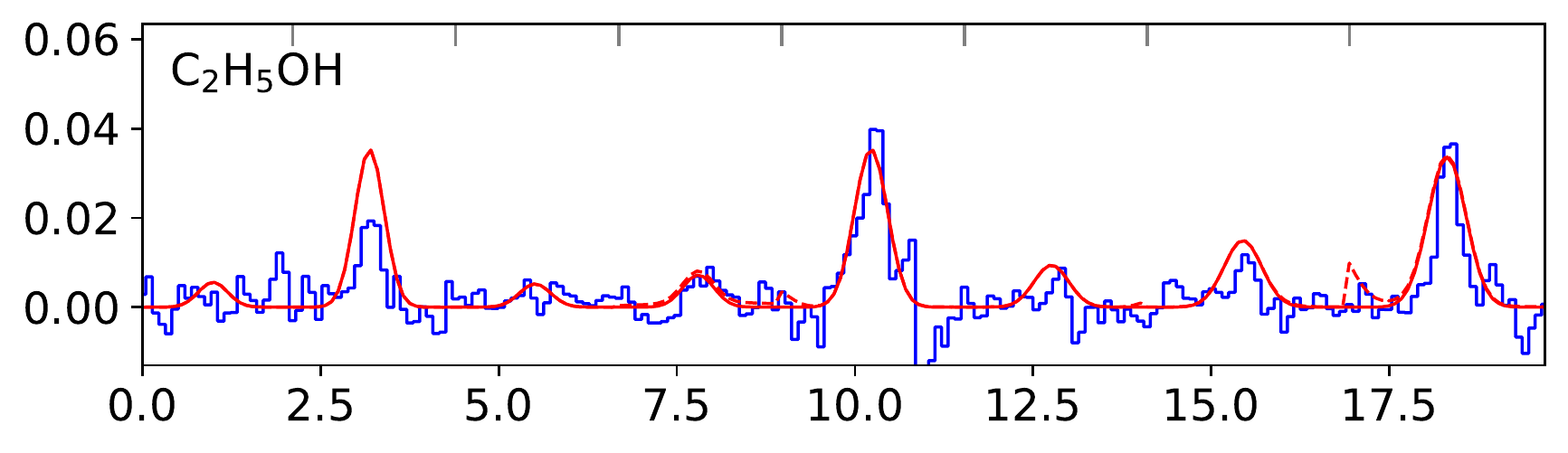}
\\
\includegraphics[width=0.485\linewidth]{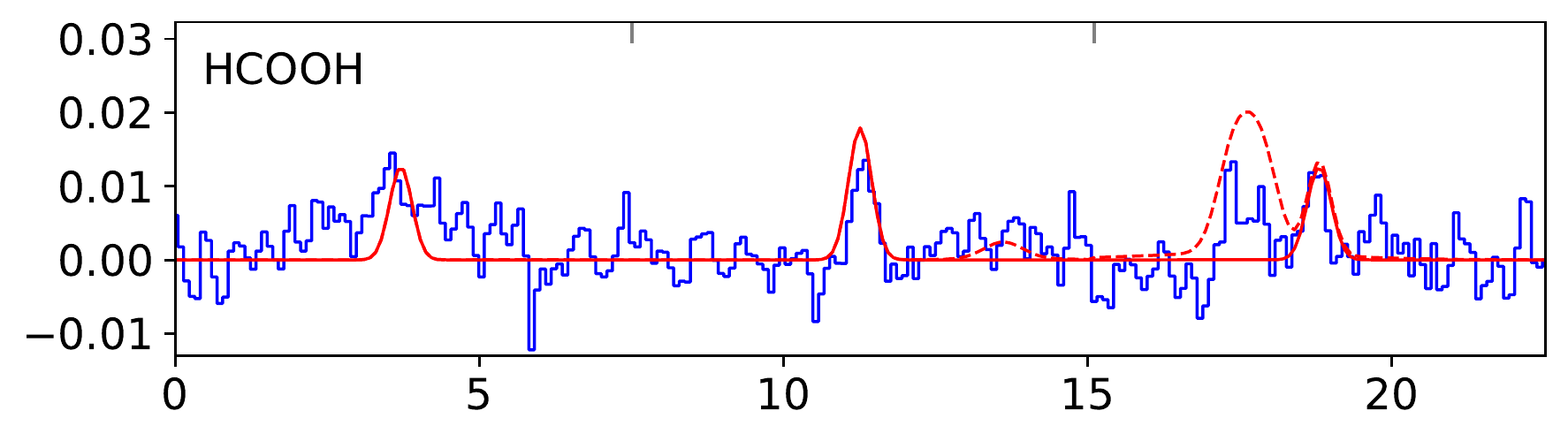}
\includegraphics[width=0.485\linewidth]{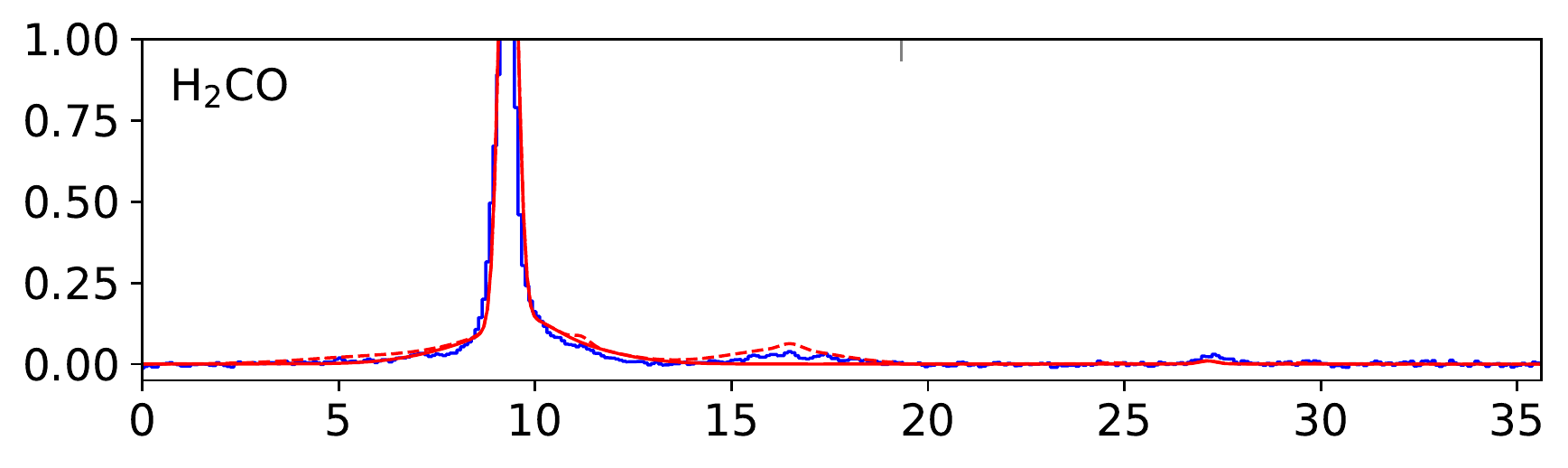}
\\
\includegraphics[width=0.485\linewidth]{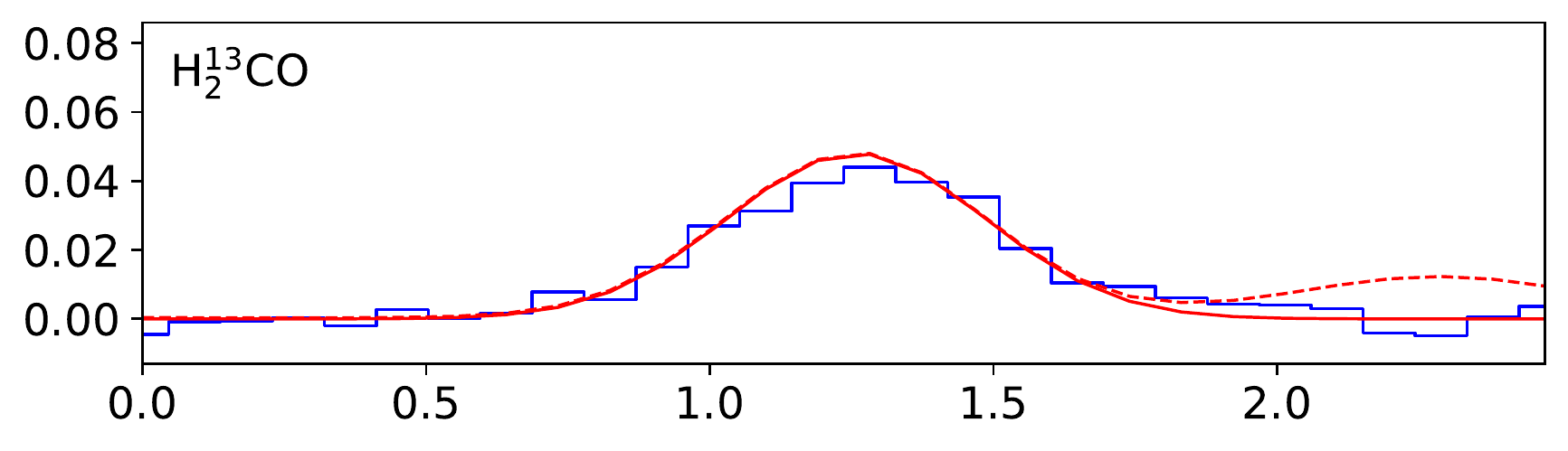}
\includegraphics[width=0.485\linewidth]{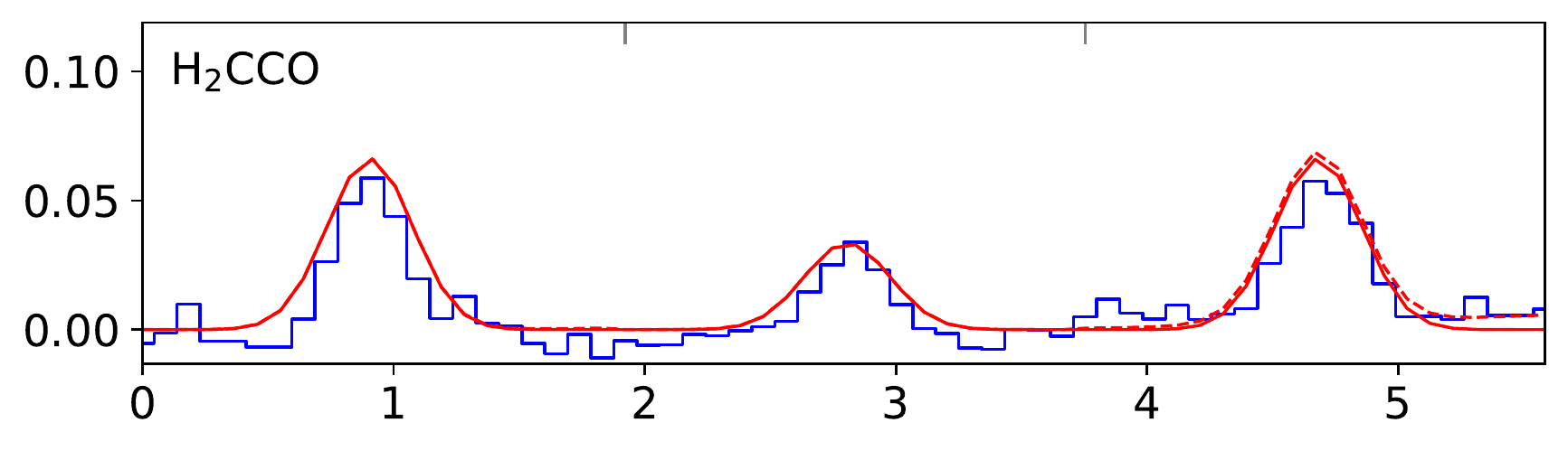}
\\
\includegraphics[width=0.485\linewidth]{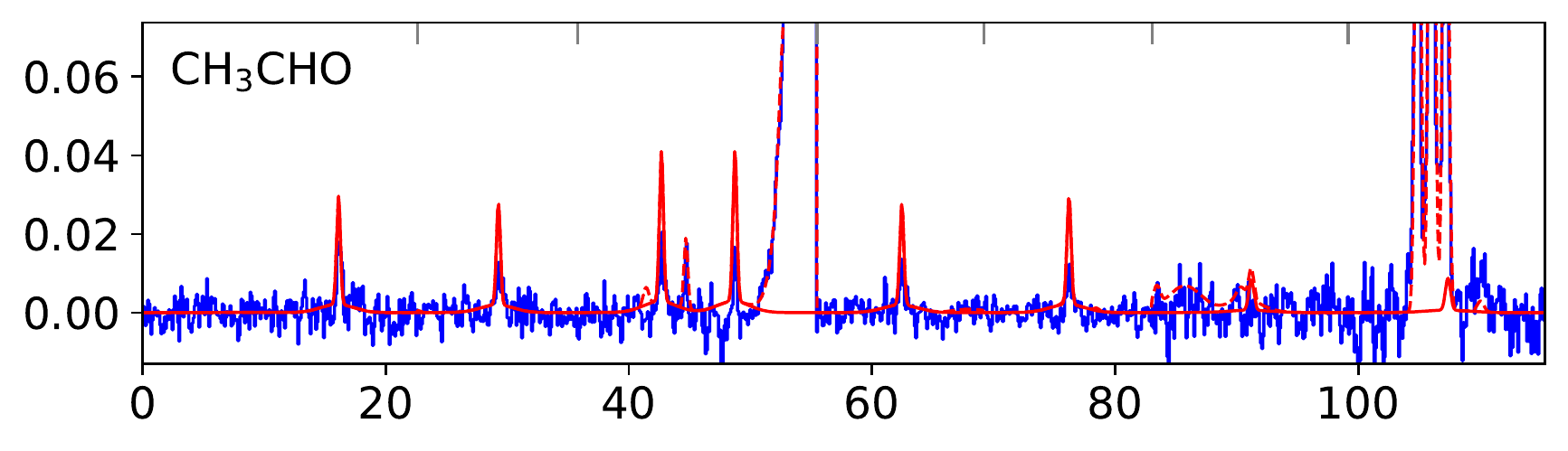}
\includegraphics[width=0.485\linewidth]{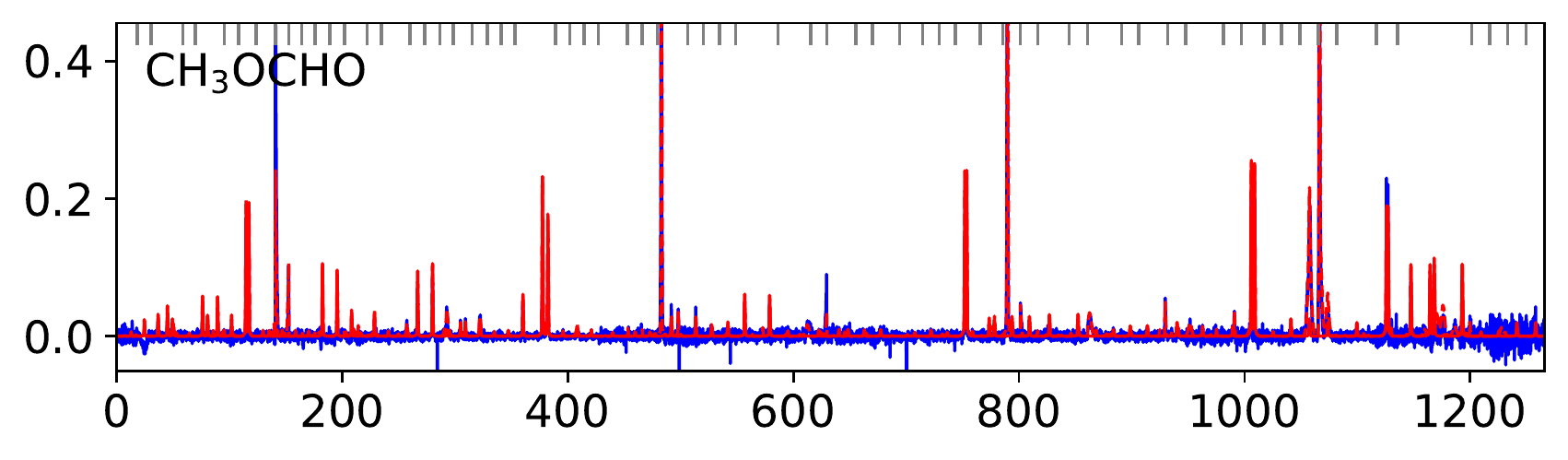}
\\
\includegraphics[width=0.485\linewidth]{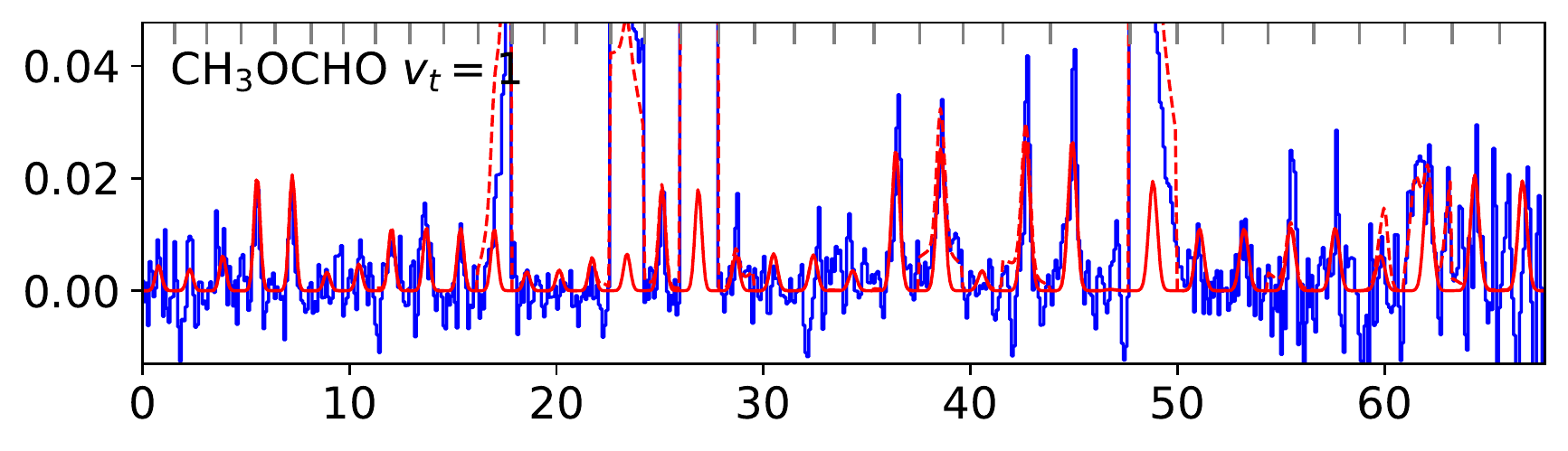}
\includegraphics[width=0.485\linewidth]{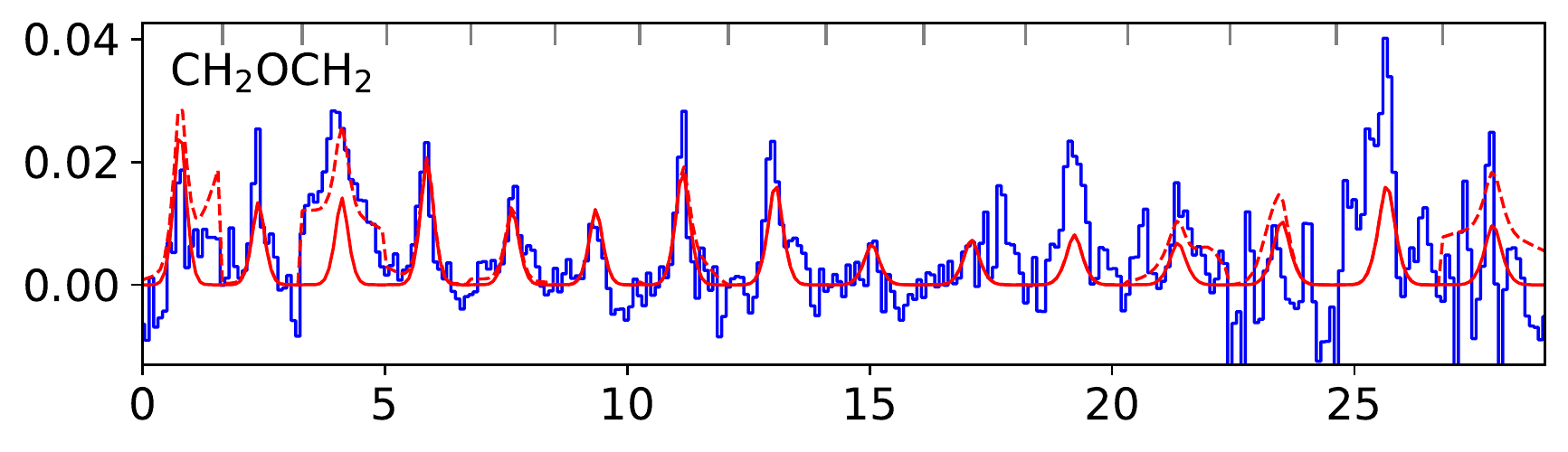}
\\
\includegraphics[width=0.485\linewidth]{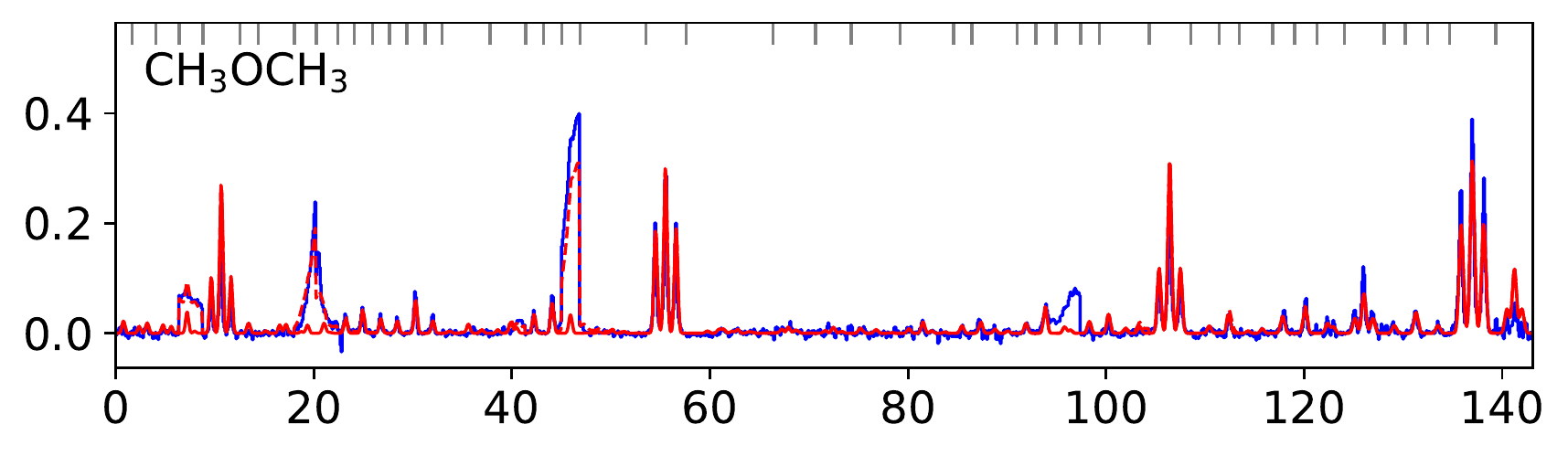}
\includegraphics[width=0.485\linewidth]{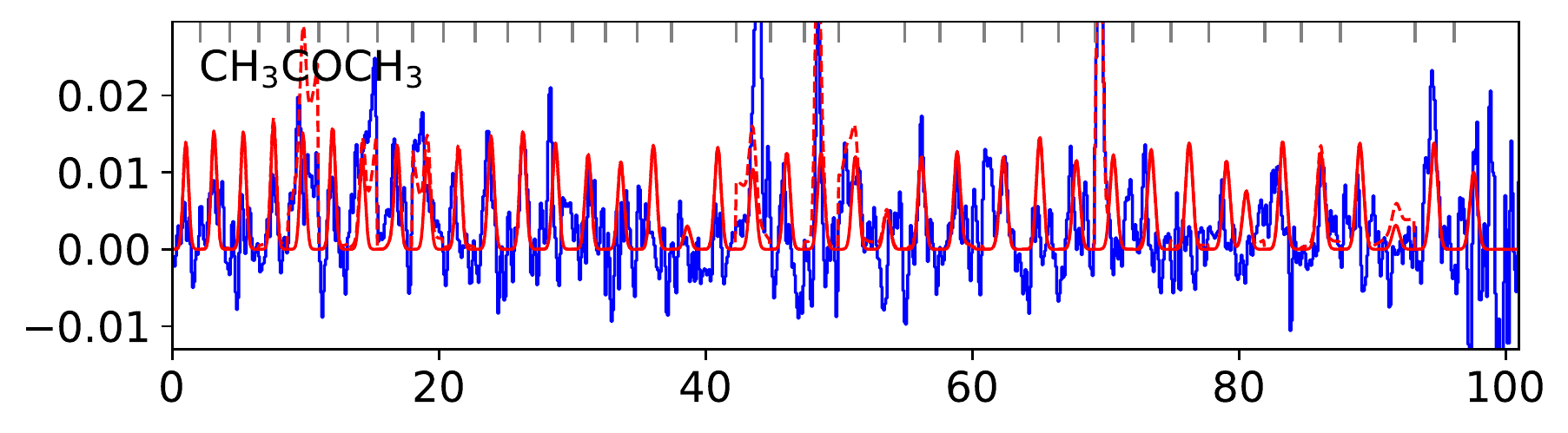}
\\
\includegraphics[width=0.485\linewidth]{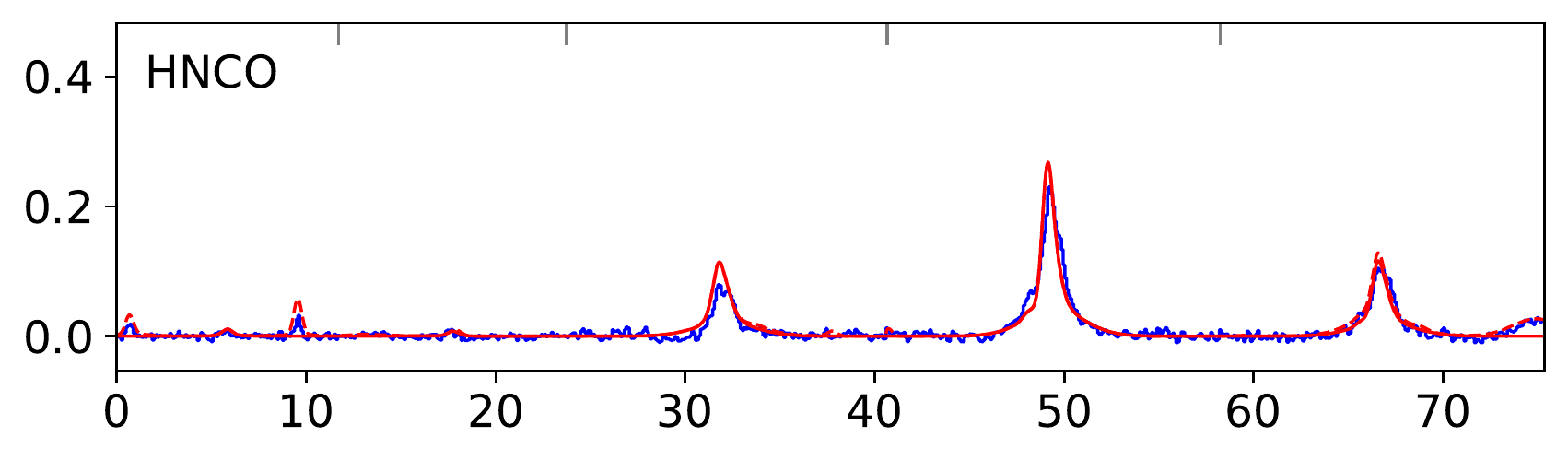}
\includegraphics[width=0.485\linewidth]{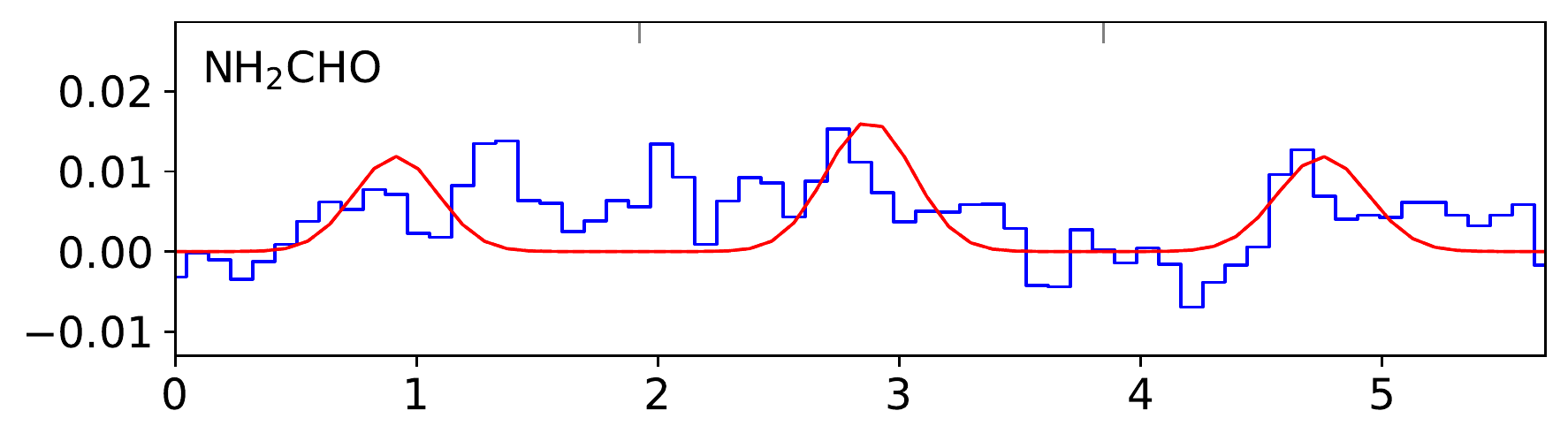}
\\
\includegraphics[width=0.485\linewidth]{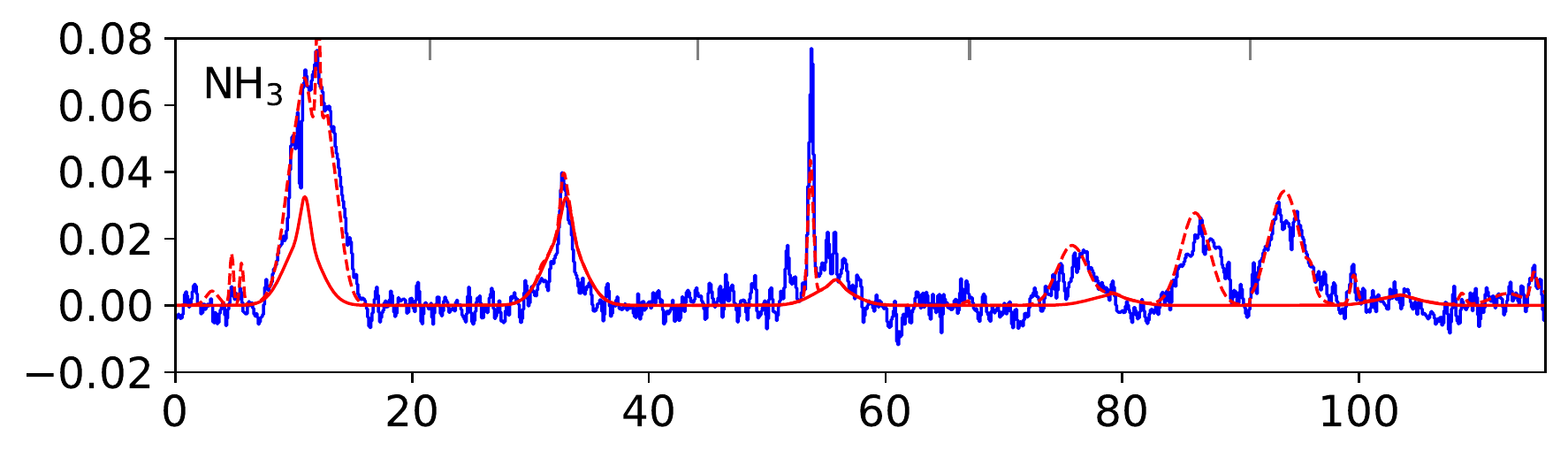}
\includegraphics[width=0.485\linewidth]{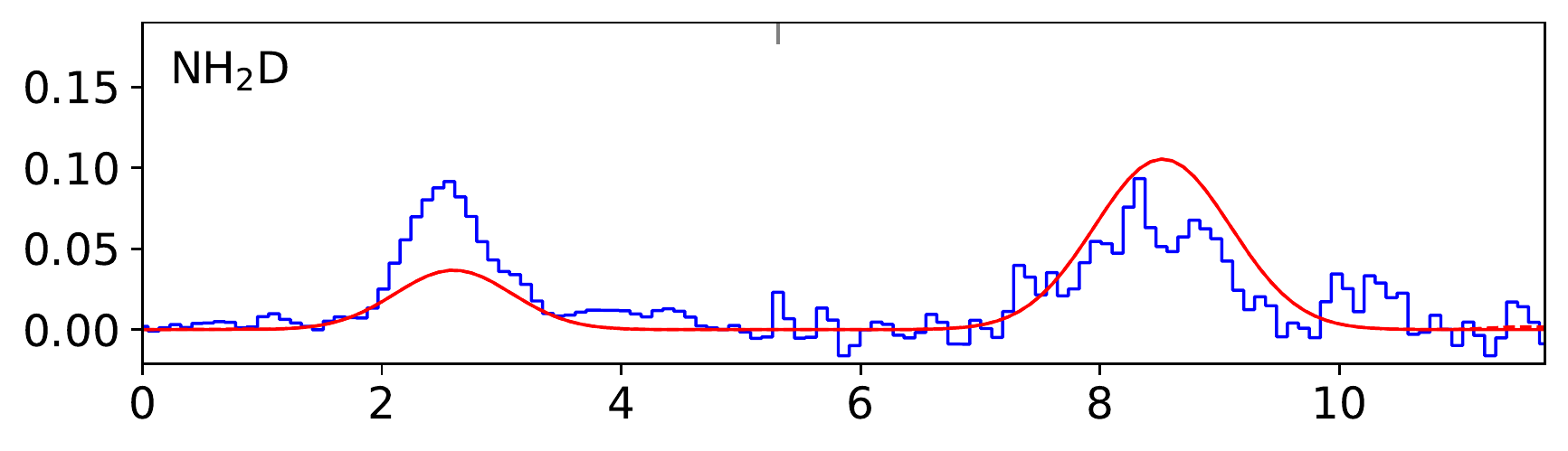}
\\
{\raggedright \centering  \textbf{Figure \thefigure} {\it continued} \par}
\end{figure*}

\clearpage  
\section{The spectrum of Orion KL}
\begin{figure*}[!htb] \caption{The Orion KL spectrum \label{allspectra}}
\includegraphics[width=0.99\linewidth]{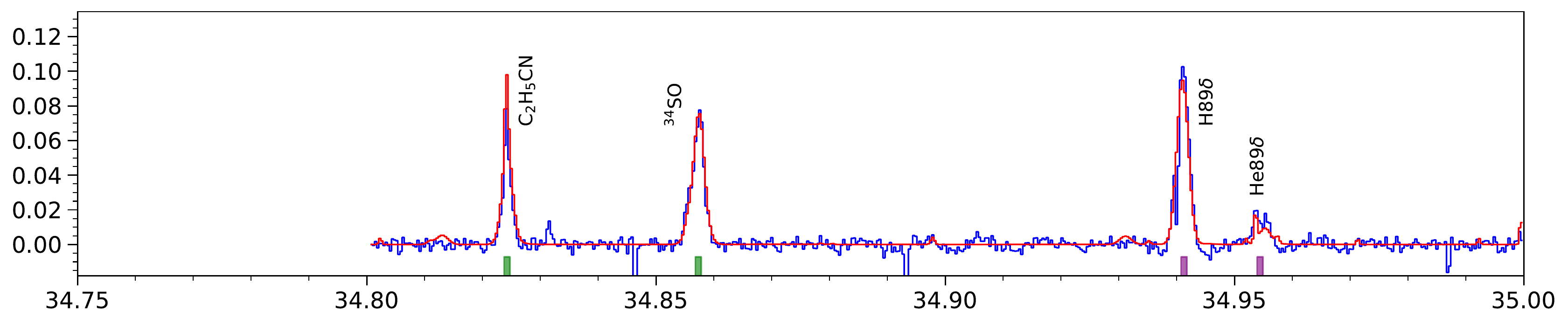}\\
\includegraphics[width=0.99\linewidth]{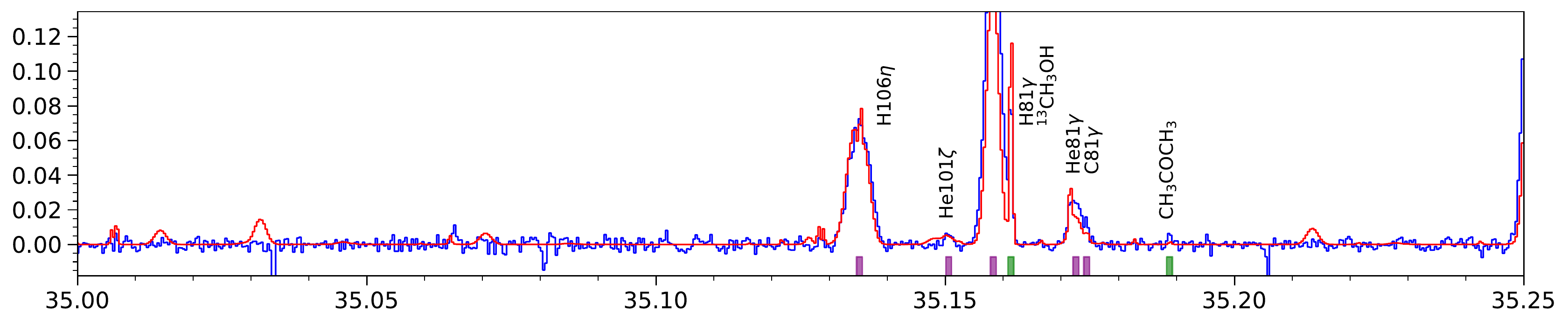}\\
\includegraphics[width=0.99\linewidth]{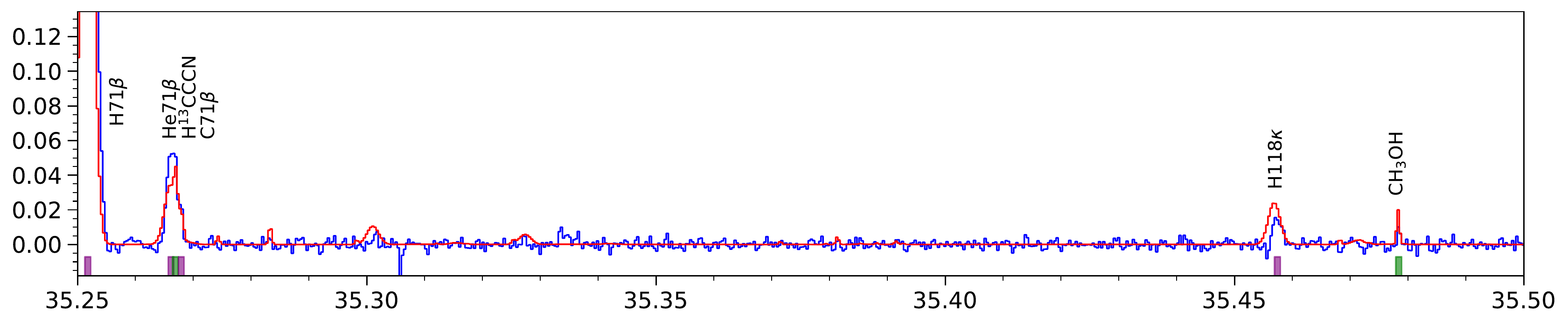}\\
\includegraphics[width=0.99\linewidth]{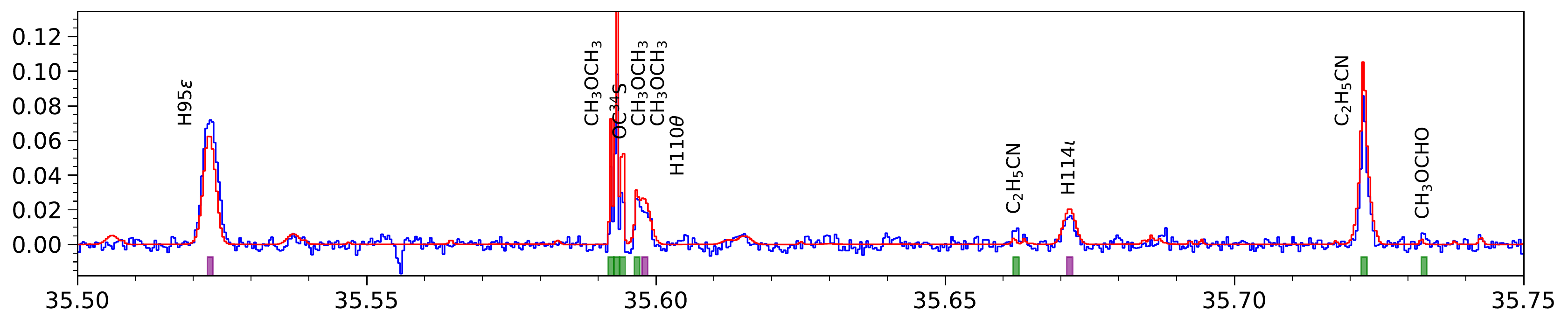}\\
\includegraphics[width=0.99\linewidth]{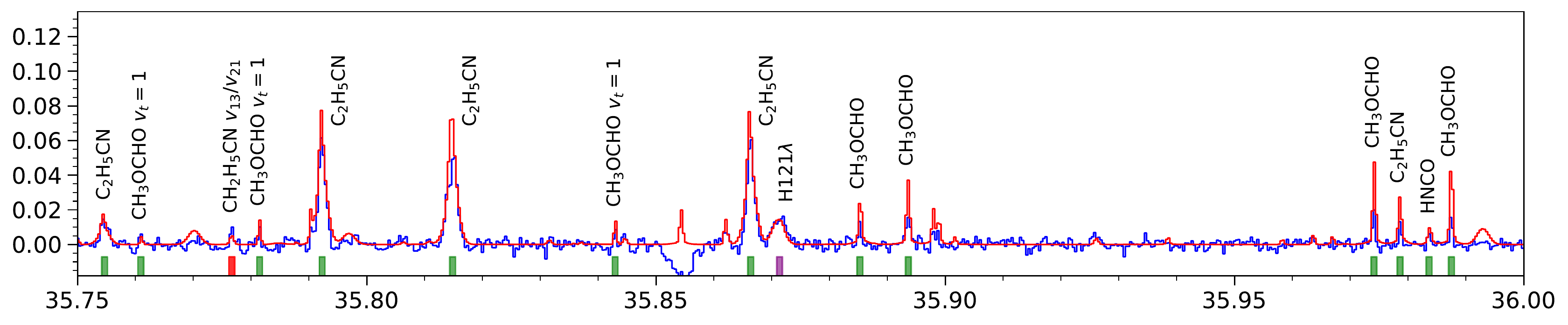}\\
{\raggedright \footnotesize \textbf{Note}: The blue line is the Orion KL spectrum observed by the TMRT 65 m,
which has been smoothed to have a frequency resolution of 364 kHz ($\sim$2.8 km s$^{-1}$ at 40 GHz). 
The red line represents the results of model fitting (Sect. \ref{modelfit_sec}). 
The purple strips denote the detected RRLs. The yellow strips 
denote the molecular lines which have also been detected and resolved by \citet{2017A&A...605A..76R}. The green strips denotes the 
molecular lines detected by TMRT 65 m but have not been detected by \citet{2017A&A...605A..76R}.
The red strips denote the lines of C$_2$H$_5$CN $\varv$13/$\varv$21. The gray strips mark the U lines.
The lines of SiO and its isotopologues are not modeled.
The $x$ axis is rest frequency in  unit of GHz (with a Doppler correction applied adopting a $V_{\rm LSR}$ of 6 km s$^{-1}$). The $y$ axis is $T_{\rm MB}$ in units of K.
\par}
{\raggedright \centering  \textbf{Figure \thefigure} {\it continued} \par}
\end{figure*} 
\begin{figure*}[!htb]
\centering
{\raggedright \centering  \textbf{Figure \thefigure} {\it (continued)} \par}
\includegraphics[width=0.99\linewidth]{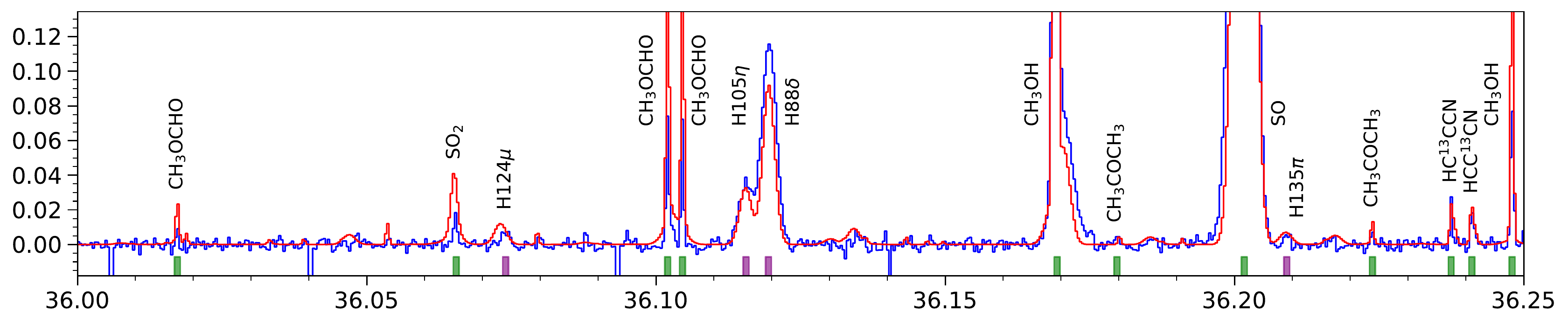}\\
\includegraphics[width=0.99\linewidth]{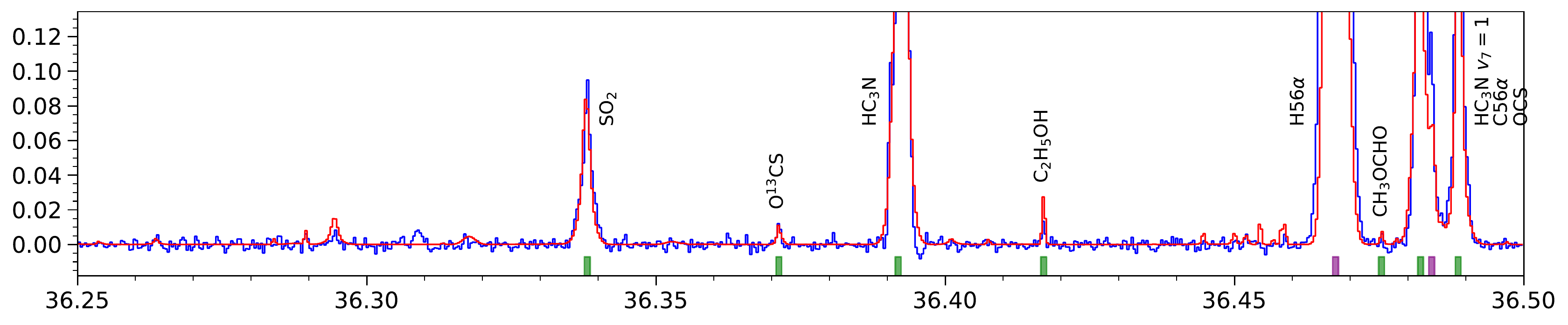}\\
\includegraphics[width=0.99\linewidth]{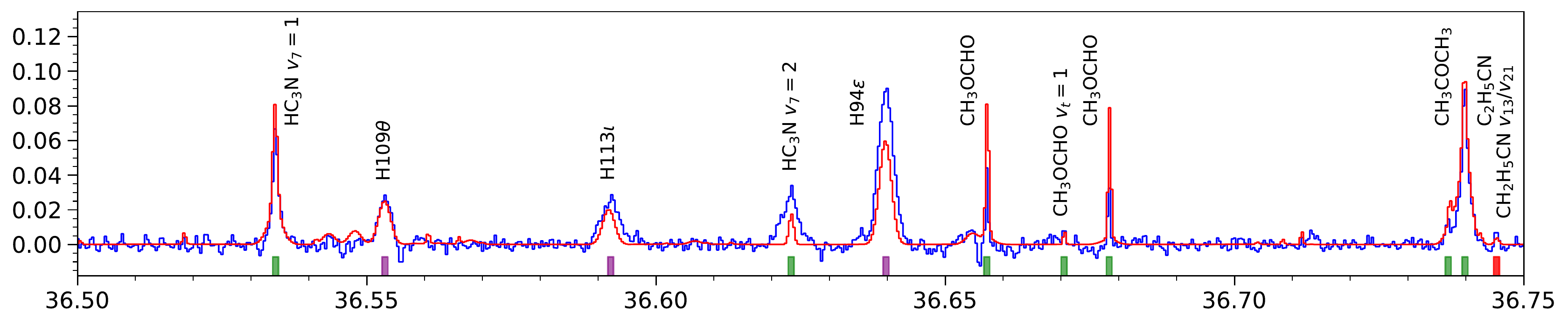}\\
\includegraphics[width=0.99\linewidth]{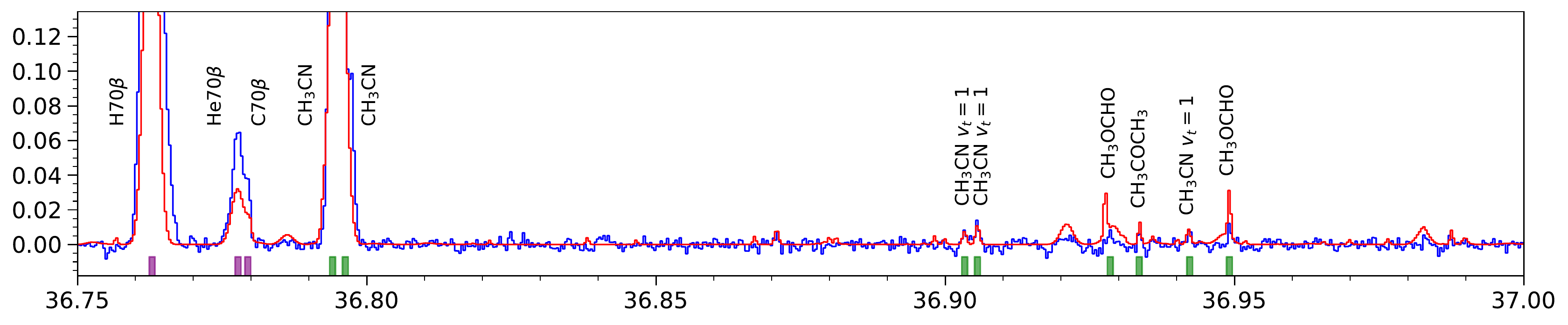}\\
\includegraphics[width=0.99\linewidth]{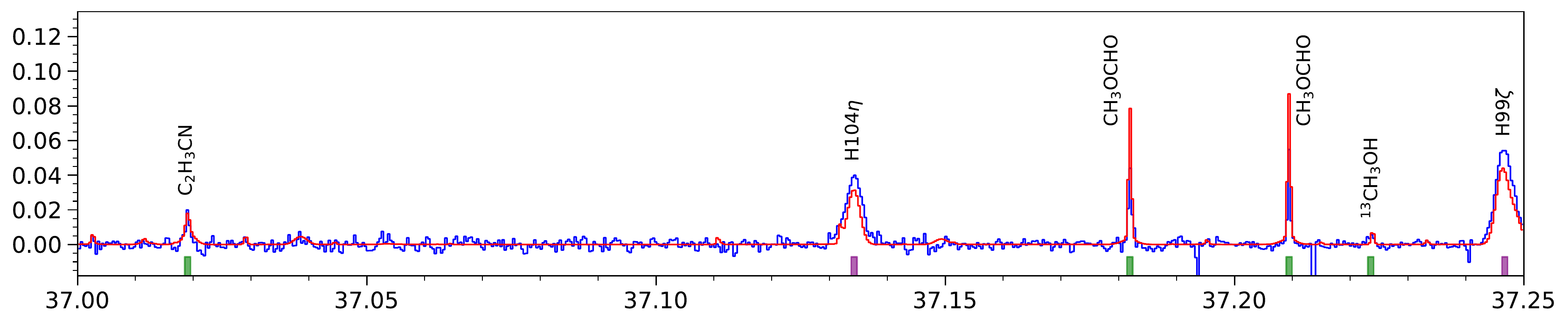}\\
\includegraphics[width=0.99\linewidth]{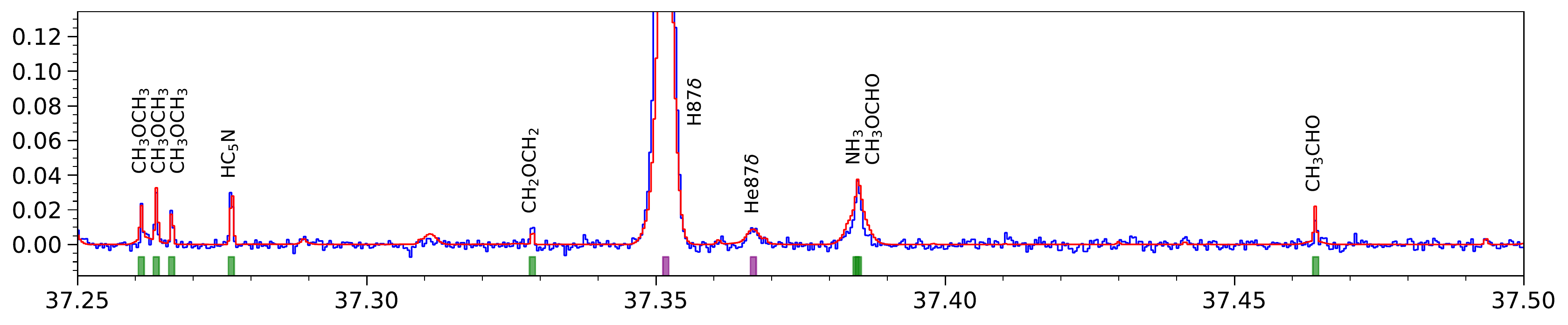}\\
{\raggedright \centering  \textbf{Figure \thefigure} {\it continued} \par}
\end{figure*}       
\begin{figure*}[!htb]
\centering
{\raggedright \centering  \textbf{Figure \thefigure} {\it (continued)} \par}
\includegraphics[width=0.99\linewidth]{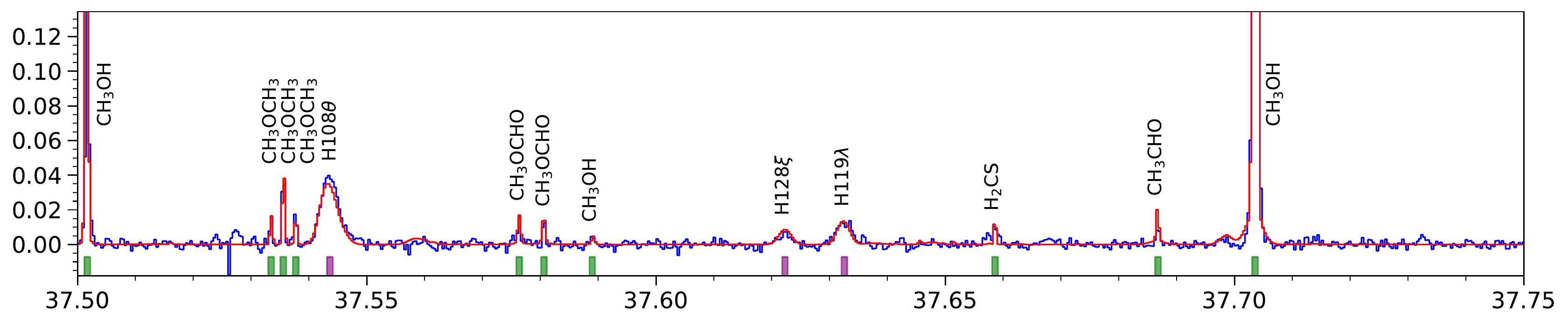}\\
\includegraphics[width=0.99\linewidth]{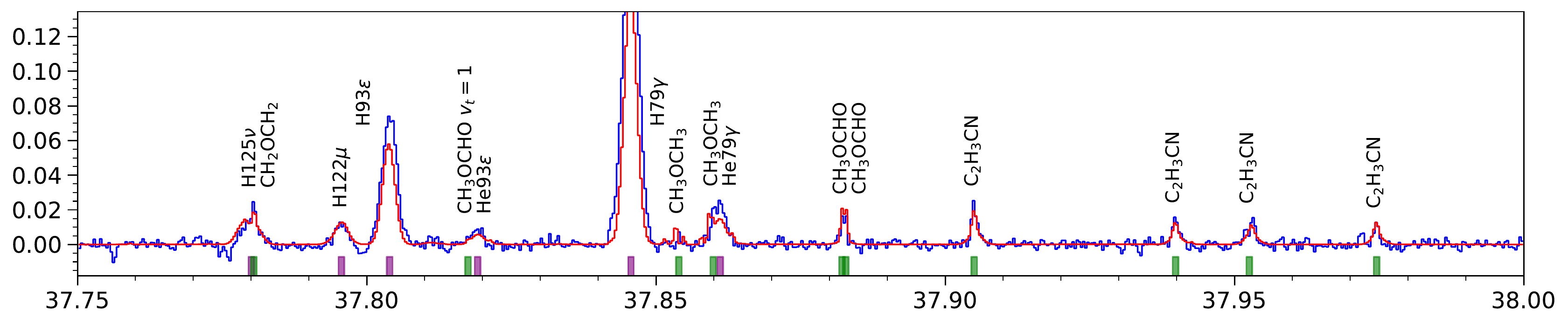}\\
\includegraphics[width=0.99\linewidth]{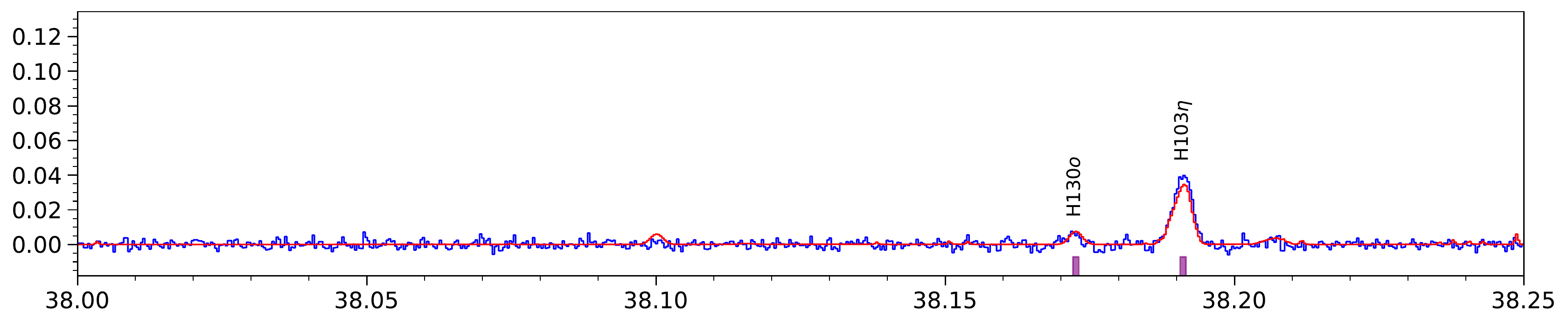}\\
\includegraphics[width=0.99\linewidth]{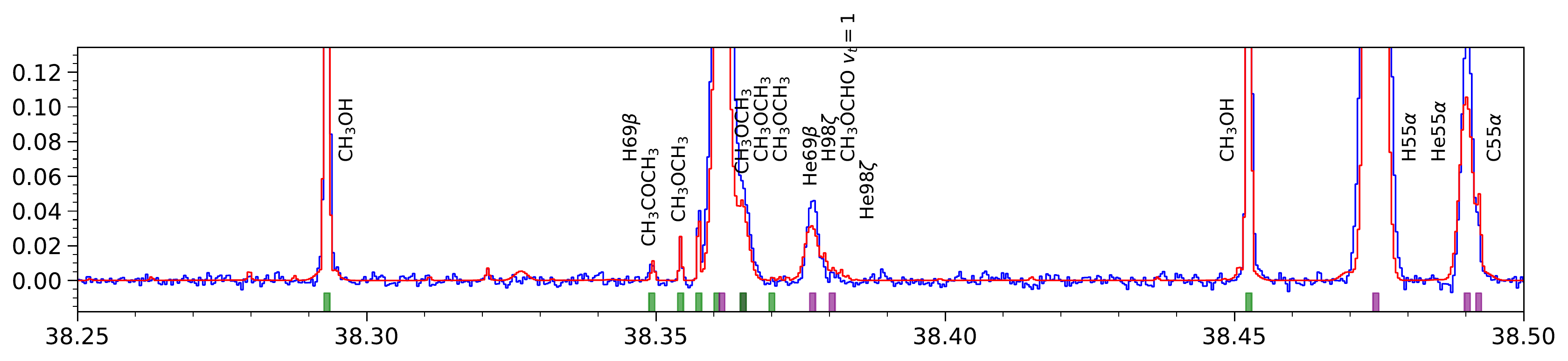}\\
\includegraphics[width=0.99\linewidth]{{38500_38750}.pdf}\\
\includegraphics[width=0.99\linewidth]{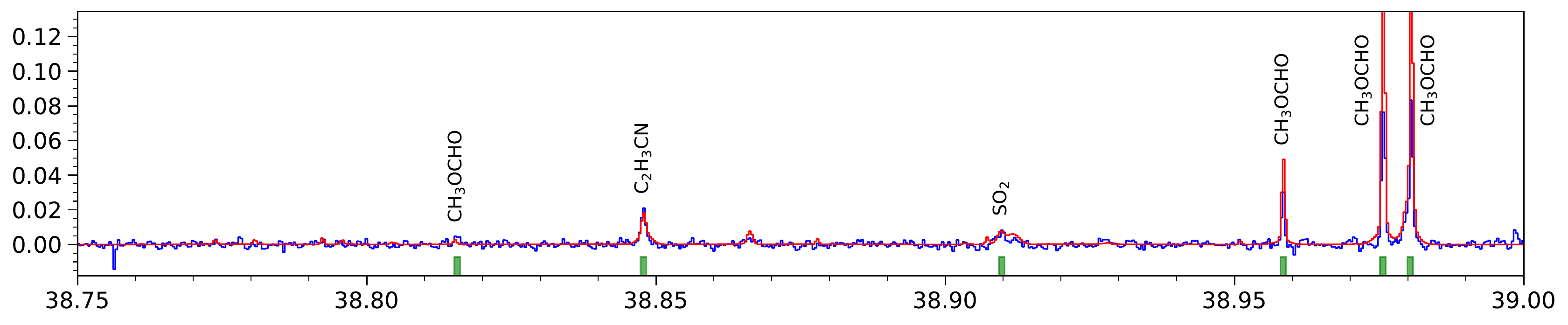}\\
{\raggedright \centering  \textbf{Figure \thefigure} {\it continued} \par}
\end{figure*}      
\begin{figure*}[!htb]
\centering
{\raggedright \centering  \textbf{Figure \thefigure} {\it (continued)} \par}
\includegraphics[width=0.99\linewidth]{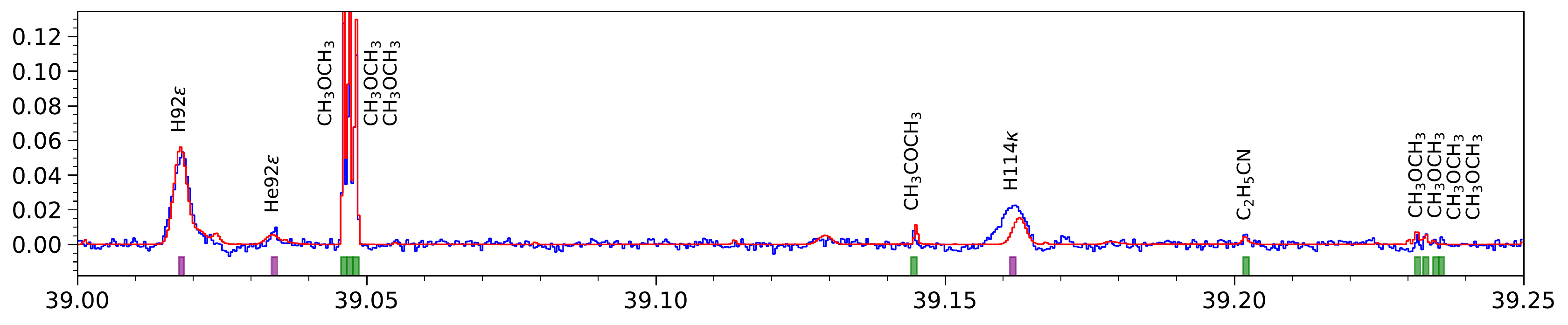}\\
\includegraphics[width=0.99\linewidth]{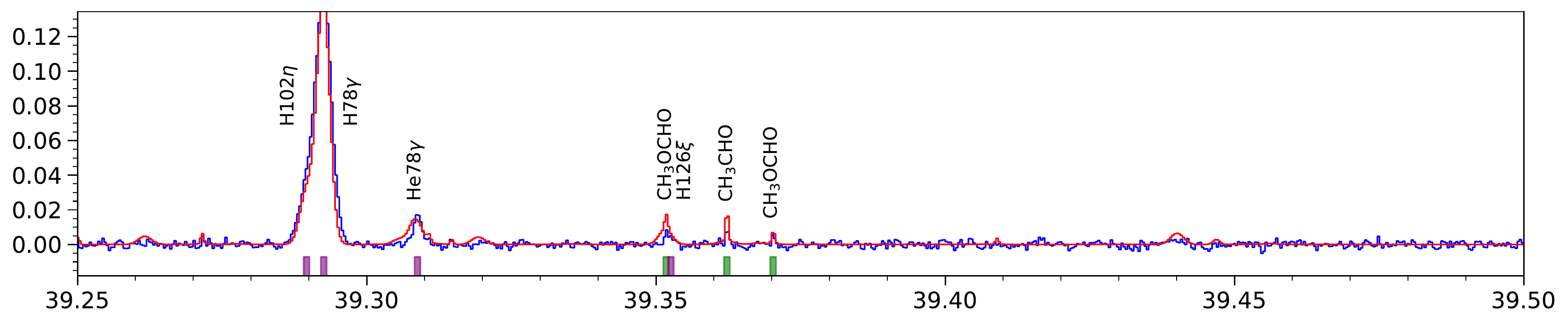}\\
\includegraphics[width=0.99\linewidth]{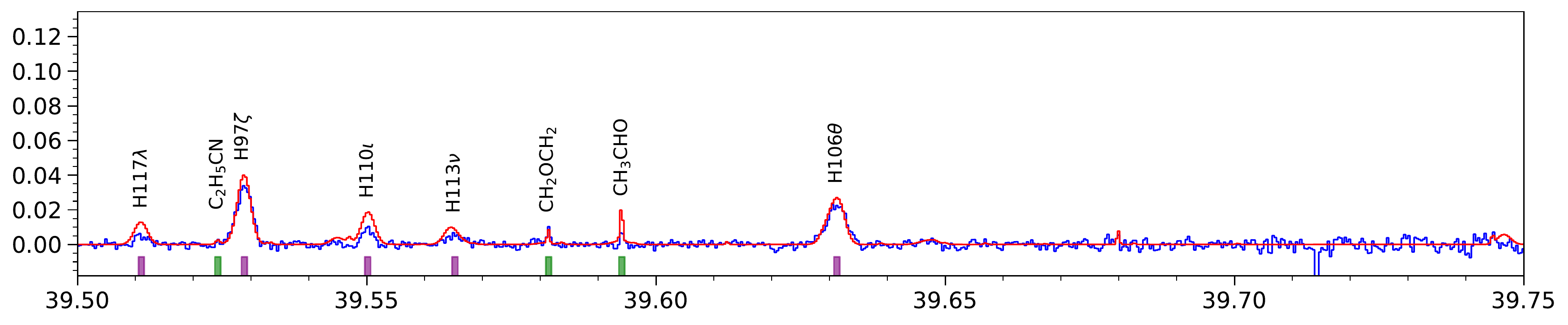}\\
\includegraphics[width=0.99\linewidth]{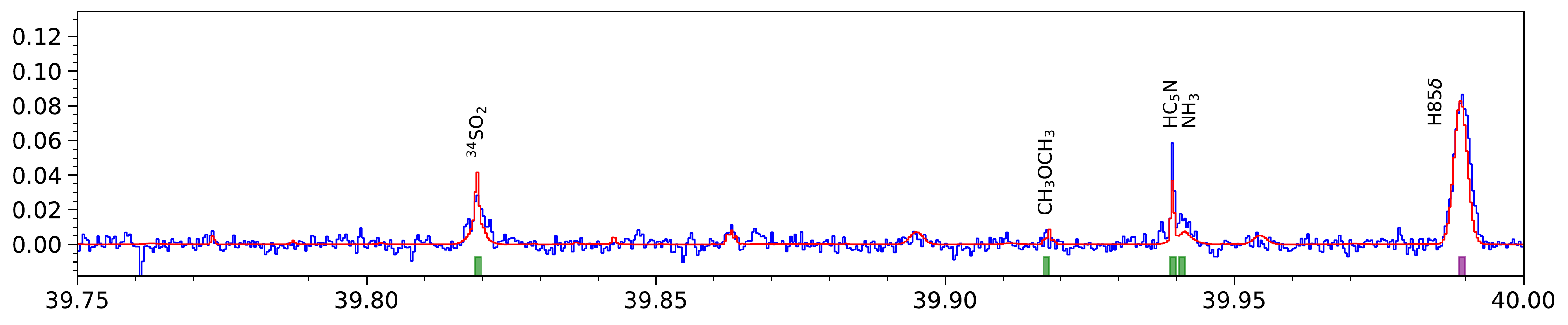}\\
\includegraphics[width=0.99\linewidth]{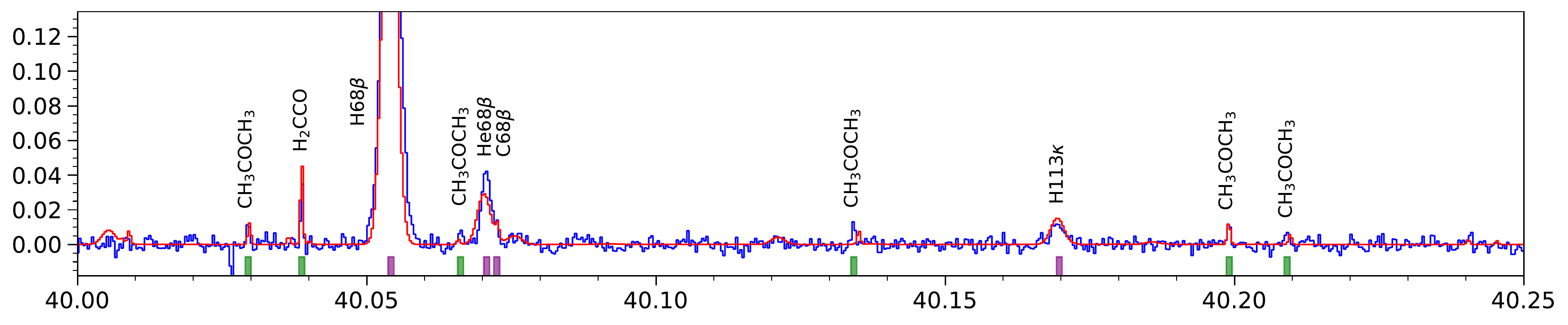}\\
\includegraphics[width=0.99\linewidth]{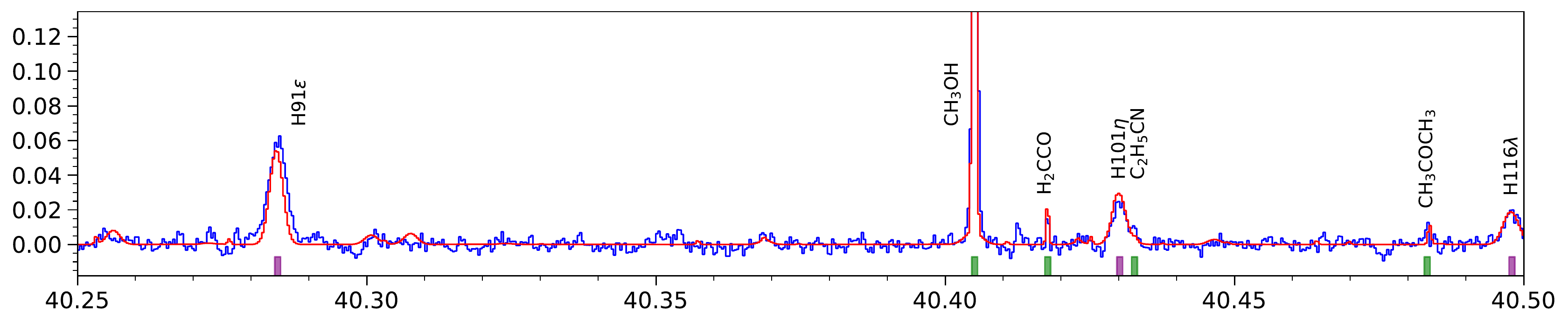}\\
{\raggedright \centering  \textbf{Figure \thefigure} {\it continued} \par}
\end{figure*} 
\begin{figure*}[!htb]
\centering
{\raggedright \centering  \textbf{Figure \thefigure} {\it (continued)} \par}
\includegraphics[width=0.99\linewidth]{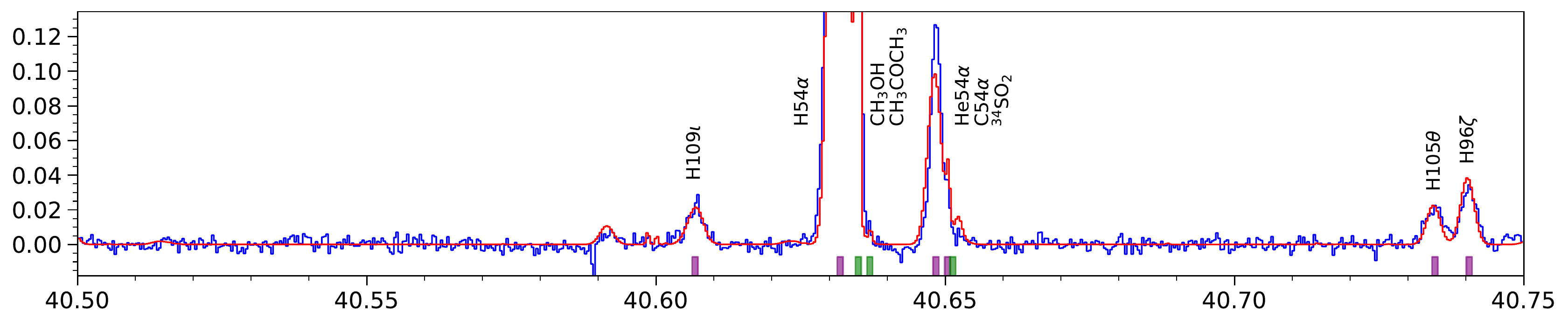}\\
\includegraphics[width=0.99\linewidth]{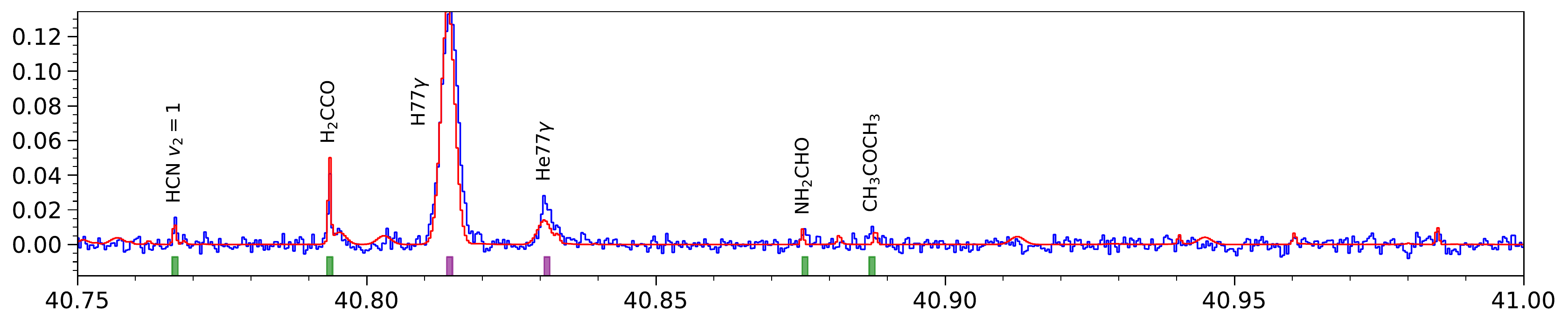}\\
\includegraphics[width=0.99\linewidth]{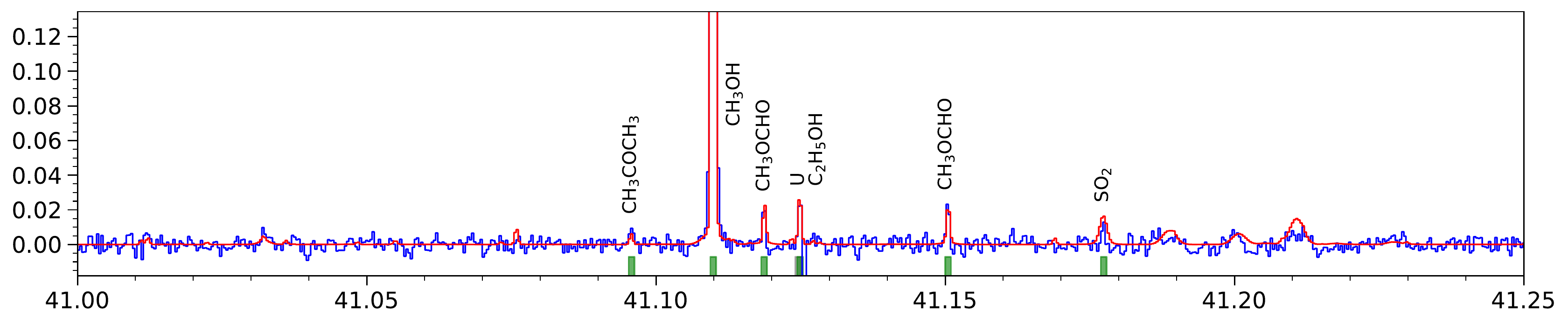}\\
\includegraphics[width=0.99\linewidth]{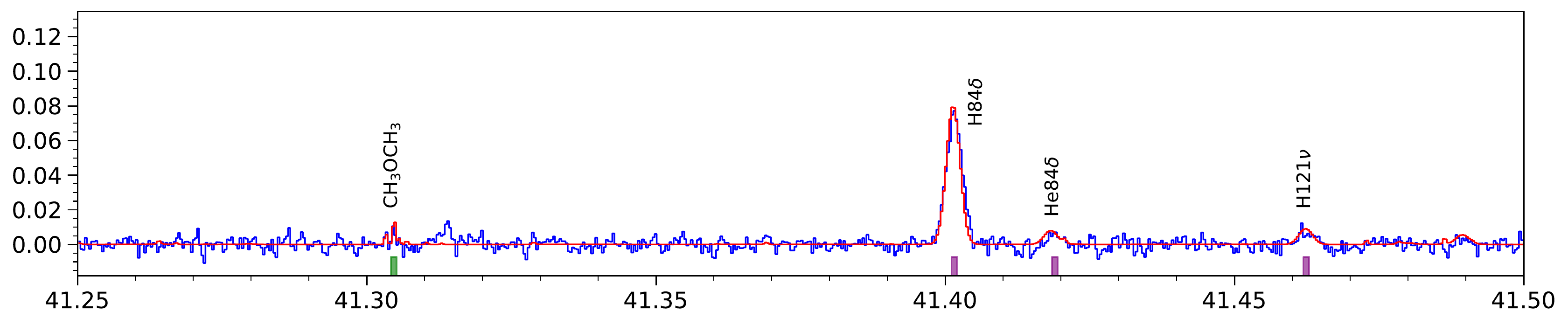}\\
\includegraphics[width=0.99\linewidth]{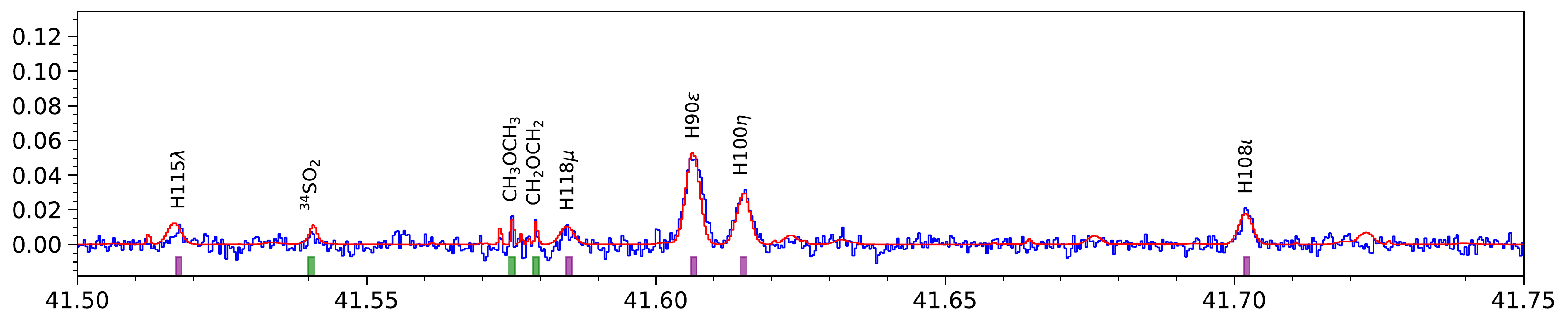}\\
\includegraphics[width=0.99\linewidth]{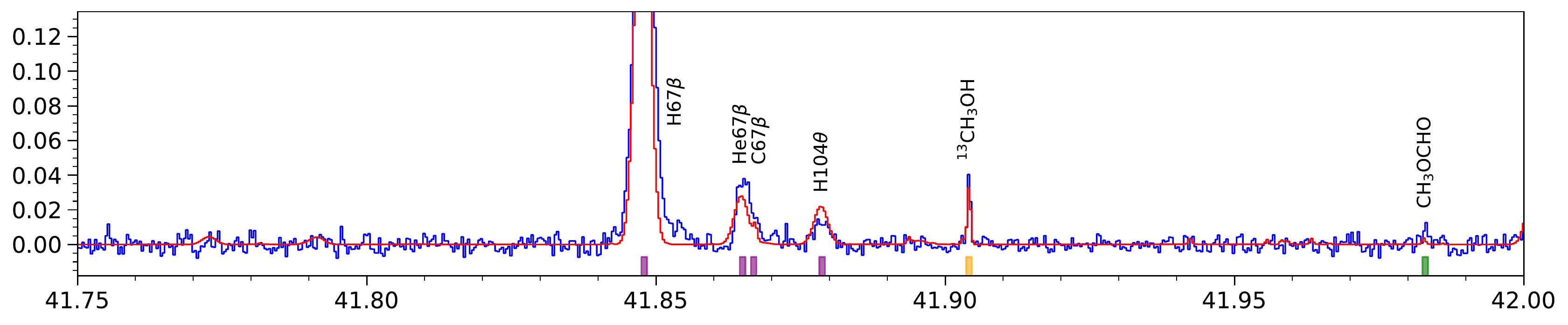}\\
{\raggedright \centering  \textbf{Figure \thefigure} {\it continued} \par}
\end{figure*}         
\begin{figure*}[!htb]
\centering
{\raggedright \centering  \textbf{Figure \thefigure} {\it (continued)} \par}
\includegraphics[width=0.99\linewidth]{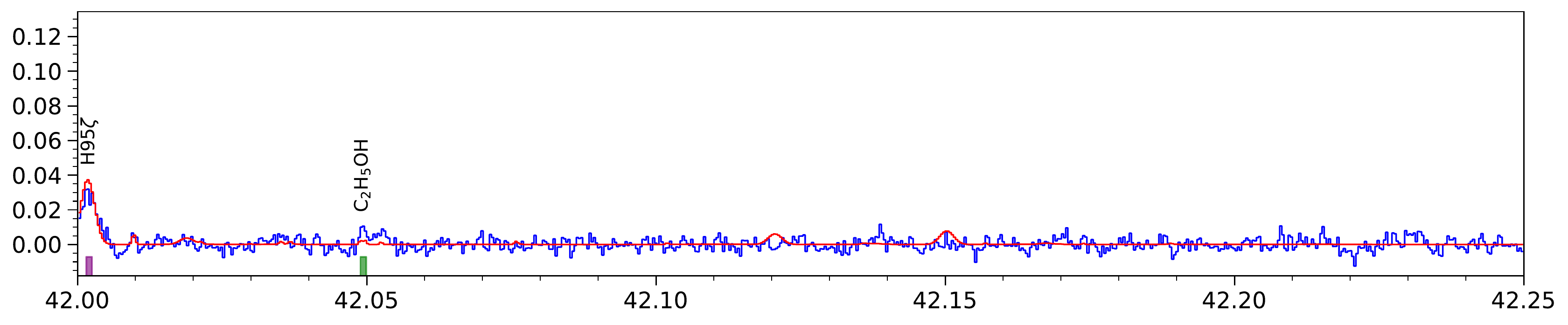}\\
\includegraphics[width=0.99\linewidth]{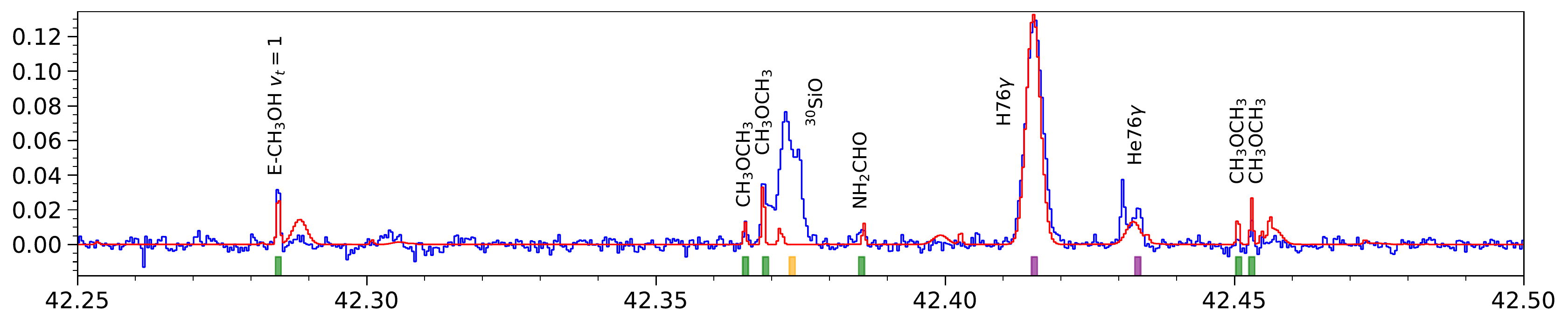}\\
\includegraphics[width=0.99\linewidth]{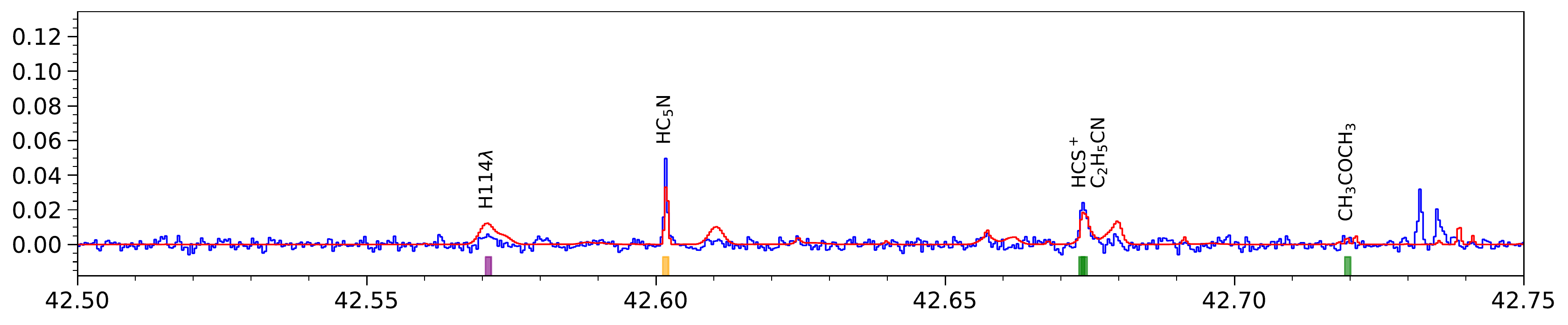}\\
\includegraphics[width=0.99\linewidth]{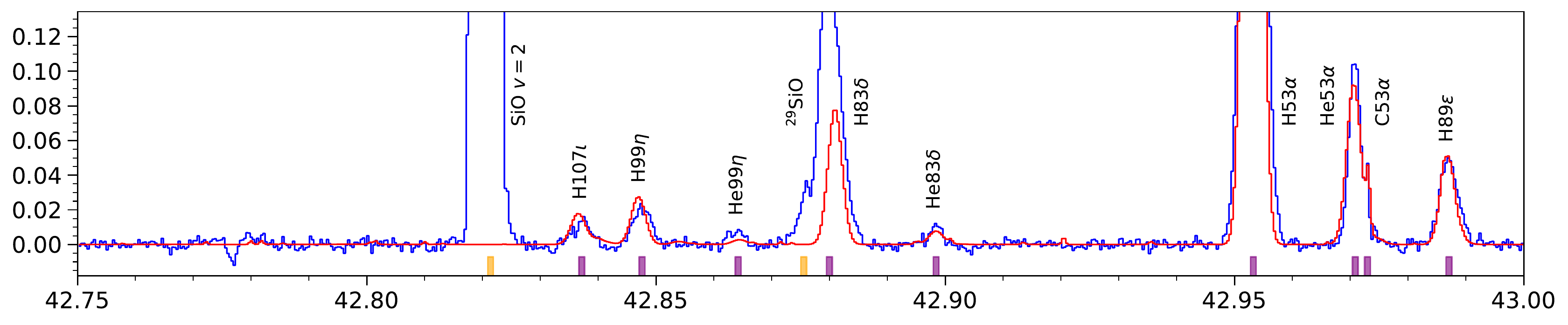}\\
\includegraphics[width=0.99\linewidth]{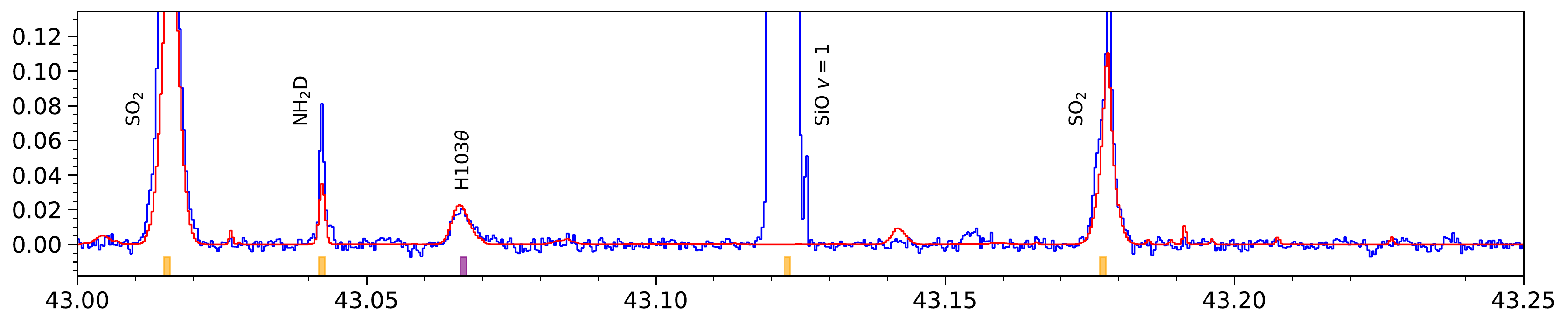}\\
\includegraphics[width=0.99\linewidth]{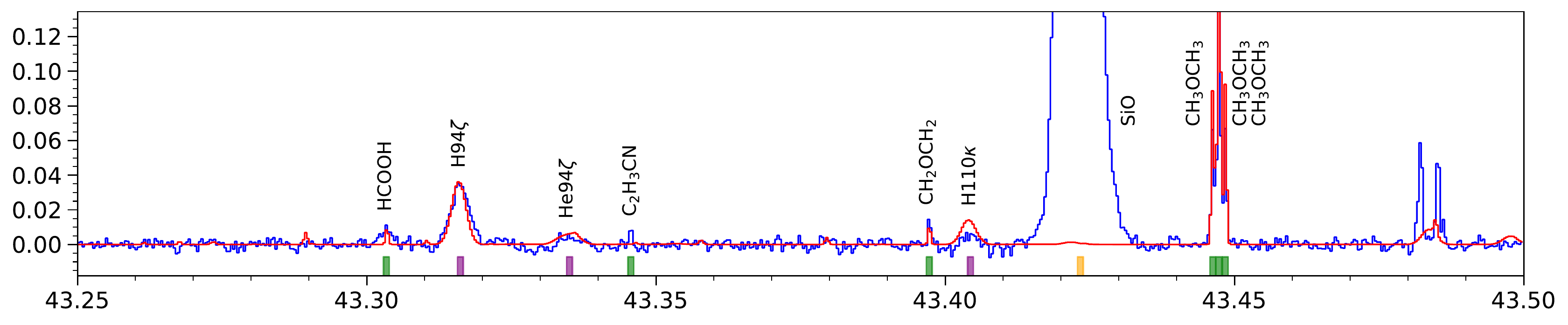}\\
{\raggedright \centering  \textbf{Figure \thefigure} {\it continued} \par}
\end{figure*}         
\begin{figure*}[!htb]
\centering
{\raggedright \centering  \textbf{Figure \thefigure} {\it (continued)} \par}
\includegraphics[width=0.99\linewidth]{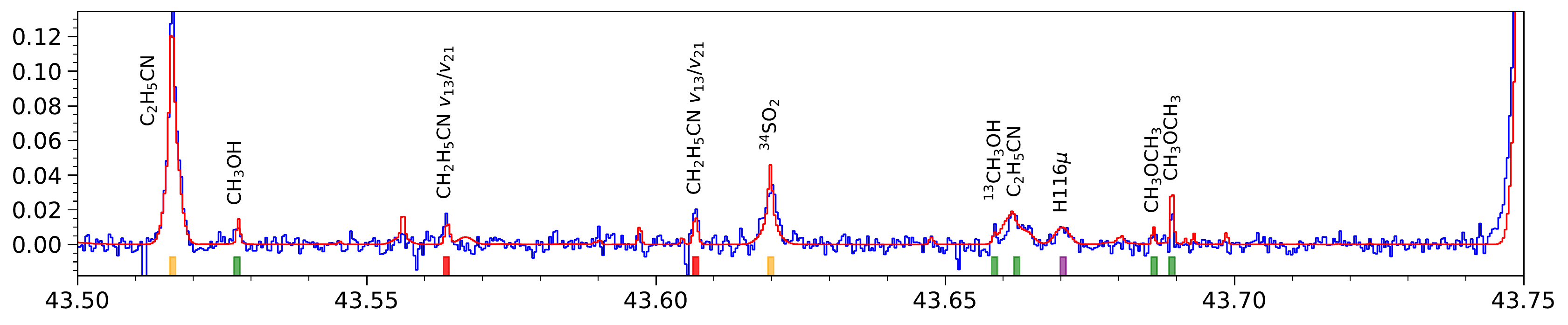}\\
\includegraphics[width=0.99\linewidth]{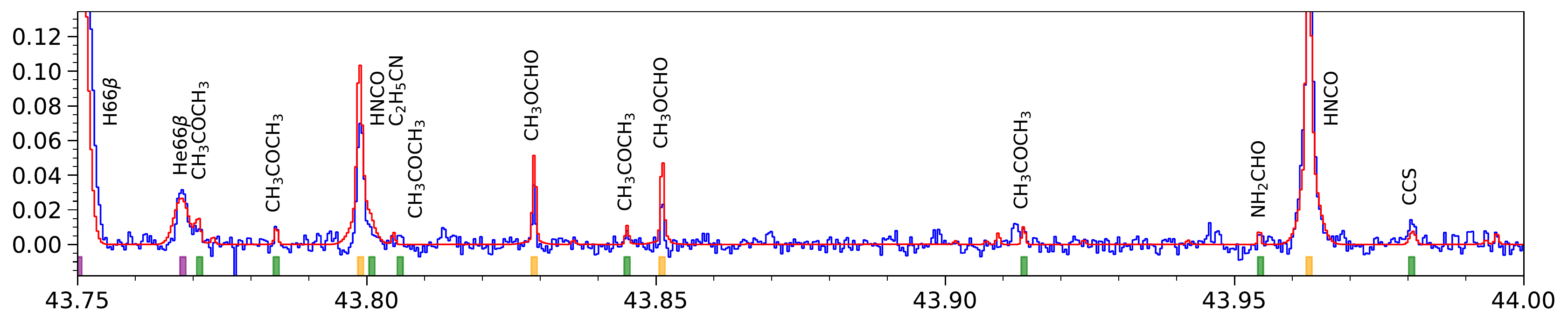}\\
\includegraphics[width=0.99\linewidth]{{44000_44250}.pdf}\\
\includegraphics[width=0.99\linewidth]{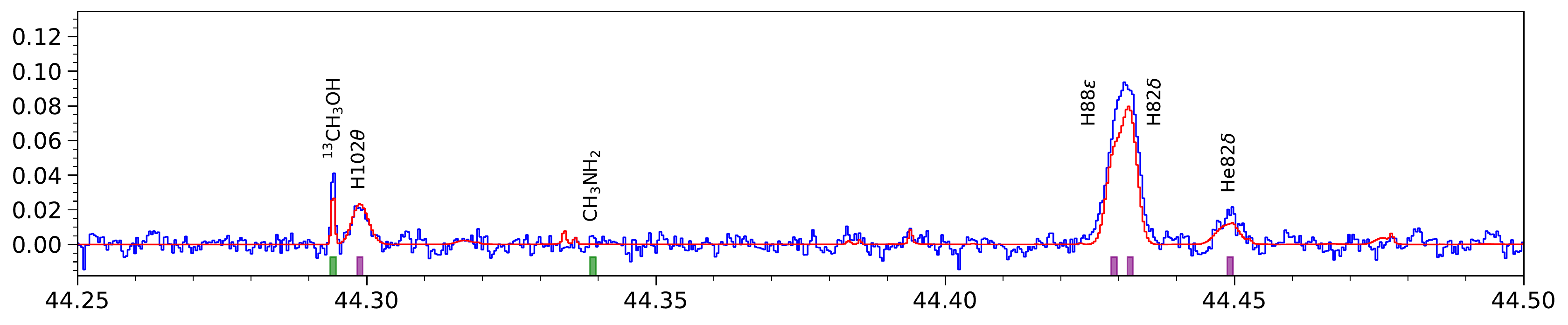}\\
\includegraphics[width=0.99\linewidth]{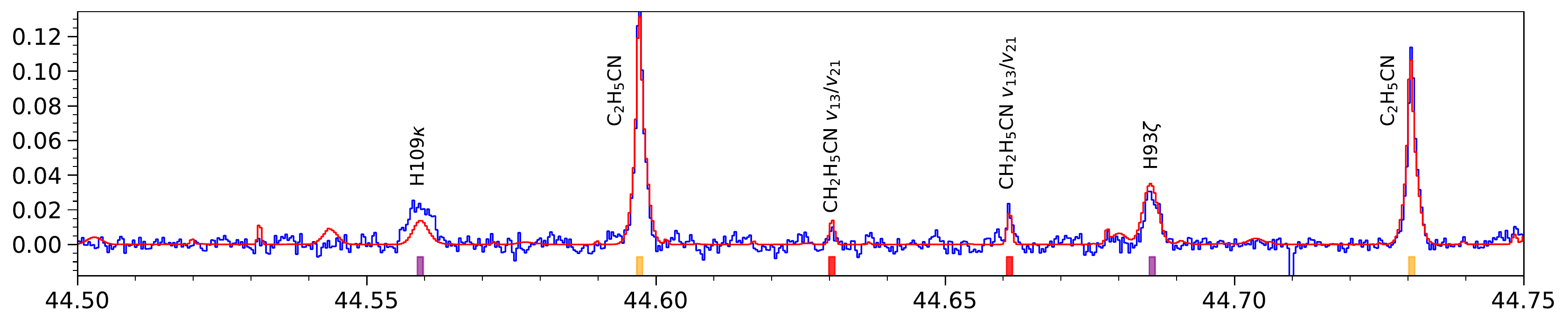}\\
\includegraphics[width=0.99\linewidth]{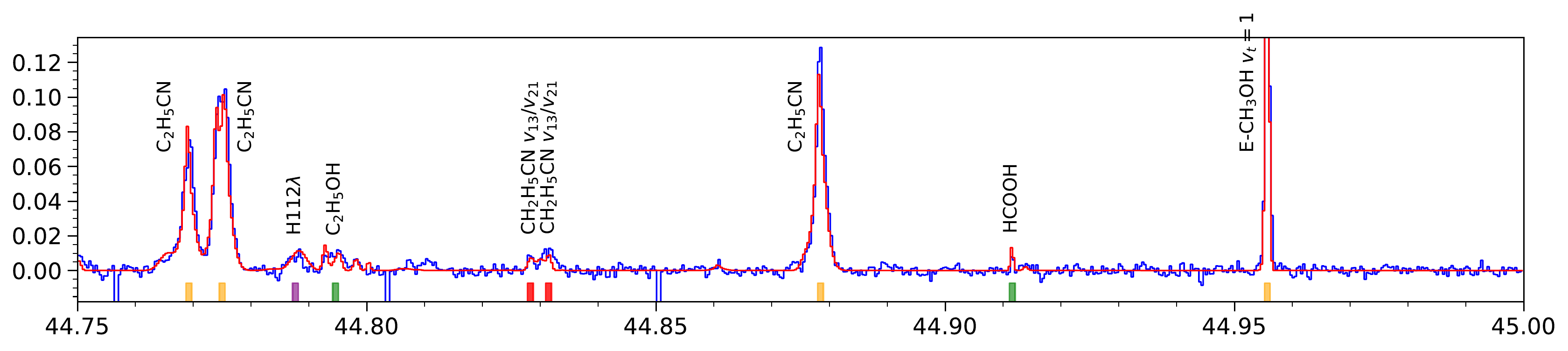}\\
{\raggedright \centering  \textbf{Figure \thefigure} {\it continued} \par}
\end{figure*}         
\begin{figure*}[!htb]
\centering
{\raggedright \centering  \textbf{Figure \thefigure} {\it (continued)} \par}
\includegraphics[width=0.99\linewidth]{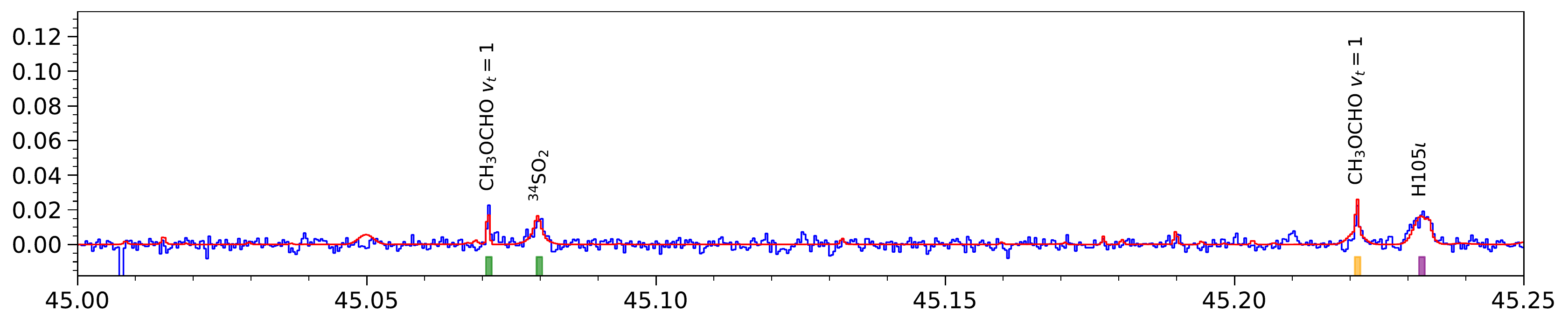}\\
\includegraphics[width=0.99\linewidth]{{45250_45500}.pdf}\\
\includegraphics[width=0.99\linewidth]{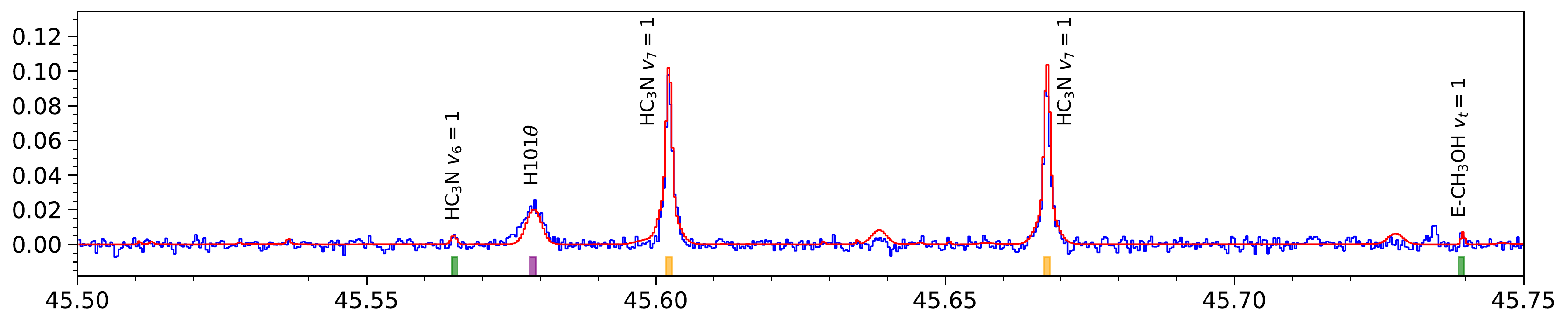}\\
\includegraphics[width=0.99\linewidth]{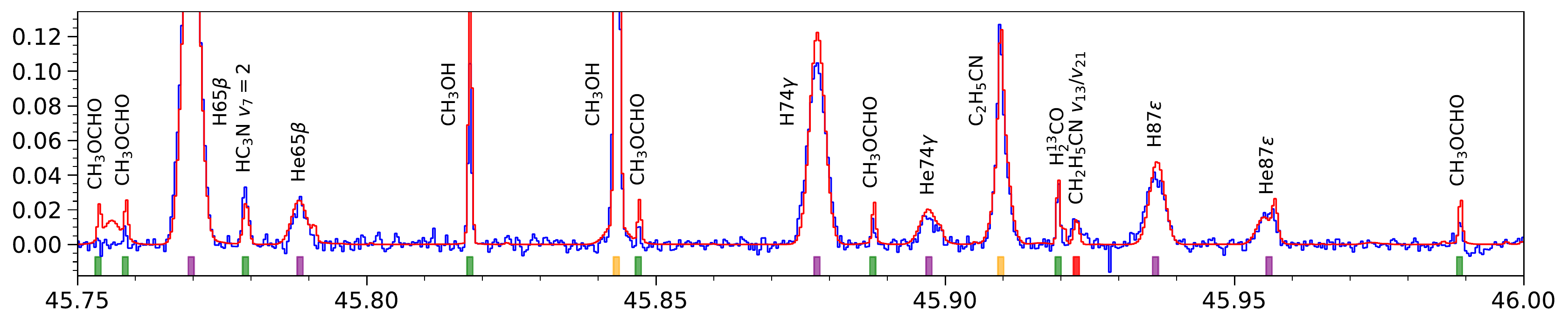}\\
\includegraphics[width=0.99\linewidth]{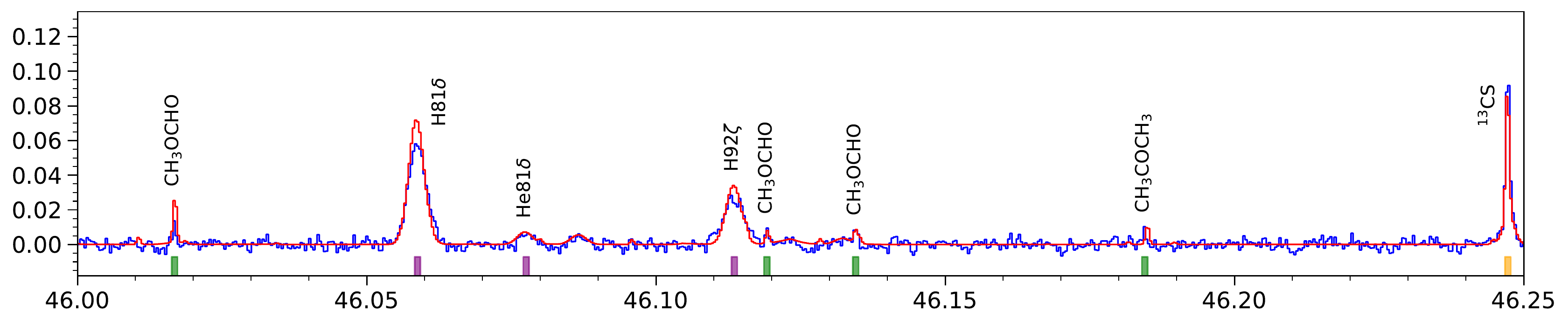}\\
\includegraphics[width=0.99\linewidth]{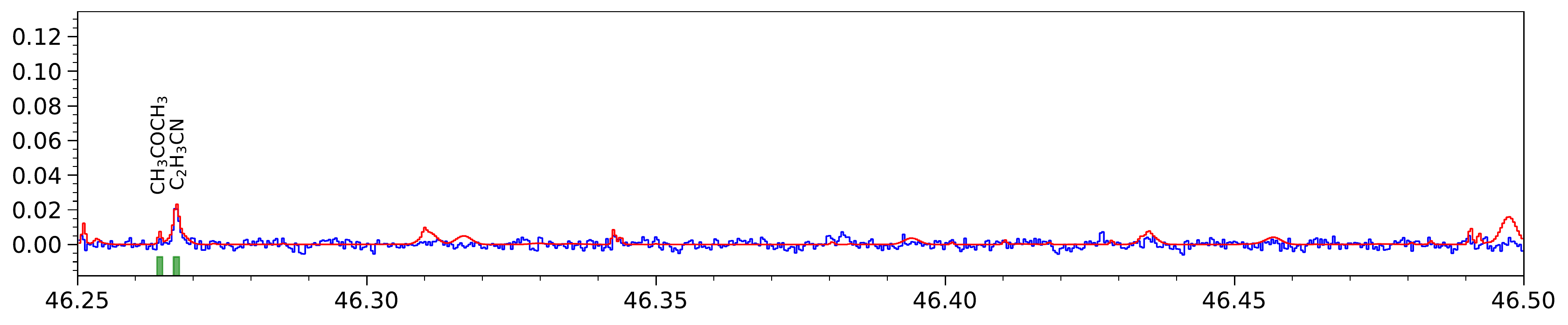}\\
{\raggedright \centering  \textbf{Figure \thefigure} {\it continued} \par}
\end{figure*}        
\begin{figure*}[!htb]
\centering
{\raggedright \centering  \textbf{Figure \thefigure} {\it (continued)} \par}
\includegraphics[width=0.99\linewidth]{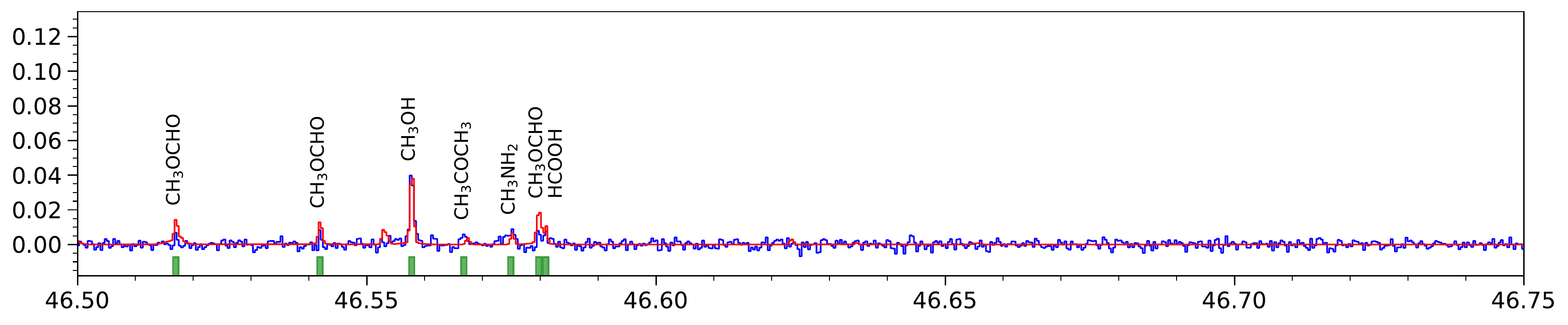}\\
\includegraphics[width=0.99\linewidth]{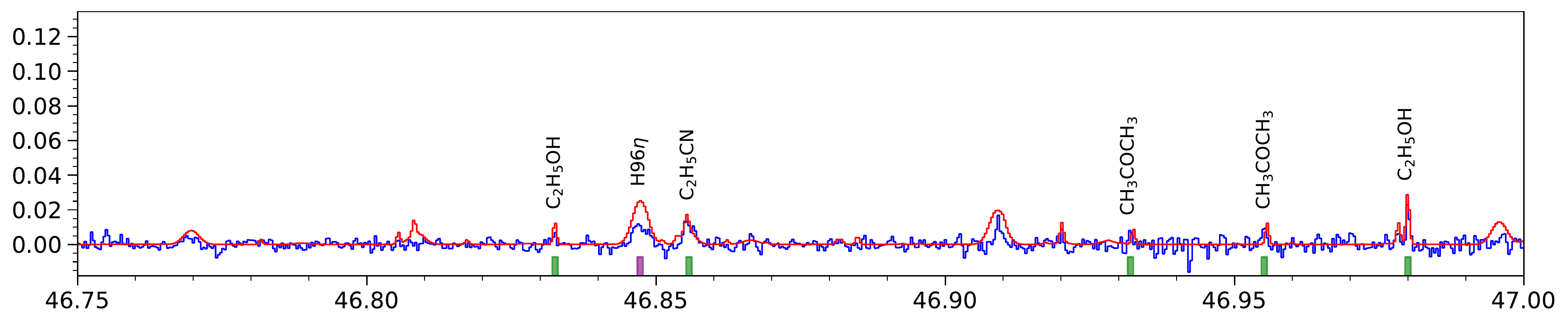}\\
\includegraphics[width=0.99\linewidth]{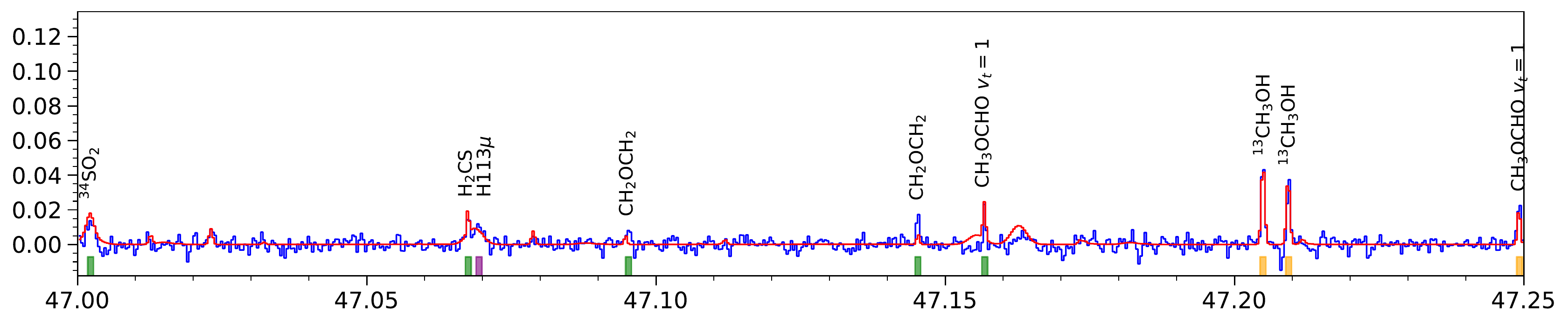}\\
\includegraphics[width=0.99\linewidth]{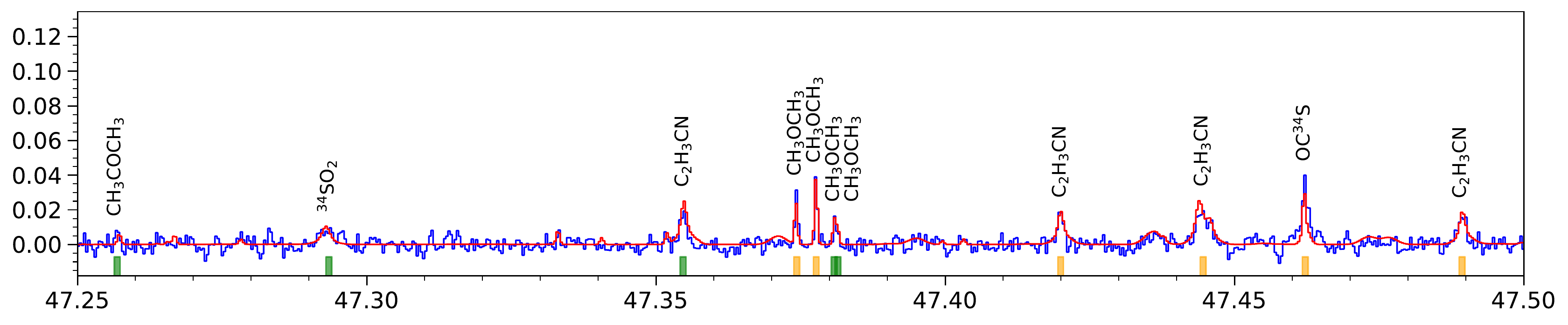}\\
\includegraphics[width=0.99\linewidth]{{47500_47750}.pdf}\\
\includegraphics[width=0.99\linewidth]{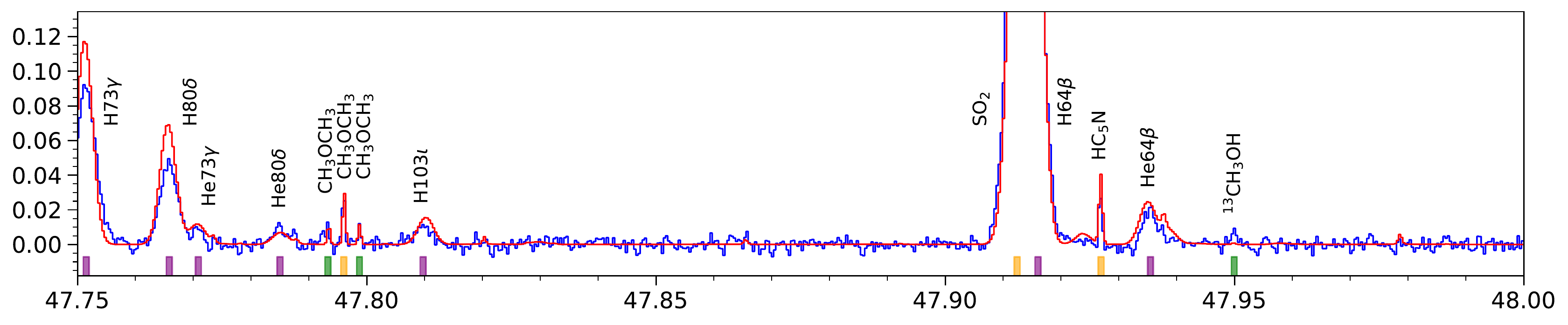}\\
{\raggedright \centering  \textbf{Figure \thefigure} {\it continued} \par}
\end{figure*}         
\begin{figure*}[!htb]
\centering
{\raggedright \centering  \textbf{Figure \thefigure} {\it (continued)} \par}
\includegraphics[width=0.99\linewidth]{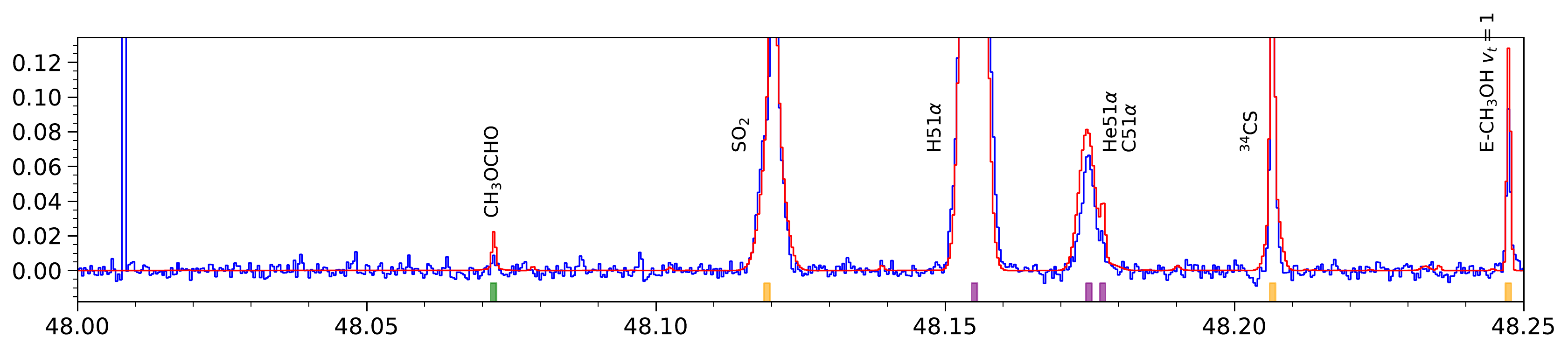}\\
\includegraphics[width=0.99\linewidth]{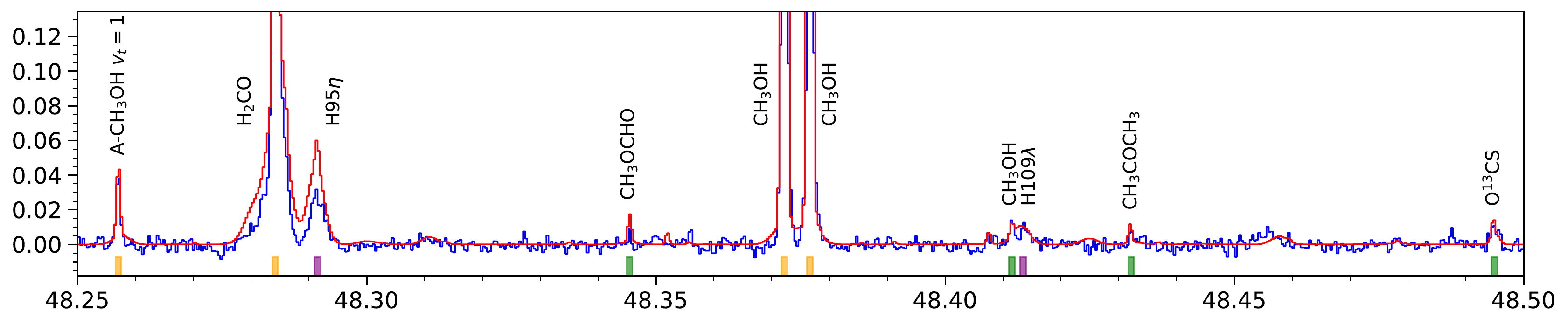}\\
\includegraphics[width=0.99\linewidth]{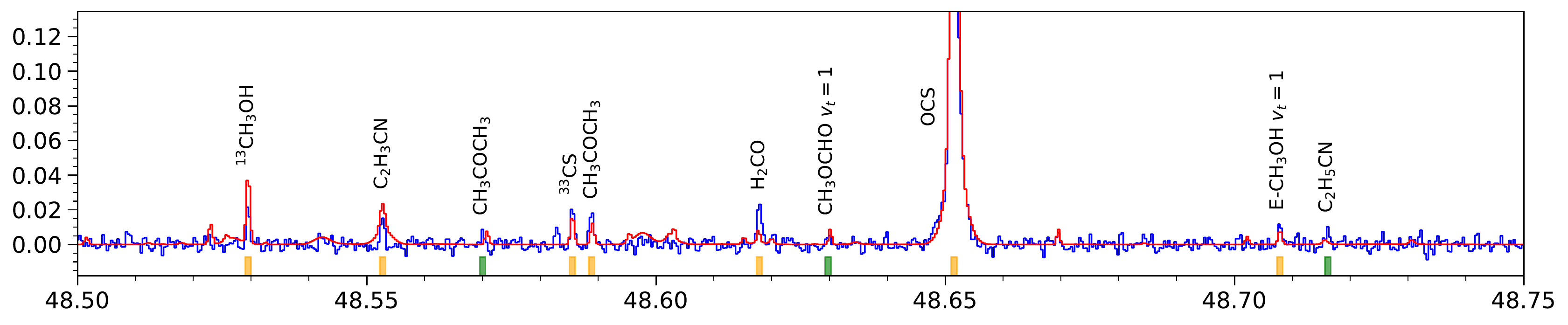}\\
\includegraphics[width=0.99\linewidth]{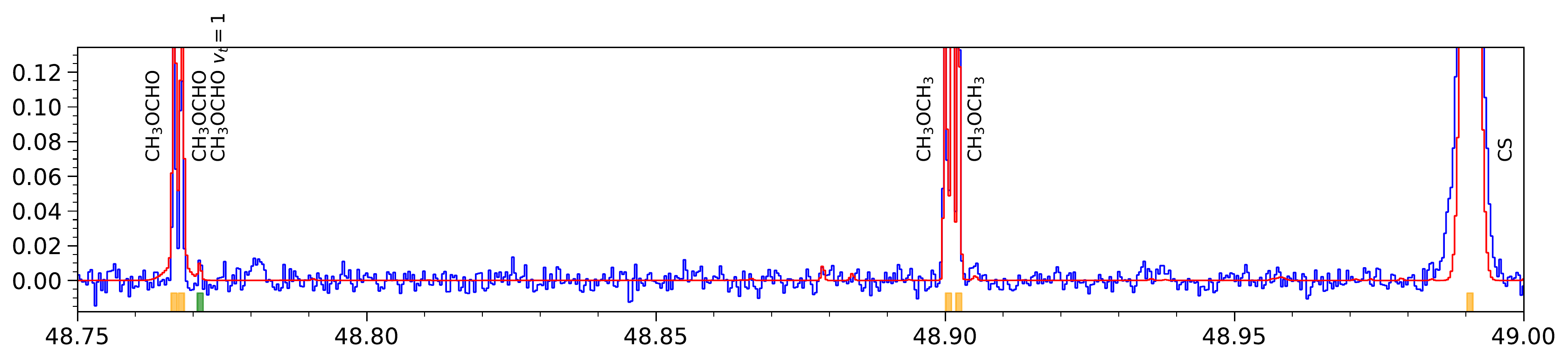}\\
\includegraphics[width=0.99\linewidth]{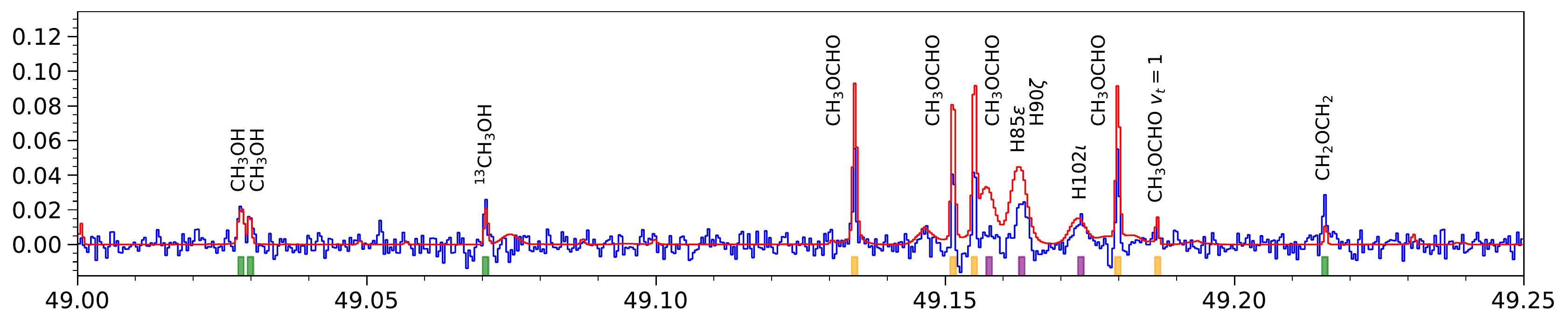}\\
\includegraphics[width=0.99\linewidth]{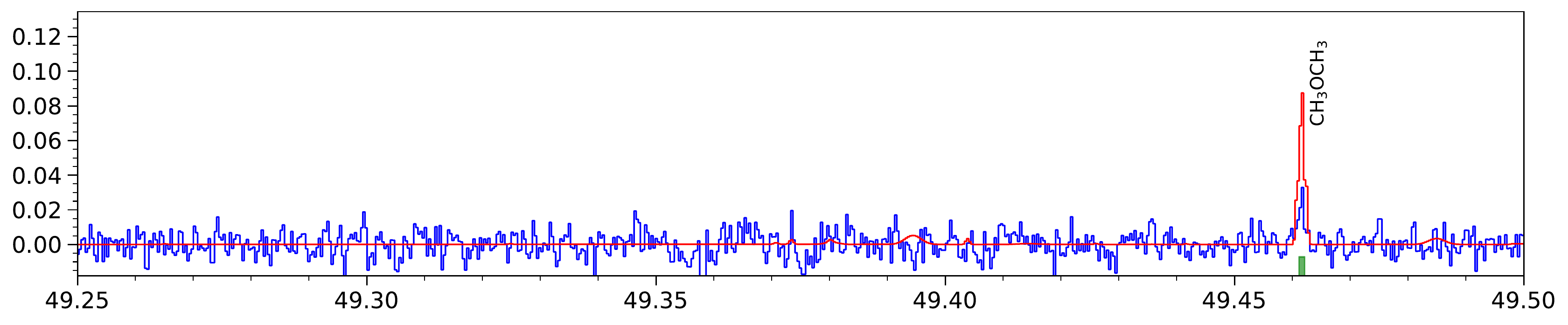}\\
{\raggedright \centering  \textbf{Figure \thefigure} {\it continued} \par}
\end{figure*}         
\begin{figure*}[!htb]
\centering
{\raggedright \centering  \textbf{Figure \thefigure} {\it (continued)} \par}
\includegraphics[width=0.99\linewidth]{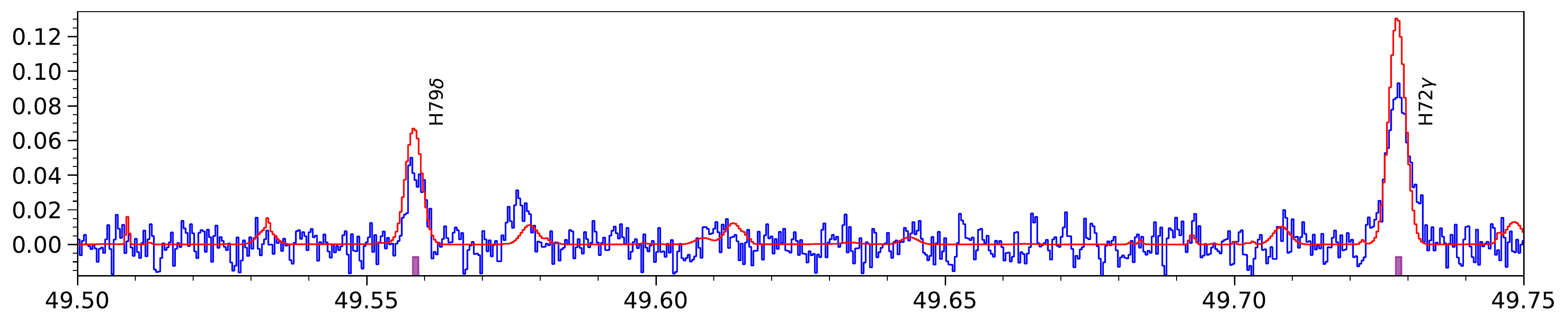}\\
\includegraphics[width=0.99\linewidth]{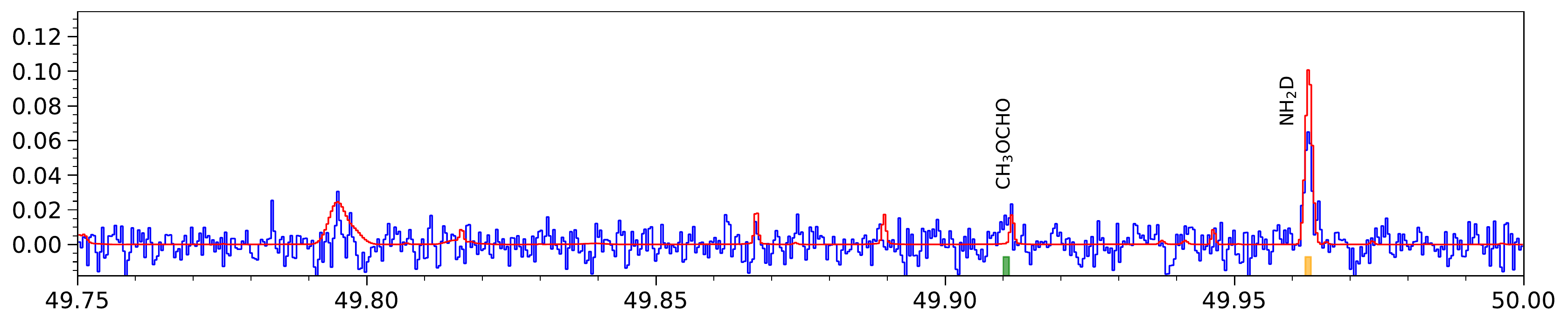}\\
{\raggedright \centering  \textbf{Figure \thefigure} {\it continued} \par}
\end{figure*}

\end{CJK*}
\end{document}